\documentclass[preprint,showpacs,preprintnumbers,amsmath,amssymb]{revtex4}

\usepackage{amsmath}
\usepackage{graphicx}
\usepackage{dcolumn}
\usepackage{bm}

\begin{document}
\title{{\bf Transport Coefficients for the Hard Sphere Granular Fluid}}
\author{Aparna Baskaran}
\altaffiliation[Present address: ]{Physics Department, Syracuse
University, Syracuse, NY 13244} \affiliation{Department of Physics,
University of Florida, Gainesville, FL 32611}
\author{James W. Dufty}
\affiliation{Department of Physics, University of Florida,
Gainesville, FL 32611}
\author{J. Javier Brey}
\affiliation{F\'{\i}sica Te\'{o}rica, Universidad de Sevilla,
Apartado de Correos 1065, E-41080, Sevilla, Spain}
\date{\today }

\begin{abstract}
In the preceding paper, linear response methods have been applied
to obtain formally exact expressions for the parameters of
Navier-Stokes order hydrodynamics. The analysis there is general,
applying to both normal and granular fluids with a wide range of
collision rules. Those results are specialized here to the case of
smooth, inelastic hard spheres with constant coefficient of normal
restitution, for further elaboration. Explicit expressions for the
cooling rate, pressure, and the transport coefficients are given
and compared with the corresponding expressions for a system of
elastic hard spheres. The scope of the results for further
analytical explorations and possible numerical evaluation is
discussed.
\end{abstract}


\maketitle

\section{Introduction}

Granular materials frequently exhibit flows similar to those of
normal fluids, and for practical purposes these flows are often
described by phenomenological hydrodynamic equations
\cite{Fluid,haff}. In the simplest cases, such equations have the
form of Navier-Stokes hydrodynamics with an energy sink representing
collisional ``cooling''. The parameters of these equations (cooling
rate, pressure, and transport coefficients) are unknown in general.
For normal fluids, methods of non-equilibrium statistical mechanics,
e.g. linear response, have been applied to derive formally exact
expressions for these parameters in a form that is suitable for
introducing  approximations. In the preceding paper \cite{DBB06},
this analysis has been extended to granular fluids with similar
results. The objective of the present work is to make those results
more explicit by specializing to the idealized model granular fluid:
a system of smooth, inelastic hard spheres or disks. As has been
shown in ref. \cite{DBB06}, this idealization is a limiting form for
more realistic collisions, where the interaction energy scale, e.g.
energy of deformation, is small compared to characteristic kinetic
energies. This can be controlled by sufficient activation of the
fluid, so the predictions of this model are more than academic
\cite{chemE}. Also, as in the case of simple atomic fluids, this
idealized model of hard particles provides the most tractable
setting for further theoretical analysis of the results obtained in
this work.

A significant advantage of this limit is a scaling of the reference
homogeneous cooling state (HCS) to which linear response is applied.
This scaling property results in several simplifications to the
formalism described in \cite{DBB06}. For example, a stationary
representation for the statistical mechanics of this granular fluid
follows from the use of dimensionless phase space variables, where
the particle velocities are scaled with the cooling thermal velocity
in the HCS ensemble \cite{Diff1,BRyM04}. Furthermore, the
hydrodynamic response functions for small spatial perturbations of
the HCS are given in terms of stationary time correlation functions
in this representation. These time correlation functions are
composed from two biorthogonal sets of observables, referred to here
as the direct and conjugate functions. The direct functions are the
microscopic observables whose ensemble averages are the hydrodynamic
fields. The conjugate functions are generated by functional
derivatives of a \textit{local }HCS with respect to the hydrodynamic
fields. For hard spheres or disks, it is possible to transform these
functional derivatives into ordinary phase space derivatives.
Generally, the simplifications made possible for hard spheres allow
a somewhat more transparent interpretation of the dynamics and the
structure involved in the formal expressions for transport
coefficients. They also provide potentially better access to
evaluation of these expressions using molecular dynamics (MD)
simulations.

The purpose of this companion paper is twofold. The first is to
specialize the results obtained in \cite{DBB06} to the particular
system of inelastic hard particles, to give explicit expressions for
the various hydrodynamic parameters, and to take advantage of this
simplified setting to interpret their content \cite{Ap06}. The
second is to record the various tools of nonequilibrium statistical
mechanics of a system of inelastic hard particles whose utility
transcends the context of their application in this work. Regarding
the first purpose, details of the analysis are suppressed in the
main text, with the reader referred to the preceding paper or to the
Appendices here. Instead, emphasis is placed on the final results,
and similarities or differences between Helfand and Green-Kubo
representations for normal and granular fluid transport
coefficients.

The second purpose is accomplished in most of the eight
Appendices, which summarize the statistical mechanics and dynamics
for inelastic hard spheres. Although much of the formalism appears
in part elsewhere, there are many new results as well. In addition
to the generators for dynamics of phase space observables and the
Liouville equation, the generator for time reversed dynamics is
identified. Also, the microscopic conservation laws for both the
forward and time reversed dynamics are given, and the two sets of
different fluxes are identified. Although straightforward to
obtain, these results do not appear in the literature even for
elastic hard spheres. Similarly, although the Helfand
representation for transport coefficients has been used in MD
simulations of elastic hard spheres, the corresponding Green-Kubo
expressions have not been discussed explicitly until recently
\cite {Dufty, ErnstBrito}. All such elastic hard sphere
coefficients are given in detail in Appendix \ref{ap6} using the
more general granular fluid formalism developed here.
Dimensionless forms of the Liouville equation, including one
suitable for MD simulation \cite{lutsko1,Kandrup}, the microscopic
conservation laws, and a new set of balance equations for the
conjugate functions noted above are derived in these Appendices.
All of these results are critical for interpreting the exact
expressions given here for the transport coefficients, but are
also the building blocks for more general future investigations of
granular fluid properties using the non-equilibrium statistical
mechanics of hard spheres.

The layout of this presentation is as follows. In the next section,
the primary results of the previous paper are specialized for the
case of inelastic hard particles whose collisional loss in energy is
characterized by a constant coefficient of normal restitution
$\alpha $. Only a broad outline of the results is given in the main
text, with the details of the dynamics and statistical mechanics of
hard spheres recorded in the Appendices. Then, explicit expressions
for the various hydrodynamic parameters are given. The cooling rate
and pressure are obtained as averages over the HCS, which also can
be expressed in terms of integrals over the associated reduced two
particle distribution function. The transport coefficients are given
both Helfand \cite{HelfandPap} and Green-Kubo \cite{McLbook}
representations, in terms of the long time limit of appropriate
stationary time correlation functions for the HCS. Next, the
relationship of these results to those obtained from kinetic
theories is sketched briefly, as well as the possibility of
numerical evaluation of the results using molecular dynamics (MD)
simulation. Finally the key points in this work are summarized and
some concluding remarks are made.

\section{Linear Response}
\label{s2}

The microscopic model considered here is a system of smooth
mono-disperse hard spheres ($d=3$) or disks ($d=2$) of mass $m$ and
diameter $\sigma$, whose collisional loss in energy is characterized
by a constant coefficient of normal restitution $\alpha $. The
details of the model, the statistical mechanics of this system, the
generators of dynamics, and time correlation functions are given in
Appendix \ref{ap1}. In this section, the simplifications to the
linear response theory developed in the previous paper, due to this
choice of collision model, are identified and the results there
translated to the case at hand.

\subsection{Homogeneous Cooling State}

The reference state with respect to which the linear response of
this system is studied, is the homogeneous hydrodynamic state whose
entire time dependence arises due to the cooling temperature. This
is the so-called homogeneous cooling state (HCS). First this
reference state is characterized in the context of both the
phenomenological macroscopic hydrodynamics, and the more fundamental
statistical mechanics. At the level of hydrodynamics, this state is
described by a solution with constant number density $n_{h}$ and
flow velocity ${\bm U}_{h},$ and a homogeneous but time dependent
temperature $T_{h}(t)$, that obeys the hydrodynamic equation
\begin{equation}
\left\{ \partial _{t}+\zeta _{0}\left[ n_{h},T_{h}\left( t\right)
\right] \right\} T_{h}\left( t\right) =0,  \label{1.1}
\end{equation}
where $\zeta _{0}\left[ n_{h},T_{h}\left( t\right) \right] $ is a
cooling rate that must be specified from the microscopic theory.
By means of a Galilean transformation, it is always possible to
consider ${\bm U}_{h}={\bm 0}$. For the case of inelastic hard
spheres or disks considered here, there is no microscopic energy
scale associated with the collision model. Therefore, the
temperature dependence of the cooling rate can be determined by
dimensional arguments as $\zeta _{0}\left[ n_{h},T_{h}\left(
t\right) \right] \propto T_{h}^{1/2}\left( t\right) $. Equation
(\ref{1.1}) then can be integrated to obtain the time dependence
of the temperature,
\begin{equation}
T_{h}\left( t\right) =T_{h}(0)\left[ 1+\frac{v_{0}(0)\zeta
_{0}^{\ast }t}{2l} \right] ^{-2} \label{1.1b}
\end{equation}
with
\begin{equation}
\zeta _{0}^{\ast } \equiv \frac{l \zeta_{0}}{v_{0}(t)}, \quad
v_{0}(t) \equiv \left[ \frac{2T_{h}(t)}{m} \right]^{1/2}
\end{equation}
being a dimensionless cooling rate and a ``thermal velocity'',
respectively. Moreover, $l $ is an appropriate length scale; for
example, the mean free path. Equation (\ref{1.1b}) is the familiar
Haff's cooling law \cite{haff} that is well established for
inelastic hard sphere fluids as one of the signatures of the HCS.

At the level of statistical mechanics, the reference ensemble
corresponding to this macrostate, the HCS ensemble, is given by a
homogeneous ``normal'' solution to the Liouville equation, i.e., one
whose entire time dependence occurs through the cooling temperature.
Furthermore, the absence of any additional microscopic energy scale
for hard spheres, implies that this temperature dependence can occur
only through the scaling of the particle velocities ${\bm v}_{r}$
with the thermal velocity $v_{0}\left[ T_{h}\left( t\right) \right]
$,
\begin{equation}
\rho _{h}\left[ \Gamma;  T_{h}\left( t\right) \right]) =\left[l
v_{0}\left( t\right)\right])^{-Nd}\rho _{h}^{\ast }\left( \left\{
\frac{{\bm q}_{rs}}{l }, \frac{{\bm v}_{r}-{\bm U}_{h}}{v_{0}(t)};
r,s=1, \dots , N\right\}\right),  \label{1.2}
\end{equation}
where $ \Gamma \equiv \left\{ {\bm q}_{r}, {\bm v}_{r}; r=1, \ldots,
N \right\}$ is a point in the phase space of the system and ${\bm
q}_{rs}\equiv {\bm q}_{r}-{\bm q}_{s}$ the relative position of
particle $r$ with respect to particle $s$. This special form of the
$N$-particle distribution function allows the temperature dependence
of many average properties such as the cooling rate, pressure, and
transport coefficients, to be determined without explicit
calculation. Another interesting consequence of the scaling nature
of the HCS ensemble is its restriction to a constant total momentum
surface in phase space,
\begin{equation}
\rho _{h}\left[ \Gamma; n_{h}, T_{h}(t) \right] =\delta \left[
\frac{{\bm P}-mN {\bm U}_{h}}{ mv_{0}\left( t\right) }\right] [\ell
v_{0}\left( t\right) ]^{-Nd}\overline{ \rho }_{h}^{\ast } \left(
\left\{ \frac{{\bm q}_{rs}}{l }, \frac{{\bm v}_{r}-{\bm
U}_{h}}{v_{0}(t)}; r,s=1, \dots , N\right\}\right). \label{1.3.4}
\end{equation}
Here $\delta (x)$ is the Dirac delta function and ${\bm P} =
\sum_{r} m {\bm v}_{r}$ the total momentum. The proof of this result
is given in Appendix \ref{ap2}.

It is useful to represent the dynamics more generally for other
states in terms of the same dimensionless scaled variables as occur
in the HCS, i.e. in terms of $\Gamma ^{\ast } \equiv \left\{ {\bm
q}_{r}^{\ast} ,{\bm v}_{r}^{\ast}; r=1, \ldots,N \right\} $, with
${\bm q}^{\ast}_{r}={\bm q}_{r}/ l$ and ${\bm v}^{\ast}_{r}={\bm
v}_{r}/v_{0}(t)$, where $v_{0}(t)$ is defined from the temperature
of a reference HCS state having the same intial total energy as the
system under consideration. In this way, it is seen that the HCS is
a stationary solution to the dimensionless Liouville equation, and
the HCS time correlation functions also become stationary. Thus,
this representation for the statistical mechanics, allows the
problem of linear response about a non-equilibrium time dependent
reference state to be mapped onto one where the reference is
stationary, in closer correspondence to the equilibrium case for
normal fluids. More details of this transformation are given in
Appendix \ref{ap2}.

\subsection{Hydrodynamic Response}

In this subsection, the response of the fluid to small spatial
perturbations about the reference state described above is
considered. The macroscopic variables of interest are the
dimensionless deviations, $\delta y^{\ast}_{\alpha}$, of the
hydrodynamic fields, $y_{\alpha}$, from their HCS values,
$y_{\alpha,h}$. They are defined by
\begin{equation}
\left\{ \delta y_{\alpha }^{\ast } \right\} \equiv \left\{ \delta
n^{\ast },\delta T^{\ast },\delta {\bm U}^{\ast }\right\},
\end{equation}
\begin{equation}
\delta n^{\ast } \equiv  \frac{\delta n}{n_{h}} =
\frac{n-n_{h}}{n_{h}}, \quad \delta T^{\ast } \equiv \frac{\delta
T}{T_{h}(t)}= \frac{ T-T_{h}\left( t\right) }{T_{h}\left( t\right)
}, \quad \delta {\bm U}^{\ast } \equiv \frac{\delta {\bm
U}}{v_{0}(t)} =\frac{{\bm U}-{\bm U}_{h}}{v_{0}\left( t\right) }\,
, \label{1.3.1}
\end{equation}
where $n({\bm r},t)$, $T({\bm r},t)$, and ${\bm U}({\bm r},t)$
denote respectively the number density, temperature, and flow
velocity fields. The dynamics of these hydrodynamic fields is given
by a response equation of the form
\begin{equation}
\delta \widetilde{y}_{\alpha }^{\ast }\left({\bm
k}^{\ast},s\right) =\sum_{\beta} \widetilde{ C}_{\alpha \beta
}^{\ast }\left( {\bm k}^{\ast},s\right) \delta \widetilde{y}
_{\beta }^{\ast }\left( {\bm k}^{\ast},0\right).  \label{1.3.2}
\end{equation}
A Fourier representation in the reduce space variable has been
introduced to take advantage of the homogeneity of the reference
state. Also, the dimensionless time scale $s$ defined through
\begin{equation}
ds=\frac{ v_{0}\left( t\right) dt }{l}
\end{equation}
is used. The matrix of response functions $\widetilde{C}^{\ast}$,
gives the effect of weak spatial perturbations of the hydrodynamic
fields on the dynamics of this fluid. Such response functions can be
identified in two ways \cite{DBB06}: either from the linearized
phenomenological hydrodynamic equations or from non-equilibrium
statistical mechanics. Equating the two in their common domain of
validity (large space and time scales), allows identification of
exact expressions for the parameters of the phenomenological
hydrodynamic equations.

The solution to the dimensionless, linearized phenomenological hydrodynamic
equations gives
\begin{equation}
\widetilde{C}^{\ast hyd}\left( {\bm k}^{\ast},s\right) = e^{-s
\mathcal{K}^{\ast hyd }\left( {\bm k}^{\ast} \right) }.
\label{1.4.1}
\end{equation}
If the components of the flow velocity are chosen to be the
longitudinal component relative to ${\bm k}^{\ast}$, $\delta
\widetilde{U}^{\ast}_{\parallel} \equiv {\bm k}^{\ast} \cdot \delta
\widetilde{\bm U}^{\ast}/k^{\ast}$ and $d-1$ orthogonal transverse
components, $\delta \widetilde{U}^{\ast}_{\perp i}$, the
dimensionless transport matrix $\mathcal{K}^{\ast hyd }$ turns is
block diagonal,
\begin{equation}
\mathcal{K}^{\ast hyd}= \left(
\begin{array}{cc}
\mathcal{K}^{\ast hyd}_{1} & 0 \\
0 & \mathcal{K}^{\ast hyd}_{2}
\end{array}
\right). \label{1.4.1a}
\end{equation}
Here $\mathcal{K}_{1}^{\ast hyd}$ is the ``longitudinal'' part,
corresponding to $\left\{ \delta \widetilde{n}^{\ast}, \delta
\widetilde{T}^{\ast}, \delta \widetilde{U}_{\parallel}^{\ast}
\right\}$, and it is given by
\begin{equation}
\mathcal{K}^{ \ast hyd }\left( {\bm k}^{\ast }\right) =\left(
\begin{array}{ccc}
0 & 0 & -ik^{\ast} \\
\zeta _{0}^{\ast }\frac{\partial \ln \zeta _{0}}{\partial \ln
n_{h}}+\left( \frac{2 \mu ^{\ast }}{d}-\zeta^{ \ast n}\right)
k^{\ast 2} & \frac{\zeta _{0}^{\ast }}{2}+\left( \frac{2 \lambda
^{\ast }}{d}-\zeta^{\ast T }\right) k^{\ast 2} & - i \left(
\frac{2 p_{h}^{\ast }}{d} +\zeta^{ \ast U }\right)
k^{\ast } \\
-\frac{i p_{h}^{\ast }}{2} \frac{\partial \ln p_{h}}{\partial \ln
n_{h}} k^{\ast } & -\frac{i p^{\ast}_{h}}{2}\, k^{\ast} &
-\frac{\zeta _{0}^{\ast }}{2}+\left[ \frac{2(d-1)}{d}\eta ^{\ast
}+\kappa ^{\ast }\right] k^{\ast 2}
\end{array}
\right),  \label{1.4.2}
\end{equation}
while $\mathcal{K}^{\ast hyd}_{2}$ is associated to the
``transverse'' components, $\delta \widetilde{U}_{\perp i}$, which
are decoupled from the former,
\begin{equation}
\mathcal{K}^{\ast hyd}_{2} ({\bm k}^{\ast})= \left( -\frac{\zeta
_{0}^{\ast }}{2}+\eta ^{\ast }k^{\ast 2} \right) I,\label{1.4.3}
\end{equation}
where $I$ is the unit matrix of dimension $d-1$.  The
dimensionless parameters and transport coefficients in the above
equations are defined by
\begin{equation*}
p^{\ast }_{h}=\frac{p_{h}}{n_{h}T_{h}}, \quad   \zeta _{0}^{\ast
}=\frac{l \zeta_{0} }{v_{0}},  \quad \lambda ^{\ast
}=\frac{\lambda}{l n_{h} v_{0} },
\end{equation*}
\begin{equation*}
\mu ^{\ast }=\frac{\mu}{l T_{h} v_{0}},  \quad \eta ^{\ast
}=\frac{\eta}{ mn_{h}l v_{0}},  \quad \kappa ^{\ast
}=\frac{\kappa}{mn_{h} l  v_{0}},
\end{equation*}
\begin{equation}
\zeta^{\ast U }=\zeta^{U}, \quad  \zeta^{\ast n }=\frac{n_{h}
\zeta^{n}}{ l v_{0}}, \quad \zeta^{ \ast T }=\frac{T_{h} \zeta^{T}}{
l v_{0}}. \label{1.4.5}
\end{equation}
Above, $p_{h}$ is the pressure of the granular fluid, $\zeta _{0}$
is the cooling rate, $\lambda $ is the thermal conductivity, $\mu $
is the new transport coefficient associated with granular fluids
that measures the contribution of density gradients to the heat
flux, $\eta $ is the shear viscosity, and $\kappa $ is the bulk
viscosity. Finally, $\zeta^{U}$, $ \zeta^{n},$ and $\zeta^{T}$ are
transport coefficients arising from the local cooling rate. The
phenomenological definition of all these quantities has been given
in ref. \cite{DBB06}. Upon introducing the associated dimensionless
quantities in Eq.\ (\ref{1.4.5}), advantage has been taken of the
already mentioned fact that the temperature dependence of all the
parameters in the hydrodynamic equations can be determined by
dimensional arguments, as a consequence of being dealing with hard
particles.

The dimensionless transport matrix $\mathcal{K}^{\ast hyd}$ is a
constant independent of time $s$ and, consequently, the response
function $\widetilde{C}^{\ast}$ is related to the transport matrix
by the simple exponential form given in Eq. (\ref{1.4.1}). This is
one of the primary simplifications that occur for this hard sphere
or disk system. Further, the eigenvalues of the $\mathcal{K}^{\ast
hyd}$ matrix give the hydrodynamic modes of the system, and its
eigenfunctions identify the particular excitations of the
hydrodynamic fields that give rise to these modes. These can be used
to get physical insight into the hydrodynamic response of this
fluid. Some of these considerations are addressed in Appendix
\ref{ap3}.

Alternatively, an exact response equation of the form of Eq.
(\ref{1.3.2}), can be identified by carrying out a linear response
analysis at the level of statistical mechanics of the system. In
this way, the response function $\widetilde{C}^{\ast}$ is found to
be a matrix of time correlation functions of the form
\begin{equation}
\widetilde{C}_{\alpha \beta }^{\ast }\left( {\bm k}^{\ast} ;s\right)
=V^{\ast -1} \int d\Gamma ^{\ast }\, \widetilde{a}_{\alpha }^{\ast
}\left( \Gamma^{\ast}; {\bm k}^{\ast} \right) e^{-s
\overline{\mathcal{L}}^{\ast }} \widetilde{\psi }_{\beta }^{\ast
}\left(\Gamma^{\ast}, -{\bm k}^{\ast} \right) . \label{1.5.1}
\end{equation}
Here, $V^{\ast}$ is the volume of the system in reduced units and
the ``direct'' functions $\widetilde{a}_{\alpha }^{\ast }$ are
dimensionless Fourier transforms of linear combinations of the
microscopic densities whose ensemble average gives the hydrodynamic
fields,
\begin{equation}
\widetilde{a}_{\alpha}^{\ast}({\bm k}^{\ast})= l^{-d} \int d {\bm
r}\, e^{i {\bm k} \cdot {\bm r}} a^{\ast}_{\alpha}({\bm r}) =\int
d{\bm r}^{\ast}\, e^{i {\bm k}^{\ast} \cdot {\bm r}^{\ast}}
a^{\ast}_{\alpha}({\bm r}), \label{1.5.1a}
\end{equation}
\begin{equation}
\left\{ a_{\alpha }^{\ast } \right\} \equiv  \left\{ {\mathcal
N}^{\ast },\frac{2}{d} \left( {\mathcal E}^{\ast
}-\frac{d}{2}{\mathcal N}^{\ast }\right) , {\bm {\mathcal G}}^{\ast
}\right\} , \label{1.5.2}
\end{equation}
where
\begin{equation}
\left\{ {\mathcal N}^{\ast },{\mathcal E}^{\ast }, {\bm  {\mathcal
G}}^{\ast }\right\}  \equiv \left\{ \frac{{\mathcal N}(\Gamma; {\bm
r})}{n_{h}} ,\frac{{\mathcal E}(\Gamma;{\bm r})}{n_{h}T_{h}(t)} ,
\frac{{\bm {\mathcal G}} (\Gamma;{\bm r})}{n_{h}m v_{0}(t)}
\right\}. \label{1.5.3}
\end{equation}
The phase functions ${\mathcal N}$, ${\mathcal E}$, and ${\bm
{\mathcal G}}$ are the microscopic number density, energy density
and momentum density, respectively. Their mathematical expressions
are given in  ref. \cite{DBB06}. The ``conjugate'' functions $
\widetilde{\psi }_{\beta }^{\ast } $ in Eq.\ (\ref{1.5.1}) are
generated by functional derivatives of the local HCS ensemble
$\rho_{lh} \left[ \Gamma|\left\{ y_{\alpha}\right\} \right] $ as
also described in \cite{DBB06},
\begin{equation}
\widetilde{\psi }_{\beta }^{\ast }\left( \Gamma^{\ast}; {\bm
k}^{\ast} \right) = M_{\beta } \int d {\bm r}\, e^{i{\bm k}\cdot
{\bm r}} \left[ \frac{\delta \rho _{lh}\left[ \Gamma |\left\{
y_{\alpha } \right\} \right]}{\delta y_{\beta }\left( {\bm r}
\right) }\right]_{\{ y_{\alpha}\}=\left\{ n_{h},T_{h}(t),{\bm
0}\right\}}. \label{1.5.4}
\end{equation}
In the above expression,  the  $M_{\beta }$'s are factors that
render the functions $ \psi _{\beta }^{\ast } $ dimensionless,
namely
\begin{equation}
M_{1 }=\left[ l v_{0}\left( T_{h} \right) \right]^{dN}n_{h} , \quad
M_{2}= \left[ l v_{0}\left( T_{h} \right) \right]^{dN}T_{h}, \quad
M_{\parallel}=M_{\perp i}= \left[ l v_{0}\left( T_{h} \right)
\right]^{dN}v_{0}\left( T_{h} \right).  \label{1.5.N}
\end{equation}
The local HCS inherits some of the scaling nature of the true HCS
ensemble, and is identified as \cite{DByB06}
\begin{equation}
\rho _{l h}\left( \Gamma |\left\{ y_{\alpha } \right\} \right)=
\left\{ \prod_{r=1}^{N} \left[ l v_{0}({\bm q}_{r})
\right]^{-d}\right\} \rho _{h}^{\ast }\left( \left\{ \frac{{\bm
q}_{rs}}{l },\frac{{\bm v}_{r}- {\bm U}\left( {\bm q}_{r}\right)
}{v_{0}({\bm q}_{r})} \right\} | n  \right), \label{1.5.LHCS}
\end{equation}
where $v_{0}({\bm q}_{r}) \equiv v_{0}[T({\bm q}_{r})]$. The
functional dependence on the density field is obtained by
considering the HCS in an external inhomogeneous field and inverting
the relation giving the density profile \cite{DByB06}. Clearly, for
uniform hydrodynamic fields this becomes the HCS of Eq.\
(\ref{1.2}).

The generator of the dynamics $\overline{\mathcal{L}}^{\ast }$ in
Eq.\, (\ref{1.5.1}), is a composition of the dimensionless
Liouville operator $\overline{L} ^{\ast }$ that generates the
trajectories in phase space plus a velocity scaling operator,
\begin{equation}
\overline{\mathcal{L}}^{\ast }=\overline{L}^{\ast }
+\frac{\zeta_{0}^{\ast}}{2}\sum_{r=1}^{N} \frac{\partial}{\partial
{\bm v}^{\ast}_{r}} \cdot {\bm v}^{\ast}_{r},   \label{1.5.5}
\end{equation}
\begin{equation}
\overline{L}^{\ast}= \frac{l}{v_{0}(t)} \overline{L} = [L]_{\{
{\bm q}_{r}={\bm q}^{\ast}_{r}, {\bm v}_{r}={\bm v}^{\ast}_{r} \}
}.
\end{equation}
The Liouville operator $\overline{L}$ associated with this system of
hard particles is identified explicitly in Appendix \ref{ap1}. The
second operator on the right hand side of Eq.\ (\ref{1.5.5})
rescales the velocities with an $ s $ dependence associated with the
thermal velocity, and represents the effects of cooling in the
reference HCS. This simplified form of the modified generator of
dynamics in the time correlation functions is another special
feature of the hard sphere collision model, or equivalently, of the
absence of any other microscopic energy scale.

Using the form of the local HCS given in Eq.\ (\ref{1.5.LHCS}),
the functional derivatives with respect to the temperature and
flow velocity in Eq.(\ref{1.5.4}), can be simplified to the forms
\begin{eqnarray}
\widetilde{\psi }_{2}^{\ast }\left(\Gamma^{\ast};{\bm k}\right)
&=&M_{2}\int d{\bm r }e^{i{\bm k}\cdot {\bm r}}\sum_{s=1}^{N}
\left\{ \frac{\partial \rho _{lh} \left[ \Gamma | \left\{
y_{\alpha} \right\} \right] }{\partial T\left( {\bm q}_{s}\right)
}\right\} _{ \left\{ y_{\alpha} \right\}= \left\{ n,T,{\bm 0}
\right\}} \delta \left( {\bm r}-
{\bm q}_{s}\right)  \nonumber \\
&=&-\frac{1}{2}\sum_{s=1}^{N} e^{i {\bm k}^{\ast }\cdot {\bm
q}_{s}^{\ast }} \frac{\partial}{\partial {\bm v}_{s}^{\ast}} \cdot
\left[ {\bm v}^{\ast}_{s} \rho^{\ast}_{h} (\Gamma^{\ast}) \right],
\label{1.6.2}
\end{eqnarray}
\begin{eqnarray}
\widetilde{\psi }_{\parallel,\perp i}^{\ast }\left(
\Gamma^{\ast};{\bm k}\right) &=& M_{\parallel} \int d {\bm r}\,
e^{i{\bm k}\cdot {\bm r}}\sum_{s=1}^{N} \left\{ \frac{\partial
\rho _{lh} \left[ \Gamma | \left\{ y_{\alpha} \right\} \right]
}{\partial U_{\parallel, \perp i }\left( {\bm q}_{s}\right)
}\right\} _{ \left\{ y_{\alpha} \right\}= \left\{ n,T,{\bm 0}
\right\}} \delta \left( {\bm r}-
{\bm q}_{s}\right)  \nonumber \\
&=& -\sum_{s=1}^{N} e^{i {\bm k}^{\ast }\cdot {\bm q}_{s}^{\ast }}
\frac{\partial}{\partial v_{s, \parallel,\perp i}^{\ast}}
\rho^{\ast}_{h} (\Gamma^{\ast}). \label{1.6.1}
\end{eqnarray}
If this linear response analysis were done for a normal fluid using
the canonical Gibbs ensemble $\rho _{c}(\Gamma)$, the functions
$\widetilde{\psi }_{\alpha }^{\ast } $ would reduce to quantities
proportional to the familiar microscopic densities $ \widetilde{
a}_{\alpha } $ multiplied by the Gibbs equilibrium ensemble $\rho
_{c}$. Hence, the different forms of the conjugate densities
appearing in Eq.\ (\ref{1.5.1}) for the case of granular fluids, are
a manifestation of the fact that the linear response is now being
measured about a non-equilibrium macrostate.

Two different representations for the hydrodynamic response matrix
$C^{\ast}$ have been identified in Eqs.\, (\ref{1.4.1}) and
(\ref{1.5.1}), respectively. The latter representation is exact and
valid for all times and for all length scales. On the other hand,
the former representation is obtained from the phenomenological
hydrodynamic equations, which are the relevant description of the
fluid on length scales long compared to the mean free path and time
scales long compared to the mean free time. In this limit, the exact
response functions (\ref{1.5.1}) must go over to that given in
Eq.(\ref{1.4.1}), thereby providing an exact identification of all
the hydrodynamic parameters  in terms of time correlation functions.
To facilitate this identification, the exact analog of the
hydrodynamic transport matrix $\mathcal{K}^{\ast hyd}$ defined in
Eqs.\, (\ref {1.4.1a})-(\ref{1.4.3}) above is identified as
\begin{equation}
\mathcal{K}^{\ast }\left({\bm k}^{\ast},s\right) =-\left[
\partial _{s}\widetilde{C}^{\ast }\left( {\bm k}^{\ast};s\right)
\right] \widetilde{C}^{\ast -1}\left({\bm k}^{\ast} ;s\right) ,
\label{1.7.1a}
\end{equation}
and the cross over to hydrodynamics stated above is
\begin{equation}
\mathcal{K}^{\ast hyd }\left({\bm k}^{\ast}\right)
=\lim_{s>>0,k^{\ast}<<1}\, \mathcal{K} ^{\ast }\left( {\bm
k}^{\ast},s\right).
\end{equation}
This is the prescription carried out in ref. \cite{DBB06} to obtain
Helfand and Green-Kubo representations for the transport
coefficients of a granular fluid. To lowest order, $k^{\ast}=0$, the
exact expression for the cooling rate of the fluid is identified. At
Euler order, linear order in $k^{\ast}$, the pressure and the Euler
transport coefficient $\zeta^{U}$ are identified. Finally, to order
$k^{\ast 2}$ the six Navier-Stokes transport coefficients are all
obtained.

\section{Hydrodynamic Parameters and Transport Coefficients}
\label{s3}
\subsection{Overview}

In this Section, the exact explicit expressions for each of the
parameters occurring in the phenomenological hydrodynamic equations
at Navier-Stokes order are given, and their technical and physical
content is noted. The dimensionless units are used throughout, and
for simplicity the length scale is now chosen by the condition
$n_{h} l^{d}=1$.

The required ``equations of state'' are expressions for the cooling
rate and the pressure as functions of the density and temperature.
These are given as averages of specific phase functions over the HCS
distribution function. The phase functions are sums of  functions
involving the phase coordinates of pairs of particles, and so the
averages can be expressed as integrals over the associated reduced
two-particle distribution function defined by
\begin{equation}
f_{h}^{ \ast \left( 2\right) }\left({\bm q}_{12}^{\ast },{\bm v}
_{1}^{\ast },{\bm v}_{2}^{\ast }\right) =N\left( N-1\right) \int
d{\bm q}_{3}^{\ast } \int d {\bm v}_{3}^{\ast}  \ldots \int d{\bm
q}_{N}^{\ast } \int d{\bm v} _{N}^{\ast }\rho _{h}^{\ast }(\Gamma
^{\ast }). \label{5.1A}
\end{equation}
This function depends on the positions of particles only through
${\bm q}_{12}^{\ast } \equiv {\bm q}_{1}^{\ast }-{\bm q}_{2}^{\ast
}$, as a consequence of the translational invariance of the HCS
(homogeneity). The occurrence of $ f_{h}^{\ast \left( 2\right)  }$
for the properties considered here, depends on ${\bm q}_{12}^{\ast
}$ only on the sphere for the pair at contact, i.e. for
$q_{12}^{\ast }=\sigma ^{\ast } \equiv \sigma / l $. The velocity
dependence can be represented in terms of the relative velocity
${\bm g} _{12}^{\ast }={\bm v}_{1}^{\ast }-{\bm v}_{2}^{\ast }$,
and the center of mass velocity ${\bm G}_{12}^{\ast }=\left({\bm
v}_{1}^{\ast }+ {\bm v}_{2}^{\ast }\right) /2$. If the center of
mass velocity can be integrated out, the result is a probability
distribution for the magnitude of the relative velocity and the
cosine of the impact angle, i.e., $x= |{\bm q}_{12}^{\ast} \cdot
{\bm g}_{12}^{\ast}|/ q_{12}^{\ast} g_{12}^{\ast}$ at contact,
\begin{equation}
F_{h}^{\ast }\left( g_{12}^{\ast },x\right) =\int d{\bm
G}_{12}^{\ast }\, f_{h}^{\left( \ast  2\right)  }\left( {\bm
q}_{12}^{\ast },{\bm v} _{1}^{\ast },{\bm v}_{2}^{\ast }\right) .
\label{5.1}
\end{equation}
As shown below, the cooling rate, the pressure, and some parts of
the transport coefficients can be expressed as low degree moments of
$F_{h}^{\ast }\left( g_{12}^{\ast },x\right) $.

The dimensionless transport coefficients $\varpi^{\ast}$ will be
given in two equivalent representations. The (intermediate) Helfand
expressions are long time limits of time correlation functions,
while  the Green-Kubo expressions are given in terms of time
integrals of related time correlation functions. These two
representations are related by the simple identities
\begin{equation}
\varpi^{\ast} =\lim \Omega _{H}(s)=\Omega _{0}+\lim
\int_{0}^{s}ds^{\prime}\, \Omega _{G}(s^{\prime}),  \label{5.2}
\end{equation}
with
\begin{equation}
\Omega _{0}=\Omega _{H}(s=0), \quad \Omega _{G}(s)=
\frac{\partial}{\partial s} \Omega _{H}(s).  \label{5.3}
\end{equation}
The first equality in Eq.\ (\ref{5.2}) identifies the Helfand
expression, while the second one is the Green-Kubo representation.
The symbol $\lim $ denotes the ordered limits of $V^{\ast}
\rightarrow \infty $ followed by $s\rightarrow \infty $. Note that
what is called here the Helfand representation for the sake of
simplicity, was termed the intermediate Helfand representation in
ref. \cite{DBB06}. The correlation functions $\Omega _{H}(s)$
appearing in the Helfand representation have the general form
\begin{equation}
\Omega _{H}\left( s\right) = V^{\ast -1}\int d\Gamma ^{\ast }\,
F^{\ast S,f}\left( \Gamma ^{\ast }\right) e^{- s \left(
\overline{\mathcal{L}} ^{\ast }-\lambda \right) s}\mathcal{M}^{\ast}
\left( \Gamma ^{\ast }\right) . \label{5.4}
\end{equation}
The phase functions $F^{\ast S,f}$ are defined in terms of either
a volume integrated source $S$ or a flux $f$ present in the
microscopic conservation laws for the direct functions $ a_{\alpha
}^{\ast } $ defined in Eq.\ (\ref{1.5.2}). The phase functions
$\mathcal{M}^{\ast}(\Gamma^{\ast})$ are first space moments of the
functional derivatives of the local HCS distribution function.
They are obtained from Eq.\ (\ref{1.5.4}) to first order in $k$,
then having the general structure
\begin{equation}
\mathcal{M}^{\ast}\left( \Gamma ^{\ast }\right) =M  \int d{\bm r}\,
\widehat{\bm k } \cdot {\bm r} \left[ \frac{\delta \rho _{lh}\left[
\Gamma |\left\{ y_{\alpha } \right \} \right] }{\delta y\left( {\bm
r}\right) } \right]_{ \left\{ y_{\alpha} \right\} =\left\{
n_{h},T_{h}(t),{\bm 0} \right\}}. \label{5.4a}
\end{equation}
Finally, the dynamics in Eq.\ (\ref{5.4}) is generated by the
modified operator $\overline{ \mathcal{L}}^{\ast }$ of the Liouville
equation, minus one of its eigenvalues $\lambda_{\alpha}$ determined
from
\begin{equation}
\left( \overline{\mathcal{L}}^{\ast }-\lambda_{\alpha} \right)
\widetilde{\psi } _{\alpha }^{\ast }\left( \Gamma^{\ast}; {\bm
0}\right) =0. \label{5.5}
\end{equation}
As discussed in refs. \cite{DBB06} and \cite{DByB06}, the
functions $ \widetilde{ \psi }_{\alpha }^{\ast }\left(
\Gamma^{\ast}; {\bm 0}\right) $ are the exact eigenfunctions of
the modified Liouville operator, representing \textit{
homogeneous} perturbations of the HCS hydrodynamics. The
eigenvalues $ \lambda_{\alpha} = 0,\zeta _{0}^{\ast }/2,-\zeta
_{0}^{\ast }/2$, the last one being $d$-fold degenerate, are
therefore the same as those of the transport matrix
$\mathcal{K}^{\ast hyd }\left( {\bm 0} \right) $. Since the
transport coefficients characterize the response to
\textit{spatial} variations, it is reasonable that the dynamics in
their representation as time correlation functions does not
include such homogeneous excitations.

The correlation functions $\Omega _{G}(s)$ present in the Green-Kubo
representation have the corresponding forms
\begin{equation}
\Omega _{G}(s)= V^{\ast -1} \int d\Gamma ^{\ast }\, F^{\ast
S,f}\left( \Gamma ^{\ast }\right) e^{- s\left(
\overline{\mathcal{L}}^{\ast }-\lambda \right) }\Upsilon^{\ast}
\left( \Gamma ^{\ast }\right) , \label{5.6}
\end{equation}
with the new ``fluxes'' $\Upsilon ^{\ast} \left( \Gamma ^{\ast
}\right) $ given by
\begin{equation}
\Upsilon^{\ast}  \left( \Gamma ^{\ast }\right) =-\left(
\overline{\mathcal{L}}^{\ast }-\lambda \right) \mathcal{M} ^{\ast}
\left( \Gamma ^{\ast }\right) .  \label{5.7}
\end{equation}
The terminology ``flux'' for these functions is justified in
Appendix \ref{ap5}, where it is shown that they are the volume
integrals of fluxes occurring in the microscopic balance equations
for the conjugate functions $\widetilde{\psi }_{\alpha }^{\ast }
$. In detail, it is found that the $F^{\ast S,f}\left( \Gamma
^{\ast }\right) $ are projected orthogonal to the set of
eigenfunctions $\left\{ \widetilde{\psi }_{\alpha }^{\ast }\left(
\Gamma^{\ast};{\bm 0} \right) \right\} $, so that the zeros of the
operator $\overline{\mathcal{L}}^{\ast }-\lambda $ do not occur in
the dynamics. Consequently, the limit on the time integral in Eq.\
(\ref{5.2}) is expected to exist. This is the analogue of the
``subtracted fluxes'' present for the same reason in the
Green-Kubo expressions for normal fluids \cite{McLbook}.

\subsection{The cooling rate and the pressure}

The cooling rate was identified in ref.\ \cite{DBB06} as
proportional to the average (negative) rate of change of the total
energy in the HCS. Its dimensionless value is given by
\begin{equation}
\zeta _{0}^{\ast } = V^{\ast -1}\int d\Gamma ^{\ast }\, W^{\ast}
\left( \Gamma ^{\ast }\right) \rho _{h}^{\ast }\left( \Gamma ^{\ast
}\right) \label{5.1AA}
\end{equation}
where
\begin{equation}
W^{\ast}(\Gamma^{\ast})= \frac{2}{d} \int d{\bm r}^{\ast} w^{\ast}
(\Gamma^{\ast};{\bm r}^{\ast}), \label{5.1AAB}
\end{equation}
$w^{\ast}$ being the source term in the dimensionless balance
equation for the energy, Eq.\ (\ref{E9}). Then, using Eqs.\
(\ref{E9.2}) and (\ref{D22}), it is found that
\begin{equation}
W\left( \Gamma ^{\ast }\right) =\frac{1-\alpha ^{2}}{2d}
\sum_{r=1}^{N} \sum_{s \neq r}^{N}  \delta (q_{rs}^{\ast}-
\sigma^{\ast}) \Theta \left( -\widehat{\bm q}_{rs}^{\ast }\cdot {\bm
g}_{rs}^{\ast } \right) |\widehat{\bm q}_{rs}^{\ast }\cdot {\bm
g}_{rs}^{\ast }|^{3}. \label{5.1.aaa}
\end{equation}
Here, $\widehat{\bm q}_{rs}^{\ast} \equiv  {\bm
q}_{rs}^{\ast}/q_{rs}$ and $\Theta (x)$ is the Heaviside step
function. Therefore, $\zeta_{0}^{\ast}$ is proportional to the third
moment of the normal component of the relative velocities of all
pairs at contact, averaged over one of the collision hemispheres, in
the HCS, weighted by the fractional loss in the kinetic energy of
the two particles that is proportional to $\left( 1-\alpha
^{2}\right) $. Since the cooling rate is determined by the average
of a sum of phase functions involving only phase coordinates of
pairs of particles, it can be expressed in the reduced form
\begin{eqnarray}
\zeta _{0}^{\ast } &=&\frac{1-\alpha^{2}}{2 V^{\ast} d}\int d{\bm q}
_{1}^{\ast } \int d{\bm v}_{1}^{\ast } \int d{\bm q}_{2}^{\ast }\int
d{\bm v}_{2}^{\ast }\,  \delta (q^{\ast}_{12}-\sigma^{\ast}) \Theta
(
\widehat{\bm q}_{12}^{\ast} \cdot {\bm g}_{12}^{\ast}) \nonumber  \\
&&\times  |\widehat{\bm q}_{12}^{\ast } \cdot {\bm g}_{12}^{\ast
}|^{3}f_{h}^{\ast \left( 2\right)  }\left( {\bm q}_{12}^{\ast },{\bm
v}_{1}^{\ast },{\bm v}_{2}^{\ast }\right). \label{5.1.3}
\end{eqnarray}
Transforming to relative and center of mass coordinates, allows
further simplification to
\begin{equation}
\zeta _{0}^{\ast }=\frac{(1-\alpha ^{2})\pi^{d/2} \sigma^{\ast
d-1}}{\Gamma (d/2) d}\int d{\bm g}^{\ast}_{12} \Theta (x) g^{\ast
3}_{12} x^{3}   F^{\ast}_{h}(g_{12}^{\ast},x), \label{5.1.5}
\end{equation}
where $x=\cos \phi$ is now the projection of $\widehat{\bm g}_{12}$
along an arbitrary direction. This rather simple result is still
exact, and retains all of the two particle (position and velocity)
correlations representative of the HCS. The two forms (\ref{5.1AA})
and (\ref{5.1.5}) are suitable for complementary molecular dynamics
simulations of the cooling rate.

Similarly, the pressure is identified as the HCS average of the
trace of the microscopic momentum flux
\begin{equation}
p^{\ast }=\frac{2}{V^{\ast }d}\int d\Gamma ^{\ast }\, \rho
_{h}^{\ast }\left( \Gamma ^{\ast }\right) \text{tr } {\sf
H}^{\ast}(\Gamma^{\ast}). \label{5.2.A}
\end{equation}
Here  the phase tensor ${\sf H}^{\ast}$ is the volume integral of
the dimensionless flux defined in Eq. (\ref{E9.0}),
\begin{eqnarray}
{\sf H}^{\ast }(\Gamma^{\ast})& = & \int d{\bm r}^{\ast} {\sf
h}^{\ast}(\Gamma^{\ast};{\bm r}^{\ast}) \nonumber \\
& = & \sum_{r=1}^{N} {\bm v}_{r}^{\ast }{\bm v}_{r}^{\ast
}+\frac{(1+\alpha)\sigma^{\ast}}{4} \sum_{r=1}^{N} \sum_{s \neq
r}^{N} \delta \left( q_{rs}^{\ast }-\sigma ^{\ast }\right) \Theta
\left( -\widehat{\bm q}_{rs}^{\ast }\cdot {\bm g}_{rs}^{\ast
}\right)
\nonumber \\
&& \times \left( \widehat{\bm q}_{rs}^{\ast }\cdot {\bm
g}_{rs}^{\ast }\right) ^{2}\widehat{\bm q}_{rs}^{\ast}
\widehat{\bm q} _{rs}^{\ast}. \label{5.2.B}
\end{eqnarray}
The pressure can be partitioned into two terms,
\begin{equation}
p^{\ast }=p_{K}^{\ast }+p_{C}^{\ast },  \label{5.2.10y}
\end{equation}
where
\begin{equation}
p_{K}^{\ast }=\frac{2}{d}\int d\Gamma^{\ast}\, \rho_{h}^{\ast}
(\Gamma^{\ast}) v_{1}^{\ast 2}=1 \label{5.2.10z}
\end{equation}
and
\begin{eqnarray}
p_{C}^{\ast } &=&\frac{(1+\alpha) \sigma ^{\ast }}{2 V^{\ast}d}
\int d{\bm q}_{1}^{\ast } \int d{\bm v}_{1}^{\ast } \int d{\bm
q}_{2}^{\ast } \int d{\bm v}_{2}^{\ast }\, \delta \left(
q_{12}^{\ast }-\sigma ^{\ast }\right) \Theta \left( -\widehat{\bm
q}_{12}^{\ast }\cdot {\bm g} _{12}^{\ast }\right)
\nonumber \\
&& \times \left( \widehat{\bm q}_{12}^{\ast }\cdot {\bm
g}_{12}^{\ast }\right) ^{2}f_{h}^{\ast \left(  2\right)}\left({\bm
q}_{12}^{\ast },{\bm v}_{1}^{\ast },{\bm v}_{2}^{\ast }\right)  \nonumber \\
&=& \frac{\pi^{d/2}(1+\alpha) \sigma^{\ast d}}{\Gamma (d/2) d} \int
d{\bm g}^{\ast }_{12} g^{\ast 2}_{12} \Theta (x) x^{2} F_{h}^{\ast}
(g^{\ast}_{12},x). \label{5.2.11}
\end{eqnarray}
The first term on the right hand side of Eq.\ (\ref{5.2.10y}) is
the kinetic part of the pressure. It arises purely from the
transport of momentum associated with the free streaming of the
particles and implies that $p_{K}=n_{h}T_{h}\left( t\right) $. \
The second term, determined by the two particle distribution
function at contact, is the collisional transfer part of the
pressure. Like the cooling rate, it is given by a moment of the
simple distribution $F_{h}^{\ast }\left( g_{12}^{\ast },x\right)
$. Again, Eqs.\ (\ref{5.2.A}) and (\ref {5.2.10y})-(\ref{5.2.11})
provide different, complementary expressions for the evaluation of
the pressure by means of molecular dynamics simulations.

The pressure and the cooling rate, as functions of the
hydrodynamic fields $ n$ and $T$, are the required equations of
state for the hydrodynamics equations which are determined
entirely by the HCS. This is analogous to the case of normal
fluids where the dependence of the pressure on $n$ and $T$ is
given by the equilibrium Gibbs state. Note that this definition is
independent of the presence or absence of other normal stresses,
which appear as additional independent dissipative contributions
to the hydrodynamic equations. Similarly, the cooling rate has
additional independent dissipative contributions characterized by
transport coefficients.

\subsection{Euler Transport Coefficient $\protect\zeta^{U}$}

As already indicated, a new transport coefficient, $\zeta^{U}$,
occurs at Euler order in the temperature equation of the granular
fluid. It is a measure of the contribution to heat transport from
divergences in the flow velocity, and does not occur in the
``perfect fluid'' Euler equations for a normal fluid that have no
dissipation. In Appendix \ref{ap6}, the intermediate Helfand
representation for the dimensionless form of this transport
coefficient is identified as
\begin{equation}
\zeta^{\ast U }=\lim \Omega _{H}^{\zeta^{U}}\left( s\right) ,
\label{5.3.A}
\end{equation}
where the time correlation function is
\begin{equation}
\Omega _{H}^{\zeta^{U}}\left( s\right) = V^{\ast -1} \int d\Gamma
^{\ast }\, W^{S}\left( \Gamma ^{\ast }\right) e^{-s \left(
\overline{\mathcal{L}} ^{\ast }+\frac{\zeta _{0}^{\ast }}{2}\right)
}\mathcal{M}^{\ast}_{\zeta ^{U}}(\Gamma^{\ast}). \label{5.8}
\end{equation}
Here $W^{S}(\Gamma^{\ast})$ is the volume integrated source term
$W\left( \Gamma ^{\ast }\right) $ given in  Eq.\ (\ref{5.1.aaa}),
projected orthogonal to the set $ \left\{ \widetilde{\psi }_{\alpha
}^{\ast }\left(\Gamma^{\ast}; {\bm 0}\right) \right\} $, as
discussed above. More precisely, it is
\begin{equation}
W^{S}\left( \Gamma ^{\ast }\right) = (1-P^{\ast \dagger})
W(\Gamma^{\ast}) = W\left( \Gamma ^{\ast }\right) -\frac{3
\zeta_{0}^{\ast}}{d} \left( E^{\ast } -\frac{d}{2} N\right) -\zeta
_{0}^{\ast }\left( \frac{\partial \ln \zeta _{0}}{\partial \ln
n_{h}}+1\right) N. \label{5.3.B}
\end{equation}
The projection gives the additional terms proportional to the total
number of particles $N$ and the total energy $E^{\ast
}=\sum_{r}v_{r}^{\ast 2}$. The moment $\mathcal{M}^{\ast}_{\zeta
^{U}}$ in Eq.\ (\ref{5.8}) is given by
\begin{equation}
\mathcal{M}^{\ast}_{\zeta ^{U}}(\Gamma^{\ast})=-\sum_{r=1}^{N} {\bm
q}_{r}^{\ast }\cdot \frac{\partial}{\partial {\bm v}_{r}^{\ast}}
\rho_{h}^{\ast}(\Gamma^{\ast}), \label{5.3.5a}
\end{equation}
that is a space first moment of the velocity derivative of the HCS
distribution. To provide some interpretation for
$\mathcal{M}^{\ast}_{\zeta^{U}}$, note that for a normal fluid, when
the canonical Gibbs ensemble $\rho_{c}(\Gamma)$ is considered as the
reference state,
\begin{equation}
\mathcal{M}^{\ast}_{\zeta^{U}} = 2 \sum_{r=1}^{N}{\bm q}_{r}^{\ast
}\cdot {\bm v}_{r}^{\ast }\rho _{c}^{\ast},  \label{5.3.Z}
\end{equation}
which is the space moment of the momentum density of the fluid. This
is modified for granular fluids to account for the non-equilibrium
nature of the HCS.

The generator of the dynamics in Eq.\ (\ref{5.8}) consists of the
scaled Liouville operator $\overline{\mathcal{L}}^{\ast }$ together
with the factor $\zeta _{0}^{\ast }/2$. The latter is one of the
eigenvalues $\left( \lambda =-\zeta _{0}^{\ast }/2\right) $ solution
of Eq.\ (\ref{5.5}), and occurs here because $\mathcal{M}^{\ast}
_{\zeta ^{U}}$ has a non-vanishing component along the corresponding
eigenfunction in this set of zeros for $
\overline{\mathcal{L}}^{\ast }-\lambda  $. The projected form of
$W^{S}\left( \Gamma ^{\ast }\right) $ assures that this component
does not contribute to $\zeta ^{\ast U}$. Further discussion of this
effect for the other transport coefficients is given below.

The equivalent Green-Kubo representation is
\begin{equation}
\zeta ^{\ast U }=\Omega _{0}^{\zeta ^{U}}+\lim
\int_{0}^{s}ds^{\prime }\, \Omega _{G}^{\zeta^{U}}\left( s^{\prime
}\right),  \label{5.3.GK}
\end{equation}
with $\Omega_{0}^{\zeta^{U}} =\Omega_{H}^{\zeta^{U}}(0)$ and
\begin{equation}
\Omega _{G}^{\zeta^{U}}\left( s\right) =V^{\ast -1}\int d\Gamma
^{\ast }\, W^{S}(\Gamma^{\ast}) e^{- s \left(
\overline{\mathcal{L}}^{\ast }+\frac{\zeta _{0}^{\ast }}{2}\right)
}\Upsilon^{\ast} _{\zeta ^{U}}(\Gamma^{\ast}), \label{5.9}
\end{equation}
where
\begin{equation}
\Upsilon^{\ast} _{\zeta^{U}}=-\left( \overline{\mathcal{L}}^{\ast
}+\frac{\zeta _{0}^{\ast }}{2}\right) \mathcal{M}^{\ast}_{\zeta^{U}}
\label{5.3.AF}
\end{equation}
is a  ``conjugate flux'', associated with the new conserved
quantities of the dynamics of this system \cite{DBB06}. The usual
integrands of the time integral in Green-Kubo expressions are
flux-flux correlation functions. Here, one of the fluxes has been
replaced by the source in the energy equation due to collisional
loss of energy. This is characteristic of the Green-Kubo expressions
for transport coefficients arising from the cooling rate. For all
other transport coefficients, the familiar flux-flux correlation
form occurs, although involving both a direct and a conjugate flux.
This difference in the two sets of fluxes is due to both the
dissipation in the collision rule and the singular nature of hard
spheres. The occurrence of the instantaneous part to the Green-Kubo
relation, $\Omega _{0}^{\zeta ^{U}}$, has similar origins and is not
present for normal fluids with non-singular forces.

The instantaneous part of the Green-Kubo representation can be simplified
further to
\begin{equation}
\Omega _{0}^{\zeta^{U}}= (V^{\ast} d)^{-1}\int d\Gamma ^{\ast }\,
W^{S}(\Gamma^{\ast}) \mathcal{M}^{\ast}_{\zeta^{U}}
(\Gamma^{\ast})=- \frac{3}{d} \left( 1-\alpha \right) p_{C}^{\ast },
\end{equation}
where $p_{C}^{\ast }$ is the collisional part of the pressure given
in Eq.\ (\ref {5.2.11}). Hence, $\zeta^{\ast U}$ is the contribution
of the source to what would physically constitute the effects of
hydrostatic pressure at the Euler order, where it enters in the
combination $\left( 2p^{\ast }/d\right) +\zeta ^{\ast U}$. If a
small volume element of the fluid is considered, then the amount of
pressure that the fluid element can exert on its boundaries is
decreased by the energy lost locally due to collisions. Part of the
effect of this transport coefficient is to decrease the effective
pressure in the system. At the level of linear hydrodynamics, the
two coefficients are indistinguishable in their physical
consequences.

\subsection{Shear Viscosity}

The (intermediate) Helfand and Green-Kubo expressions for the
dimensionless shear viscosity $\eta^{\ast}$ are also elaborated in
Appendix \ref{ap6} to the the form
\begin{equation}
\eta ^{\ast }=\lim \Omega _{H}^{\eta }\left( s\right) =\Omega
_{0}^{\eta }+\lim \int_{0}^{s}ds^{\prime }\, \Omega _{G}^{\eta
}\left( s^{\prime }\right) , \label{5.10}
\end{equation}
where the  correlation functions are defined by
\begin{equation}
\Omega _{H}^{\eta }\left( s\right) =- \frac{V^{\ast -1}}{d^{2}+d-2}
\sum_{i=1}^{d} \sum_{j=1}^{d} \int d\Gamma ^{\ast }\ {\sf
H}_{ij}^{\ast }(\Gamma^{\ast}) e^{-s \left(
\overline{\mathcal{L}}^{\ast }+\frac{\zeta _{0}^{\ast }}{2}\right)
}\mathcal{M}^{\ast}_{\eta ,ij}(\Gamma^{\ast}), \label{5.11}
\end{equation}
\begin{equation}
\Omega _{G}^{\eta }\left( s\right) =- \frac{V^{\ast -1}}{d^{2}+d-2}
\sum_{i=1}^{d} \sum_{j=1}^{d} \int d\Gamma ^{\ast}\, {\sf
H}_{ij}^{\ast }(\Gamma^{\ast}) e^{-s \left(
\overline{\mathcal{L}}^{\ast }+\frac{\zeta _{0}^{\ast }}{2}\right)
}\Upsilon^{\ast} _{\eta ,ij} (\Gamma^{\ast}), \label{5.12}
\end{equation}
and $\Omega_{0}^{\eta}= \Omega_{H}^{\eta}(0)$. In the above
expressions, ${\sf H}_{ij}^{\ast }(\Gamma^{\ast}) $ is the volume
integrated momentum flux given by Eq.\ (\ref{5.2.B}),
$\mathcal{M}^{\ast} _{\eta ,ij}$ is the traceless tensor
\begin{equation}
\mathcal{M}_{\eta
,ij}^{\ast}(\Gamma^{\ast})=-\frac{1}{2}\sum_{r=1}^{N} \left(
q_{r,i}^{\ast } \frac{\partial}{\partial
v^{\ast}_{r,j}}+q_{r,j}^{\ast } \frac{\partial}{\partial
v^{\ast}_{r,i}} - \frac{2}{d}\delta _{ij}{\bm q}_{r}^{\ast }\cdot
\frac{\partial}{\partial {\bm v}^{\ast}_{r}} \right)
\rho^{\ast}(\Gamma^{\ast}),  \label{5.13}
\end{equation}
and the associated Green-Kubo conjugate flux $\Upsilon^{\ast} _{\eta
,ij }$ is
\begin{equation}
\Upsilon^{\ast} _{\eta ,ij}=-\left( \overline{\mathcal{L}}^{\ast
}+\frac{\zeta _{0}^{\ast }}{2}\right) \mathcal{M}^{\ast}_{\eta
,ij}(\Gamma^{\ast}). \label{5.14}
\end{equation}
The generator of the dynamics is now $\left(
\overline{\mathcal{L}}^{\ast }+\frac{ \zeta _{h}^{\ast }}{2}\right)
$ and, therefore,  has the $k=0$ mode of Eq.\ (\ref{5.5}) with $
\lambda =-\zeta _{0}^{\ast }/2$ subtracted out.

In order to better understand the differences from the corresponding
expression for a normal fluid, note that for the latter with the
canonical Gibbs reference state $\rho_{c}(\Gamma)$, it is
\begin{equation}
\mathcal{M}^{\ast}_{\eta ,ij}(\Gamma^{\ast}) =  \sum_{r=1}^{N}
\left( q_{r,i}^{\ast }v_{r,j}^{\ast }+q_{r,j}^{\ast }v_{r,i}^{\ast
}-\frac{2}{d}\delta _{ij}{\bm q}_{r}^{\ast }\cdot {\bm v}_{r}^{\ast
}\right) \rho _{c}^{\ast}(\Gamma^{\ast}), \label{5.15}
\end{equation}
and
\begin{equation}
\Upsilon_{\eta ,ij}^{\ast} (\Gamma^{\ast})= -\overline{L}^{\ast}
\mathcal{M}^{\ast}_{\eta,ij}(\Gamma^{\ast}). \label{4.16a}
\end{equation}
Using the property (\ref{A26}) and taking into account that
$\overline{L}^{\ast} \rho_{c}^{\ast}=0$, it is obtained that, for a
normal fluid,
\begin{equation}
\Upsilon_{\eta ,ij}^{\ast} (\Gamma^{\ast})=- 2 \left( {\sf
H}_{ij}^{\ast -}-\frac{1}{d}\delta _{ij} \text{tr } {\sf H}^{\ast
-}\right) \rho^{\ast} _{c}(\Gamma^{\ast}), \label{5.16}
\end{equation}
where ${\sf H}_{ij}^{\ast -}$ is the volume integrated momentum flux
for the time reversed microscopic conservation laws identified in
Appendix \ref{ap4}, namely
\begin{eqnarray}
{\sf H}^{\ast -}(\Gamma^{\ast}) &  =  & \sum_{r=1}^{N} {\bm
v}_{r}^{\ast} {\bm v}_{r}^{\ast} +\frac{\sigma^{\ast}}{2}
\sum_{r=1}^{N} \sum_{ s \neq r}^{N} \delta ( q_{rs}^{\ast
}-\sigma^{\ast}) \Theta \left( \widehat{\bm q}_{rs}^{\ast} \cdot
{\bm g}_{rs} \right)  \nonumber \\
& & \times ( \widehat{\bm q}_{rs}^{\ast} \cdot {\bm g}_{rs})^{2}
\widehat{\bm q}_{rs} \widehat{\bm q}_{rs}.
\end{eqnarray}

Therefore, the normal fluid version of Eq.(\ref{5.10}) is
\begin{equation}
\eta ^{\ast } =  \Omega _{0}^{\eta }+\lim \frac{2 V^{\ast
-1}}{d^{2}+d-2} \sum_{i=1}^{d} \sum_{j=1}^{d} \int_{0}^{s}ds^{\prime
} \int d\Gamma ^{\ast }\, {\sf H}_{ij}^{\ast }e^{-s
\overline{L}^{\ast }}\left[ {\sf H}_{ij}^{\ast
-}(\Gamma^{\ast})-\frac{1}{d}\delta _{ij} \text{tr } {\sf H}^{\ast
-}\right] \rho _{c}^{\ast}(\Gamma^{\ast}). \label{5.17}
\end{equation}
The normal fluid shear viscosity is given by the long time
integral of the equilibrium time correlation functions of the
forward momentum flux $ {\sf H}^{\ast}$ with the time reversed
momentum flux ${\sf H}^{\ast -}$, together with an instantaneous
part. This instantaneous term remains nonzero in the elastic
limit, as an artifact of the singular nature of the hard particle
interaction. Note that in the elastic limit, the hydrodynamic
modes defined through Eq.\ (\ref{5.5}) all have the same
eigenvalue $\lambda =0$, so the generator for the dynamics in both
the elastic and inelastic case can be thought of as having the
$k^{\ast}=0$ mode subtracted out.

The instantaneous contribution $\Omega _{0}^{\eta }$ for the
granular fluid can be expressed in terms of the reduced pair
distribution function, just as in the case of the pressure above,
with the result
\begin{equation}
\Omega _{0}^{\eta }=- \frac{V^{\ast -1}}{d^{2}+d-2} \int d\Gamma
^{\ast }\, {\sf H}_{ij}^{\ast }(\Gamma^{\ast})
\mathcal{M}^{\ast}_{\eta ,ij}(\Gamma^{\ast})
=\frac{1+\alpha}{4(d^{2}+2d)}\sigma ^{\ast 2}\nu _{av}, \label{5.18}
\end{equation}
where $\nu _{av}$ is the average collision frequency as determined
by the loss part of the right hand side of the hard sphere BBGKY
hierarchy \cite{McLbook} for the HCS
\begin{equation}
\nu _{av}=\frac{4 \pi^{d/2} \sigma ^{\ast d-1}}{\Gamma \left(d/2
\right)} \int d{\bm g}_{12} ^{\ast}\, \Theta (\widehat{\bm
q}_{12}^{\ast} \cdot {\bm g}^{\ast} _{12}) \widehat{\bm
q}^{\ast}_{12} \cdot {\bm g}_{12}^{\ast} F_{h}^{\ast }\left(
g_{12}^{\ast },x\right) . \label{5.19}
\end{equation}
This  is a purely collisional quantity, arising from the boundary
condition associated with hard sphere dynamics at the point of
contact.

\subsection{Bulk Viscosity}

The bulk viscosity has forms similar to those reported above for the
shear viscosity, although it represents resistance to volume
dilation rather than shear,
\begin{equation}
\kappa ^{\ast }=\lim \Omega _{H}^{\kappa }\left( s\right) =\Omega
_{0}^{\kappa }+\lim \int_{0}^{s}ds^{\prime }\, \Omega _{G}^{\kappa
}\left( s^{\prime }\right) .  \label{5.20}
\end{equation}
The correlation functions in the above expressions are shown in
Appendix \ref{ap6} to be given by
\begin{equation}
\Omega _{H}^{\kappa }\left( s\right) =- (V^{\ast}d^{2})^{-1} \int
d\Gamma ^{\ast }\, \text{tr } {\sf H}^{\ast f}(\Gamma^{\ast})e^{-s
\left( \overline{\mathcal{L}}^{\ast }+\frac{\zeta _{0}^{\ast
}}{2}\right) }\mathcal{M}^{\ast}_{\kappa }(\Gamma^{\ast}),
\label{5.21}
\end{equation}
\begin{equation}
\Omega _{G}^{\kappa }\left( s\right) =- (V^{\ast} d^{2})^{-1}\int
d\Gamma ^{\ast }\, \text{tr } {\sf H}^{\ast f} (\Gamma^{\ast})
e^{-s\left( \overline{\mathcal{L}}^{\ast }+\frac{\zeta _{0}^{\ast
}}{2}\right) }\Upsilon^{\ast} _{\kappa } ( \Gamma^{\ast}),
\label{5.22}
\end{equation}
and $\Omega _{0}^{\kappa }=\Omega _{H}^{\kappa }\left( 0\right)$.
Here ${\sf H}^{\ast f}$ is the projection of the integrated momentum
flux, whose trace is
\begin{equation}
\text{tr } {\sf H}^{\ast f}(\Gamma^{\ast})= \text{tr } {\sf
H}^{\ast}(\Gamma^{\ast})-\frac{p^{\ast }_{h}d}{2} \frac{\partial \ln
p_{h}}{
\partial \ln n_{h}}N-\left( E^{\ast
}- \frac{d}{2}N \right) p^{\ast }_{h} ,  \label{5.5.DF}
\end{equation}
where is ${\sf H}^{\ast}(\Gamma^{\ast})$ is given by Eq.\
(\ref{5.2.B}). The additional terms on the right hand side of the
above equation, proportional to $N$ and $E^{\ast }$, result from
projection of $ \text{tr } {\sf H}^{\ast }$ orthogonal to the
eigenfunctions defined in Eq.\ (\ref{5.5}). Further details can be
seen in Appendix \ref{ap6} and ref. \cite{DBB06}. The moment
$\mathcal{M}_{\kappa }^{\ast}$ is the same as that for the Euler
transport coefficient $\zeta^{\ast U}$, i.e.,
\begin{equation}
\mathcal{M}_{\kappa}^{\ast}=\mathcal{M}_{\zeta^{U}}^{\ast}=
-\sum_{r=1}^{N} {\bm q}_{r}^{\ast }\cdot \frac{\partial}{\partial
{\bm v}_{r}^{\ast}} \rho_{h}^{\ast}(\Gamma^{\ast}). \label{5.23}
\end{equation}
Then, also for the associated Green-Kubo conjugate flux $\Upsilon
_{\kappa }^{\ast}$,
\begin{equation}
\Upsilon _{\kappa }^{\ast}= \Upsilon _{\zeta^{U}}^{\ast}= -\left(
\overline{\mathcal{L}}^{\ast }+\frac{\zeta _{0}^{\ast }}{2}\right)
\mathcal{M}_{\kappa }^{\ast}. \label{5.24}
\end{equation}

The instantaneous part of the bulk viscosity, $\Omega _{0}^{\kappa }$, is
simply related to that for the shear viscosity by
\begin{equation}
\Omega _{0}^{\kappa }=\frac{d+2}{d}\Omega _{0}^{\eta
}=\frac{(1+\alpha)\sigma ^{\ast 2}}{4d^{2}}\nu _{av}, \label{5.25}
\end{equation}
with the average collision frequency,  $\nu_{av}$, given by Eq.\,
(\ref{5.19}).

\subsection{Thermal Conductivity}

The expressions for the dimensionless thermal conductivity
$\lambda^{\ast}$ of the granular fluid of hard spheres or disks
obtained in Appendix \ref{ap6} are
\begin{equation}
\lambda ^{\ast }= \lim \Omega _{H}^{\lambda }\left( s\right)
=\Omega _{0}^{\lambda }+\lim \int_{0}^{s}ds^{\prime }\, \Omega
_{G}^{\lambda }\left( s^{\prime }\right),  \label{5.26}
\end{equation}
where
\begin{equation}
\Omega _{H}^{\lambda }\left( s\right) = - (V^{\ast}d)^{-1}\int
d\Gamma ^{\ast }\, {\bm S}^{\ast f}(\Gamma^{\ast}) \cdot e^{- s
\left( \overline{\mathcal{L}}^{\ast }-\frac{ \zeta _{0}^{\ast
}}{2}\right) } \bm{\mathcal{M}}_{\lambda }^{\ast}(\Gamma^{\ast}),
\label{5.27}
\end{equation}
\begin{equation}
\Omega _{G}^{\lambda }\left( s\right) = - (V^{\ast}d)^{-1} \int
d\Gamma ^{\ast }\, {\bm S}^{\ast f} (\Gamma^{\ast}) \cdot e^{- s
\left( \overline{\mathcal{L}}^{\ast }-\frac{ \zeta _{0}^{\ast
}}{2}\right) s} \bm{\Upsilon} _{\lambda }^{\ast}(\Gamma^{\ast}),
\label{5.28}
\end{equation}
and $ \Omega _{0}^{\lambda }=\Omega _{H}^{\lambda }\left( 0\right)$.
In the above expressions, the projected energy flux ${\bm S}^{\ast
f}$ is
\begin{equation}
{\bm S}^{\ast f}(\Gamma^{\ast})={\bm S}^{\ast
}(\Gamma^{\ast})-\left( \frac{d}{2}+ p^{\ast }_{h} \right) {\bm
P}^{\ast },  \label{5.6.6}
\end{equation}
with ${\bm S}^{\ast }(\Gamma^{\ast}) \equiv {\bm
s}^{\ast}(\Gamma^{\ast};{\bm 0})$ being the volume integrated heat
flux obtained from Eqs.\ (\ref{D21}) and (\ref{E9.1}),
\begin{eqnarray}
{\bm S}^{\ast }(\Gamma^{\ast}) &=&\sum_{r=1}^{N} v_{r}^{\ast 2}{\bm
v}_{r}^{\ast }+ \frac{(1+\alpha)\sigma^{\ast}}{2} \sum_{r=1}^{N}
\sum_{s \neq r}^{N} \delta (q_{rs}^{\ast }-\sigma ^{\ast }) \Theta
\left( -\widehat{\bm q}_{rs}^{\ast }\cdot {\bm g}_{rs}^{\ast
}\right) \nonumber \\
& &  (\widehat{\bm q}_{rs}^{\ast } \cdot {\bm
g}_{rs}^{\ast})^{2}\left( \widehat{\bm q}_{rs}^{\ast} \cdot {\bm
G}_{rs}^{\ast } \right) \widehat{\bm q}_{rs}^{\ast }. \label{5.6.4}
\end{eqnarray}
As with the previous transport coefficients, the additional term
proportional to the total momentum ${\bm P}^{\ast }$ in Eq.\
(\ref{5.6.6}) results from projection of ${\bm S}^{\ast }\
$orthogonal to the eigenfunctions defined by Eq.\  (\ref{5.5}). The
moment $\bm{\mathcal{M}}_{\lambda }^{\ast}$ is
\begin{equation}
\bm{\mathcal{M}}_{\lambda }^{\ast}(\Gamma^{\ast})
=-\frac{1}{2}\sum_{r=1}^{N} {\bm q}_{r}^{\ast}
\frac{\partial}{\partial {\bm v}_{r}^{\ast}} \cdot \left[ {\bm
v}_{r}^{\ast} \rho_{h}^{\ast} (\Gamma^{\ast})\right] \label{5.29}
\end{equation}
and the expression of the associated Green-Kubo conjugate flux
$\bm{\Upsilon} _{\lambda }^{\ast}$ is
\begin{equation}
\bm{\Upsilon} _{\lambda }^{\ast} (\Gamma^{\ast})=-\left(
\overline{\mathcal{L}}^{\ast }-\frac{\zeta _{0}^{\ast }}{2}\right)
\bm{\mathcal{M}}_{\lambda }^{\ast}. \label{5.30}
\end{equation}
The relevant hydrodynamic excitation that is subtracted from the
generator of the dynamics is now that for $\lambda =\zeta _{0}^{\ast
}/2$.

The instantaneous part of the thermal conductivity in the Green-Kubo
representation can be simplified to
\begin{eqnarray}
\Omega _{0}^{\lambda } &=&\frac{(1+\alpha)  \sigma^{\ast d+1}}{16 d}
\int d \widehat{\bm \sigma }^{\ast} \int d{\bm g}_{12}^{\ast }\int
d{\bm G}_{12}^{\ast }\,  \Theta \left( \widehat{\bm \sigma }^{\ast}
\cdot {\bm g}_{12}^{\ast }\right) \left( \widehat{\bm \sigma}^{\ast
} \cdot {\bm g}_{12}^{\ast} \right)
\nonumber \\
&&\left[ 8 \left( \widehat{\bm \sigma } \cdot {\bm G}_{12}^{\ast}
\right) ^{2}+ ( \widehat{\bm \sigma} \cdot {\bm g}_{12}^{\ast} )^{2}
\right] f_{h}^{\ast \left( 2\right)}\left({\bm \sigma } ^{\ast
},{\bm v}_{1}^{\ast },{\bm v}_{2}^{\ast }\right), \label{5.ThC.I}
\end{eqnarray}
with ${\bm \sigma}^{\ast}= \sigma^{\ast} \widehat{\bm \sigma}^{\ast}
$. Again, the instantaneous contribution to the transport
coefficient is purely collisional, reflecting its origin as the
discontinuity in time for colliding configurations at contact. Also
note that this is the first transport coefficient where the center
of mass momentum of the pair, in addition to the relative velocity,
occurs in the final average.

The corresponding normal fluid expression for this transport
coefficient can be analyzed along the lines done for the viscosities
above. The Green-Kubo integrand for the time integral is a flux-flux
correlation function as in Eq.\ (\ref{5.17}), with both the direct
flux ${\bm S}^{\ast}$ and an conjugate flux, the volume integrated
time reversed energy flux ${\bm S} ^{\ast -} $. This and the
non-vanishing instantaneous part is a peculiarity of the singular
hard sphere interactions and neither effect occurs for nonsingular
forces in the elastic limit. For the granular fluid, the conjugate
flux is further modified by its generation from the HCS.

\subsection{The $\protect\mu $-Coefficient}

The $\mu $ coefficient is a measure of the contribution to heat
transport in the fluid due to spatial gradients in the density
field. In granular fluids, density and temperature are intimately
coupled through the cooling rate. At constant cooling rate, a local
density variation gives rise to a variation in the amount of energy
lost locally and hence can give rise to a heat flux. Since this
transport coefficient is directly related to the inelasticity of the
collisions, it is unique to granular fluids and has no analogue for
normal fluids. The simplest form for this transport coefficient is
obtained when considering the linear combination
\begin{equation}
\overline{\mu }^{\ast }=\mu ^{\ast }- 2\frac{\partial \ln \zeta _{0}}{%
\partial \ln n_{h}}\,  \lambda ^{\ast },  \label{5.7.H}
\end{equation}
where $\lambda ^{\ast }$ is the dimensionless thermal conductivity
discussed in the previous subsection. The intermediate Helfand and
Green-Kubo representations for $\overline{\mu }^{\ast }$ are
identified in Appendix \ref{ap6} as
\begin{equation}
\overline{\mu }^{\ast }=\lim \Omega _{H}^{\overline{\mu }}\left(
s\right) =\Omega _{0}^{\overline{\mu }}+\lim \int_{0}^{s}ds^{\prime
}\Omega _{G}^{\overline{\mu }}\left( s^{\prime }\right) .
\label{5.31}
\end{equation}
The expression of the Helfand correlation function is
\begin{equation}
\Omega _{H}^{\overline{\mu }}\left( s\right) = -(V^{\ast} d)^{-1}
\int d\Gamma ^{\ast }\, {\bm S}^{\ast f}(\Gamma^{\ast}) \cdot
e^{-s\overline{\mathcal{L}}^{\ast }}
\bm{\mathcal{M}}^{\ast}_{\overline{\mu }}(\Gamma^{\ast}),
\label{5.32}
\end{equation}
where ${\bm S}^{\ast f}$ is the subtracted heat flux given in Eq.\,
(\ref{5.6.6}).  The moment $\bm{\mathcal{M}}^{\ast}_{\overline{\mu
}}$ is
\begin{eqnarray}
\bm{\mathcal{M}}^{\ast}_{\overline{\mu }} &=&\left[ l v_{0}\left(
T\right) \right] ^{Nd}\int d{\bm r}\, \widehat{\bm k} \cdot {\bm
r}\left[ n \frac{\delta \rho _{lh}}{\delta n\left({\bm r}\right) }-2
\frac{
\partial \ln \zeta _{0}}{\partial \ln n}T \frac{\delta \rho _{lh}}{
\delta T\left({\bm r} \right) } \right] _{ \{y_{\alpha}\}=\{n,T,{\bm 0}\}} \nonumber \\
&=&\left[ \ell v_{0}\left( T\right) \right] ^{Nd} n \int d{\bm r}\,
\widehat{{\bm k}}^{\ast} \cdot {\bm r}^{\ast} \left[ \left (
\frac{\delta \rho _{lh}[\Gamma|\{y_{\alpha}\}]}{\delta n({\bm r})}
\right)_{\zeta_{0}} \right] _{ \{y_{\alpha}\}=\{n,T,{\bm 0}\}} .
\label{5.34}
\end{eqnarray}
As indicated, in the last expression the functional derivative with
respect to the density field is to be carried out at constant
cooling rate. This suggests that $\overline{\mu }^{\ast }$ is
characterizing density gradients under conditions of constant
cooling rate.

The Green-Kubo correlation functions are
\begin{equation}
\Omega _{0}^{\overline{\mu }}=\Omega _{H}^{\overline{\mu }}\left(
0\right),
\end{equation}
\begin{equation}
\Omega _{G}^{\overline{\mu }}\left( s\right) = - (V^{\ast}d)^{-1}
\int d\Gamma ^{\ast }\, {\bm S}^{\ast f}(\Gamma^{\ast}) \cdot e^{-s
\overline{\mathcal{L}}}\bm{\Upsilon} _{\overline{\mu }}^{\ast}
(\Gamma^{\ast}), \label{5.33}
\end{equation}
with  the associated flux $\bm{\Upsilon} _{\overline{\mu
}^{\ast}}$ having the expression
\begin{equation}
\bm{\Upsilon} _{\overline{\mu }}^{\ast}
=-\overline{\mathcal{L}}^{\ast} \bm{\mathcal{M}}^{\ast}_{
\overline{\mu }}.  \label{5.35}
\end{equation}
The hydrodynamic mode subtracted from the generator
$\overline{\mathcal{L}}^{\ast}$ in this case has the eigenvalue
$\lambda =0$. Unlike the previous transport coefficients discussed
so far, the moment $\bm{\mathcal{M}}_{\overline{\mu }}^{\ast}$ and
the flux $ \bm{\Upsilon} _{\overline{\mu }}^{\ast}$ are still
given implicitly in terms of the \textit{local} HCS state. Further
analysis requires determination of the density dependence of the
\textit{inhomogeneous} HCS, i.e. that in an external conservative
field. This is beyond the discussion here and remains an open
problem for applications of these representations for the $\mu
^{\ast }$ coefficient.

It is shown in Appendix \ref{ap7} that this transport coefficient
vanishes in the elastic limit, as expected. However, that result
is specific to the choice of an equilibrium Gibbs state for the
reference ensemble. More generally, for non-equilibrium reference
states, it is $\mu ^{\ast }\neq 0$. As noted above, the linear
combination $\overline{\mu }^{\ast }=\mu ^{\ast }- 2\frac{
\partial \ln \zeta _{0}}{\partial \ln n_{h}}\,  \lambda ^{\ast }$ is the
transport coefficient measuring the effects of density gradients
on the heat flux when the cooling rate instead of the temperature
is held fixed. The Helfand moment $\bm{\mathcal{M}}_{\overline{\mu
}}^{\ast}$ is the space moment associated with $\widetilde{\psi
}_{1}^{\ast }(\Gamma^{\ast};{\bm 0}) - 2\frac{
\partial \ln \zeta _{0}}{\partial \ln n_{h}} \widetilde{\psi }
_{2}^{\ast }(\Gamma^{\ast};{\bm 0})$. It is shown in Appendix
\ref{ap3}  that this same combination determines the eigenfunction
of the specific hydrodynamic mode solution of Eq.\ (\ref{5.5})
with eigenvalue $\lambda= 0$. Hence, it appears that in some
respects the pair $\delta \zeta _{0},\delta n$ are more natural
independent variables than $\delta T,\delta n$.

\section{Evaluation of Transport Coefficients}

The (intermediate) Helfand and Green-Kubo expressions for the
various transport coefficients obtained above, are formal results in
that the full $N$ particle problem has not yet been solved. Instead,
the linear response method gives a valuable exact and direct
relationship of the macroscopic properties to the microscopic
statistics and dynamics. Their utility is determined by the fact
that they are the ideal setting to explore controlled analytic
approximations and exact numerical evaluations. The transport
coefficients have been obtained in the form of stationary time
correlation functions. This suggests the possible generalization to
granular fluids of the extensively developed many-body tools for the
analysis of time correlation functions of a normal fluid, such as
short time expansions, mode coupling theories, and formal kinetic
theories. Also, for normal fluids such expressions have proved
particularly suitable for evaluation by MD simulation to extend the
studies of hydrodynamic descriptions to otherwise inaccessible
domains in the density of the fluid. In this section, this potential
for future studies is illustrated in two directions. First, the
applicability of a kinetic theory of time correlation functions is
outlined and discussed. Then, a scheme for implementing the
stationary representation of the dynamics considered in this work in
an MD simulation is discussed.

\subsection{Kinetic Theory}

The results for the various transport coefficients obtained in the
present work are expressed in terms of time correlation functions
over an $N$ particle distribution. However, an equivalent expression
involving two particle time-dependent reduced functions is possible
as well. This reduced representation serves as the appropriate
starting point for construction of a kinetic theory for the
correlation functions, and application of formal cluster expansion
techniques. Consider the Helfand representation for a generic
transport coefficient $\varpi^{\ast}$ as given by Eqs.\ (\ref{5.2})
and (\ref{5.4})
\begin{equation}
\varpi^{\ast} =\lim V^{\ast -1}\int d\Gamma ^{\ast }\, F^{\ast
S,f}\left( \Gamma ^{\ast }\right) e^{- s \left(
\overline{\mathcal{L}}^{\ast }-\lambda \right) }
\mathcal{M}^{\ast}\left( \Gamma ^{\ast }\right) .  \label{6.1.1}
\end{equation}
The projected source or flux, $F^{\ast S,f}\left( \Gamma ^{\ast
}\right) $, is the sum of single particle or pair functions in all
cases, so it has the form
\begin{equation}
F^{\ast S,f}\left( \Gamma ^{\ast }\right) =\sum_{r=1}^{N}F_{1}\left(
x_{r}^{\ast }\right) +\sum_{r=1}^{N} \sum_{s \neq r}^{N} F_{2}\left(
x_{r}^{\ast },x_{s}^{\ast }\right) , \label{6.1.2}
\end{equation}
with $x_{r}^{\ast } \equiv \{{\bm q}_{r}^{\ast },{\bm v}_{r}^{\ast
}\}$ denoting the one particle phase point for particle $r$, and
$F_{1}$ and $ F_{2}$  being one and two particle functions of the
phase point, respectively. Define a set of ``reduced correlation
functions'', $\mathcal{M}^{\ast (m)}\left( x_{1}^{\ast
},...,x_{m}^{\ast },s\right) $ by
\begin{equation}
\mathcal{M}^{\ast \left( m\right) }\left( x_{1}^{\ast
},...,x_{m}^{\ast },s\right) \equiv  \frac{N !}{(N-m)!}\int
dx_{m+1}^{\ast } \ldots \int dx_{N}^{\ast }\, e^{-s \left(
\overline{\mathcal{L}}^{\ast }-\lambda \right) } \mathcal{M}^{\ast}
\left( \Gamma ^{\ast }\right) .  \label{6.1.5}
\end{equation}
Then the generic expression for the transport coefficient
(\ref{6.1.1}) becomes
\begin{equation}
\varpi^{\ast} =\lim V^{\ast -1} \left[ \int dx_{1}^{\ast }\,
F_{1}\left( x^{\ast }_{1} \right) \mathcal{ M}^{\ast \left(
1\right) }\left( x_{1}^{\ast },s\right) +\int dx_{1}^{\ast } \int
dx_{2}^{\ast }\, F_{2}\left( x_{1}^{\ast },x_{2}^{\ast }\right)
\mathcal{M} ^{\ast \left( 2\right) }\left( x_{1}^{\ast
},x_{2}^{\ast },s\right) \right] . \label{6.1.4}
\end{equation}
In this way, the full $N$ body correlation function is mapped onto
one that involves only the one and two particle reduced correlation
functions.

It follows by direct differentiation with respect to $s$ of Eq.\
(\ref{6.1.5}) that the functions $\mathcal{M}^{\ast (m)}$ obey a
hierarchy of equations,
\begin{eqnarray}
\left( \frac{\partial}{\partial s}-\lambda +
\overline{\mathcal{L}}^{\ast (m)}\right) && \mathcal{M}^{\ast
\left( m\right) }\left( x_{1}^{\ast }, \ldots ,x_{m}^{\ast
},s\right)
\nonumber \\
& & =\sum_{r=1}^{m}\int dx_{m+1}^{\ast }\overline{T}^{\ast }\left(
r,m+1\right) \mathcal{M}^{\ast \left( m+1\right) }\left(
x_{1}^{\ast }, \ldots ,x_{m+1}^{\ast },s\right), \label{6.1.6}
\end{eqnarray}
where $\overline{\mathcal{L}}^{\ast (m)}$ is the same operator as
$\overline{ \mathcal{L}}^{\ast}$ but now defined for only $m$
particles, and the $\overline{T}^{\ast}(r,s)$ operator is the
dimensionless form of the operator defined in Eq.(\ref{A21a}). This
hierarchy is formally the same as the dimensionless BBGKY hierarchy
of the reduced distribution functions representing the
non-equilibrium state of the fluid \cite{McLbook,ResiboisBook}. Only
the initial conditions are different.

Formally, a kinetic theory represents a closure of the hierarchy at
the level of the first equation, through the identification of a
representation for the two body function $\mathcal{M} ^{\ast \left(
2\right) }\left( x_{1}^{\ast },x_{2}^{\ast },s\right) $ as a
functional of the one body function $\mathcal{M}^{\ast \left(
1\right) }\left( x_{1}^{\ast },s\right) $. With such a functional
relationship, the first hierarchy equation becomes a closed,
autonomous equation for $\mathcal{M} ^{\ast \left( 1\right) }\left(
x_{1}^{\ast },s\right) $. This equation is called a kinetic
equation, and the functional relationship defines the kinetic
theory. It is important to recognize that this concept of a kinetic
theory does not imply any \textit{a priori }restrictions to low
density, weak inelasticity, or absence of correlations. Such
restrictions occur only in the justification of specific choices for
approximate closure functionals.

To obtain a functional relationship for $\mathcal{M}^{\ast \left(
2\right) }\left( x_{1}^{\ast },x_{2}^{\ast },s\right) $, it is
postulated that there exists a functional expressing the two body
reduced distribution function associated with the local HCS in terms
of the one body distribution function,
\begin{equation}
f_{lh}^{\left( 2\right) }\left( x_{1},x_{2},t\right)
=f_{lh}^{\left( 2\right) }\left[ x_{1},x_{2},t|f^{\left(
1\right)}_{lh}  \right] , \label{6.1.7}
\end{equation}
where the reduced local HCS distributions are defined by
\begin{equation}
f_{lh}^{\left( m\right) }\left( x_{1} \ldots ,x_{m},t\right) \equiv
\frac{N !}{(n-m)!} \int dx_{m+1} \ldots \int dx_{N}\, e^{-t
\overline{L}}\rho _{lh}\left( \Gamma \right) . \label{6.1}
\end{equation}
The formal construction of a functional like  in  Eq.\
(\ref{6.1.7}) for normal fluids, where $\rho _{lh}$ is replaced by
the local equilibrium ensemble, is a well-studied problem, e.g.
inversion of formal cluster expansions for the reduced
distribution functions. This is the first point at which
extensions of kinetic theory to granular fluids can be explored.
From the definition of $\mathcal{M}^{\ast}$ in Eq.\ (\ref{5.4a}),
it follows that Eq.\ (\ref{6.1.7}) implies a similar but linear
functional relationship for $\mathcal{M}^{\ast \left( 2\right) }$
,
\begin{equation}
\mathcal{M}^{\ast \left( 2\right) }\left( x_{1}^{\ast },x_{2}^{\ast
},s\right) =\int dx^{\ast \prime}\, K\left( x_{1}^{\ast
},x_{2}^{\ast },x^{\ast \prime },s\right) \mathcal{M}^{\ast \left(
1\right) }\left( x^{\ast \prime },s\right), \label{6.1.8}
\end{equation}
with
\begin{equation}
K\left( x_{1}^{\ast },x_{2}^{\ast },x^{\ast \prime },s\right)
=\left[ lv_{0}\left( T_{h}\right) \right] ^{d} \left[ \frac{\delta
f_{lh}^{\left( 2\right) }\left[ x_{1},x_{2},t|f_{lh}^{\left(
1\right) }\right] }{\delta f_{lh}^{\left( 1\right) }\left( x^{\prime
},t\right) }\right]_{\{y_{\alpha}\}=\{n_{h},T_{h},{\bm 0} \}}.
\label{6.1.9}
\end{equation}
The linear kernel $K$ is a functional derivative of the two body
distribution function with respect to the one body distribution,
evaluated in the homogeneous limit. This latter evaluation leads to
the linearity of this functional. The pre-factor in the above
expression just transforms to dimensionless variables as in Eq.\
(\ref {1.5.4}). Substitution of  this form for the two body
correlation function in the first equation of the hierarchy leads
the kinetic equation
\begin{equation}
\left( \partial _{s}-\lambda +\overline{\mathcal{L}}^{\ast \left(
1\right) }- \mathcal{I}\left( s\right) \right) \mathcal{M}^{\left(
1\right) }\left( x_{1}^{\ast },s\right) =0,  \label{6.1.10}
\end{equation}
with the formal linear ``collision operator''
\begin{equation}
\mathcal{I}\left( s\right) \mathcal{M}^{\ast \left( 1\right)
}\left( x_{1}^{\ast },s\right) \equiv \int dx_{2}^{\ast }\,
\overline{T}^{\ast} \left( 1,2\right) \int dx^{\ast \prime
}K\left( x_{1}^{\ast },x_{2}^{\ast },x^{\ast \prime },s\right)
\mathcal{M}^{\ast \left( 1\right) }\left( x^{\ast \prime
},s\right). \label{6.1.9a}
\end{equation}
This is the kinetic equation governing the dynamics of
$\mathcal{M}^{\ast \left( 1\right) }$. Since this quantity also
determines $\mathcal{M} ^{\ast \left( 2\right) }$ through the
functional relationship given by Eq.\ (\ref{6.1.8}), the transport
coefficient can be calculated from Eq.\ (\ref{6.1.4}).

The above is exact but formal, and the aim here is only to provide a
flavor for the potential for developing a practical kinetic theory
for time correlation functions. Details of this method and the
explicit evaluation of the various transport coefficients an
Enskog-like approximation will be given elsewhere \cite{kinetic}.
The latter is obtained by estimating $ f^{\left( 2\right) }\left[
x_{1},x_{2},t|f^{\left( 1\right) } \right] \approx f^{\left(
2\right) }\left[ x_{1},x_{2},0|f^{\left( 1\right) } \right] $, which
is exact at $t=0$, and furthermore neglecting velocity correlations.
The resulting kinetic theory for the correlation functions leads to
cooling rate, pressure, and transport coefficients identical to
those from the Chapman-Enskog solution to the Enskog kinetic theory
\cite{GyD02,Lu05}. At low density, these results agree with those
from the Boltzmann equation \cite{BDKS}.

\subsection{Molecular Dynamics Simulation}

The formal expressions for the various transport coefficients
derived in this work, are given in the dimensionless representation
for the phase space dynamics that leave the HCS ensemble stationary
and defined by the thermal velocity associated to the HCS.   But for
practical purposes, this is a complicated representation as it must
be solved self consistently with the cooling equation. It is then
convenient to consider a different scaling, in which the function
$v_{0}(t)$ is replaced by a known function chosen in a appropriate
way. A useful choice is defined by
\begin{equation}
{\bm q}^{**}_{r}= {\bm q}^{\ast}_{r}= \frac{{\bm q}^{\ast}_{r}}{l},
\quad {\bm v}^{**}_{r} = \frac{\xi_{0} t}{2 l}\, {\bm v}_{r}
\label{6.2.1}
\end{equation}
where $ \xi_{0}$ is an arbitrary time-independent dimesnionless
frequency. The rational for his choice and some more details are
given in Appendix \ref{ap8}. The particle dynamics in the new
variables consists of an acceleration streaming between collisions,
\begin{equation}
\frac{\partial {\bm q}{**}(\tau)}{\partial \tau}= {\bm
v}^{**}(\tau), \label{6.2.2}
\end{equation}
\begin{equation}
\frac{\partial {\bm v}^{**}(\tau)}{\partial \tau}=
\frac{\xi_{0}}{2}\, {\bm v}^{**}(\tau), \label{8.2.3}
\end{equation}
while the effect of a collision between two particles $r$ and $s$ is
to instantaneously alter their relative velocity according to the
same rule as in the original scale. This invariance of the
interation law is another consequence of the singular character of
hard particle collisions.

The quantity $\xi_{0}$ and the dimensionless cooling rate
$\zeta_{0}^{\ast}$ are related through the steady state temperature
obtained in the simulations in the scaled variables
\cite{lutsko2,LByD02}. The relationship is given in Eq.\
(\ref{H11}). This method has already been implemented in
event-driven molecular dynamics simulation to measure the
self-diffusion transport coefficient of a granular fluid
\cite{LByD02}. That is a simpler case than those considered here,
since the correlation functions appearing in its Helfand and
Green-Kubo expressions are not moments of the conjugate densities,
but simply the velocities of the particles.

The low density limits of the Green-kubo expressions for a granular
gas have been evaluated by means of the direct simulation Monte
Carlo method in refs. \cite{ByR04} and \cite{BRMyG05}, also by using
the steady representation introduced above. The conjugate fluxes
were determined from the simulation themselves. Although the
technical problems are much harder beyond the low density limit,
this shows the way to evaluate the general expression derived here
at arbitrary densities.

\section{Discussion}

The parameters of Navier-Stokes hydrodynamics for a system of
smooth, inelastic hard spheres with constant restitution coefficient
have been given formally exact representations starting from the
underlying non-equilibrium statistical mechanics. General linear
response functions for small initial spatial perturbations of the
HCS are given by Eq.\ (\ref{1.5.1}). These are stationary time
correlation functions, composed of densities of the hydrodynamic
fields and corresponding conjugate densities defined by functional
derivatives of the local HCS with respect to these  fields. They are
similar to the response functions for a normal fluid,  where the
local HCS is replaced by a local equilibrium ensemble.  Such
response functions provide the basis for exploration  of a wide
range of phenomena in granular gases at the fundamental  level of
statistical mechanics, beyond the specific case of  transport
coefficients considered here \cite{Boon80}.

In the long wavelength, long time limit these response functions
exhibit the hydrodynamic behavior of the system. This is the
analogue of the Onsager regression property for normal fluids
\cite{Onsager,Kadanoff}. This limit has been extracted by defining
a general transport matrix for the response functions, and
expanding it to order $k^2$ to identify the Navier-Stokes
parameters in terms of corresponding HCS averages or stationary
time  correlation functions. The cooling rate is given as the HCS
average  rate of change of the global energy, and the pressure is
expressed as  the average fluctuation in the global momentum flux.
Such averages can be  studied directly by MD simulation.
Furthermore, they have been expressed in terms  of the reduced two
particle distribution function for pairs of particles  at contact.
This representation is suitable for inquiry about new features of
granular fluids, such as two particle velocity correlations.

The transport coefficients obtained in this way have been considered
in two representations, the (intermediate) Helfand and Green-Kubo
forms. The former is the long time limit of a correlation function
constructed from a global flux associated with the energy or
momentum densities, and a space moment of one of the conjugate
densities. This is quite similar to the corresponding Helfand
results for a normal fluid, if the  HCS ensemble is replaced by the
equilibrium Gibbs ensemble. There is an important additional
difference, however, in the generator for the time dependence. The
dimensionless generator for hard sphere trajectories
$\overline{L}^{\ast}$ is replaced by $\overline{\mathcal{L}}^{\ast}$
of Eq.\ (\ref{1.5.5}), which contains an additional velocity scaling
operator. The latter represents the cooling of the reference HCS and
it is essential for the stationary representation given here. In
addition, the time correlation functions for transport coefficients
have the further modification $ \overline{\mathcal{L}}^{\ast}
\rightarrow \overline{\mathcal{L}}^{\ast}-\lambda $, where $ \lambda
$ is one of the eigenvalues of the hydrodynamic transport matrix
$\mathcal{K} ^{\ast hyd }\left({\bm 0} \right) $. These eigenvalues
describe the dynamics of homogeneous perturbations to the HCS. Since
the transport coefficients measure only spatial correlations, it is
perhaps not surprising that this homogeneous dynamics does not occur
in their representation. These same eigenvalues occur in the exact
spectrum of $\overline{\mathcal{L}}^{\ast}$, so this generator
$\overline{\mathcal{L}}^{\ast}-\mathcal{K}^{\ast hyd }\left( {\bm 0}
\right) $ has invariants (zeros of the spectrum). In the elastic
limit, these invariants are proportional to the usual global
conserved total number, energy, and momentum.

The equivalent Green-Kubo expressions have a time independent part
plus the familiar time integral of a flux-flux correlation
function. The time independent part is absent for  normal fluids
with conservative, nonsingular forces. More generally, for either
singular forces (e.g., hard spheres) and/or nonconservative
forces, this contribution is nonzero. For normal fluids, it can be
evaluated exactly and represents the high density contribution to
the transport coefficients in the Enskog kinetic theory. Here, it
can be reduced to an average over the two particle distribution
function, just as described for the cooling rate and pressure. The
flux-flux time correlation functions in the Green-Kubo integrands
are constructed from a global flux of the energy or momentum
densities and a global flux associated with the conjugate
densities. The dynamics is generated by $\overline{
\mathcal{L}}^{\ast}-\mathcal{K}^{\ast hyd }\left( {\bm 0} \right)
$ as described above, so the associated invariants must not occur
in order for the time integral to exist in the long time limit. In
fact, the expressions obtained have the fluxes orthogonal to these
invariants. This is the analogue of ``subtracted fluxes'' for a
normal fluid \cite{McLbook}.

There are additional transport coefficients associated with
spatial gradients of the hydrodynamic fields contribution to the
cooling rate. These also have Helfand and Green-Kubo forms as
described above, but with the global flux of energy replaced by
the volume integrated source in the microscopic balance equation
for energy. One of these transport coefficients, $\zeta ^{U} $,
appears at the Euler level hydrodynamics and gives a correction to
the pressure in the coefficient of $\mathbf{\nabla }\cdot
\mathbf{U}$. The other
two give corrections to the macroscopic heat flux proportional to $\mathbf{%
\nabla }T$ and $\mathbf{\nabla }n.$ These vanish for a normal
fluid.

As with normal fluids, the value of such formal expressions is
direct access to the transport coefficients before approximations
are introduced. Often this avoids (or postpones) conceptual
difficulties associated with the particular many-body method used
for evaluation. Here, the two most common approaches for normal
fluids have been briefly discussed for extension to granular
fluids. The first is a kinetic theory for time correlation
functions. It is inherently simpler than constructing a
corresponding kinetic theory for the reduced distribution
functions, since the latter are nonlinear while that for the
correlation functions is linear. Furthermore, the special solution
required for hydrodynamics (normal solution and Chapman-Enskog
construction) has already been incorporated  implicitly in the
linear response analysis here. The necessary functional assumption
for a kinetic theory of granular time correlation functions has
been identified as a property of the local HCS. In this restricted
context, it should be possible to explore the validity or failure
of such an assumption, although the detailed investigations for
normal fluids need to be reconsidered. It is shown elsewhere
\cite{kinetic} that an ad hoc neglect of velocity correlations,
something that is justified in the case of normal fluids, leads to
transport coefficients identical to those from the Enskog kinetic
theory. The formalism developed in this paper and the preceding
one, provides the basis for critiquing that theory and correcting
it for granular fluids.

The second approach for evaluating the formal expressions given
here is MD simulation. In this case, the extension of approach
from normal fluids is not direct either. The conjugate moments
(Helfand representation) or the conjugate fluxes (Green-Kubo
representation) are not known explicitly, but rather given by
derivatives of the HCS ensemble. Thus the simulation of these
expressions requires new method for developing properties of the
HCS from the simulation itself.

\section{Acknowledgements}

The research of J.D. and A.B. was supported in part by the
Department of Energy Grant (DE-FG03-98DP00218). The research of
J.J.B. was partially supported by the Ministerio de Educaci\'{o}n
y Ciencia (Spain) through Grant No. BFM2005-01398. This research
also was supported in part by the National Science Foundation
under Grant No. PHY99-0794 to the Kavli Institute for Theoretical
Physics, UC Santa Barbara. A.B. also acknowledges a McGinty
Dissertation Fellowship and IFT Fellowship from the University of
Florida.

\appendix

\section{Hard Sphere Dynamics}
\label{ap1} In this Appendix, the details of the dynamics and
statistical mechanics of a system of hard particles is given.
Consider a system of $N$ mono-disperse smooth hard spheres ($d=3$)
or disks ($d=2$) of mass $m$ and diameter $\sigma $. A complete
specification of the initial state of the system involves knowing
a point in the $2Nd$-dimensional phase space $\Gamma \equiv \{
{\bm q} _{r},{\bm v}_{r};r=1,\ldots, N \}$, that gives the
position ${\bm q}_{r}$ and velocity ${\bm v}_{r}$ of each particle
$r$ of the system. At a later time $ t$, the state will be
characterized by another point in phase space,  $\Gamma _{t}\equiv
\left\{ {\bm q}_{r}(t),{\bm v}_{r}(t);r=1, \ldots, N \right\}$.
The dynamics of the particles consists of free streaming (straight
line motion along the direction of the velocity), until a pair of
particles, $r$ and $s$, are at contact, at which time their
velocities ${\bm v}_{r},{\bm v}_{s}$ change instantaneously to
${\bm v}_{r}^{\prime },{\bm v}_{s}^{\prime }$ according to the
collision rule
\begin{eqnarray}
{\bm v}_{r}^{\prime } & = &{\bm v}_{r}-\frac{1+ \alpha}{2}
(\widehat{\bm q}_{rs} \cdot {\bm g}_{rs} ) \widehat{\bm q}_{rs},
\nonumber \\
{\bm v}_{s}^{\prime } & = &{\bm v}_{s}+\frac{1+ \alpha}{2}
(\widehat{\bm q}_{rs} \cdot {\bm g}_{rs} ) \widehat{\bm q}_{rs},
  \label{a.7}
\end{eqnarray}
where $\widehat{\bm q}_{rs}$ is a unit vector along the relative
position ${\bm q}_{rs}={\bm q}_{r}-{\bm q}_{s}$ at contact and
$\alpha$ is the coefficient of normal restitution, defined in the
interval $ 0 < \alpha \leq  1$ and that will be taken here as
constant, i.e. velocity independent. In terms of the relative
velocity ${\bm g}_{rs}={\bm v}_{r}-{\bm v}_{s}$ and the center of
mass velocity ${\bm G}_{rs}=\left({\bm v}_{r}+ {\bm v}_{s}\right)
/2$, the above collision rule reads
\begin{equation}
{\bm G}_{rs}^{\prime }={\bm G}_{rs}, \quad {\bm g}_{rs}^{\prime
}={\bm g}_{rs}-\left( 1+\alpha \right) \left( \widehat{\bm q}_{rs}
\cdot {\bm g}_{rs}\right) \widehat{\bm q}_{rs}. \label{a.8}
\end{equation}
The energy loss on collision is
\begin{equation}
\Delta E=\frac{m}{4} \left( g_{rs}^{\prime 2}-g_{rs}^{2}\right)
=-\frac{m}{4} \left( 1-\alpha ^{2}\right) \left( \widehat{\bm
q}_{rs} \cdot {\bm g}_{rs}\right) ^{2}. \label{a.9}
\end{equation}
It is seen that $\alpha =1$ corresponds to elastic collisions.
Subsequent to the change in relative velocity for the pair $\{ r,s
\}$, the free streaming of all particles continues until another
pair is at contact, and the corresponding instantaneous change in
their relative velocities is performed. The sequence of free
streaming and binary collisions determines a unique trajectory in
phase space, $\Gamma _{t}$, for given initial conditions. The
collision rule is invertible so the trajectory can be reversed,
although with the inverted collision rule (``restituting''
collisions).

The statistical mechanics for this system \cite{Brey97} is
comprised of the dynamics just described, an initial macrostate
specified in terms of a probability density $\rho (\Gamma )$, and
a set of microscopic observables (measurables). The macroscopic
variables of interest are the expectation value for the
observables at time $t>0$ for an initial state $\rho (\Gamma )$.
Given an observable $A(\Gamma)$, its expectation value is defined
by the two equivalent forms
\begin{equation}
\langle A(t);0\rangle \equiv \int d\Gamma\, \rho (\Gamma
)e^{Lt}A(\Gamma )=\int d\Gamma\, A(\Gamma)  e^{-\overline{L}t}\rho
(\Gamma ). \label{a.9.a}
\end{equation}
In the above expression, $L$ is the generator of the dynamics of
phase functions and $\overline{L}$ is the generator of dynamics
for distribution functions. The explicit forms of these generators
will be  derived in the following.

\subsection{Generator of Trajectories}

The hard sphere dynamics described above is characterized by
piecewise constant velocities that change instantaneously (and
discontinuously) at the time of collision. This allows the generator
of trajectories to be derived using geometric arguments
\cite{lutsko1}. Consider first a system of two inelastic hard
spheres or disks. The particles move freely until they eventually
contact, at which time their velocities change instantaneously
according to Eq.\ (\ref{a.7}) above. Suppose that the two particles,
named $1$ and $2$, respectively, collide at time $\tau$. This time
is determined as a function of the initial separation ${\bm q}_{12}$
and relative velocity ${\bm g}_{12}$ from
\begin{equation}
{\bm q}_{12}+{\bm g}_{12} \tau = {\bm \sigma}, \label{a.9.b}
\end{equation}
with ${\bm \sigma}$ being an arbitrary vector of modulus $\sigma$.
Of course, if this equation does not have  a real and positive
solution for $\tau$, the particles do not collide. Then, for $ 0 < t
\leq \tau $ the trajectory is given by free streaming with the
initial velocities, i.e.  $\Gamma \left( t\right) = \Gamma_{0}(t)
\equiv \left\{ {\bm q}_{t}+{\bm v}_{r}t,{\bm v}_{r }; r=1,2
\right\}$. For $t>\tau $, it is $\Gamma \left( t\right) =
\Gamma^{\prime}(t) =\left\{ {\bm q}_{r}+ {\bm v}_{r}\tau +{\bm
v}_{r}^{\prime }\left( t-\tau \right) ,{\bm v} _{r}^{\prime };
r=1,2\right\} $, where the velocities ${\bm v}_{r}^{\prime }$ are
determined by the collision rule (\ref{a.7}). Therefore, the value
of any phase function $A\left( \Gamma \right) $ at time $t$ can be
given compactly expressed as
\begin{equation}
A\left[ \Gamma (t) \right] =\left\{ 1-\Theta \left[ t-\tau \left(
\Gamma \right) \right] \right\} A\left[ \Gamma _{0}\left( t\right)
\right] +\Theta \left[ t-\tau \left( \Gamma \right) \right] A\left[
\Gamma^{\prime} \left( t\right) \right] ,  \label{a.10}
\end{equation}
where $\Theta (x) $ is the Heaviside step function, defined by
$\Theta(x)=0$ for $x \leq 0$ and $\Theta (x)=1$ for $x>0$.
Differentiation with respect to time of the above equation leads to
\begin{eqnarray}
\frac{\partial A\left[ \Gamma \left( t\right) \right] }{\partial t}
& = & \sum_{r=1}^{2} {\bm v}_{r}\left( t\right) \cdot \frac{\partial
A\left[ \Gamma \left( t\right) \right] }{\partial {\bm q}_{r}\left(
t\right) } +\delta \left[ t-\tau \left( \Gamma \right) \right]
\left\{ A\left[ \Gamma ^{\prime}\left( t\right) \right]- A \left[
\Gamma_{0}(t) \right] \right\} \nonumber \\
& = & \sum_{r=1}^{2} {\bm v}_{r}\left( t\right) \cdot \frac{\partial
A\left[ \Gamma \left( t\right) \right] }{\partial {\bm q}_{r}\left(
t\right) } +\delta \left[ t-\tau \left( \Gamma \right) \right]
\left( b_{12}-1\right) A\left[ \Gamma \left( t\right) \right].
\label{a.11}
\end{eqnarray}
Here, $b_{12}$ is a substitution operator that changes the
velocities into their postcollisional values as given by Eq.\
(\ref{a.7}),
\begin{equation}
b_{12} X \left( {\bm v}_{1}, {\bm v}_{2} \right) =X\left( b_{12}{\bm
v}_{1}, b_{12} {\bm v}_{2} \right) =X\left({\bm v}^{\prime}_{1},{\bm
v}^{\prime}_{2} \right) . \label{a.12}
\end{equation}
Of course, the vector $\widehat{\bm q}_{12}$ to be used in the
collision rule is ${\bm q}_{12}(\tau) /q_{12}(\tau)$.

It is convenient to substitute the delta function involving the time
$t$ in Eq.\ (\ref{a.11}) by a more geometrical quantity. To begin
with, notice that not every solution of Eq.\ (\ref{a.9.b})
corresponds to a physical collision. In fact, it is easily realized
that for every contact representing a physical collision, there is
another solution to Eq.\ (\ref{a.9.b}) corresponding to the time at
which the two particles would separate if they were allowed to
stream through each other. The physical value of $\tau$ can be
identified as that for which $\widehat{\bm q}_{12}(\tau) \cdot {\bm
g}_{12} <0$, while for the unphysical solution of Eq.\ (\ref{a.9.b})
it is $\widehat{\bm q}_{12}(\tau) \cdot {\bm g}_{12} > 0$. Next,
take into account that the solutions of Eq.\ (\ref{a.9.b}) are the
same as those of $q_{12}(\tau)-\sigma=0$. Hence
\begin{eqnarray}
\delta \left[t-\tau(\Gamma) \right] & =& \Theta  \left[ -
\widehat{\bm q}_{12}(t) \cdot {\bm g}_{12}(t) \right] \delta \left[
q_{12}(t)-\sigma \right] \left| \frac{\partial
q_{12}(t)}{\partial t} \right| \nonumber \\
&=& \Theta  \left[ - \widehat{\bm q}_{12}(t) \cdot {\bm g}_{12}(t)
\right] \delta \left[ q_{12}(t)-\sigma \right] \left| \widehat{\bm
q}_{12}(t) \cdot {\bm g}_{12}(t) \right|. \label{a.14}
\end{eqnarray}
Use of this in Eq.\  (\ref{a.11}) yields
\begin{eqnarray}
\frac{\partial A\left[ \Gamma \left( t\right) \right] }{\partial
t} &=&\sum_{r=1}^{2}{\bm v}_{r}\left( t\right) \cdot
\frac{\partial A \left[ \Gamma \left( t\right) \right] }{\partial
{\bm q}_{r}\left( t\right) } +\delta \left[ q_{12}\left( t\right)
-\sigma \right] \Theta \left[ -\widehat{\bm q}_{12}\left( t\right)
\cdot {\bm g}_{12}\left( t\right) \right] \left| \widehat{\bm
q}_{12}\left( t\right) \cdot {\bm g}_{12}\left( t\right) \right|
\nonumber \\
&& \times \left( b_{12}-1\right) A\left[ \Gamma \left( t\right)
\right]. \label{a.15}
\end{eqnarray}
The generator of trajectories is defined by
\begin{equation}
A\left[ \Gamma \left( t\right) \right] =e^{Lt}A\left( \Gamma
\right) , \label{a.17}
\end{equation}
so $L$ is identified as
\begin{equation}
L=\sum_{r=1}^{2}{\bm v}_{r}\cdot \frac{\partial }{\partial {\bm
q}_{r}} +T\left( 1,2\right),  \label{a.18}
\end{equation}
with the binary collision operator $T(1,2)$ given by
\begin{equation}
T(1,2)=\delta \left( q_{12}-\sigma \right) \Theta \left(
-\widehat{\bm q}_{12} \cdot {\bm g}_{12}\right) \left|
\widehat{\bm q}_{12}\cdot {\bm g}_{12}\right| \left(
b_{12}-1\right) . \label{a.16}
\end{equation}
This expression can be rewritten in a form that is often more
suitable for explicit calculations, by employing the relation
\begin{equation}
\int d \widehat{\bm \sigma}\, \delta ({\bm q}_{12}-{\bm \sigma}) =
\sigma^{-(d-1)} \delta (q_{12}-\sigma),
\end{equation}
where $d\widehat{\bm \sigma}$ is the solid angle element
associated to ${\bm \sigma}$. In this way, it is obtained,
\begin{equation}
T(1,2) =\sigma^{d-1} \int d \widehat{\bm \sigma}\, \delta ( {\bm
q}_{12} -{\bm \sigma}) \Theta (- \widehat{\bm \sigma} \cdot {\bm
g}_{12} ) \left| \widehat{\bm \sigma} \cdot {\bm g}_{12} \right |
(b_{12}-1). \label{a.16.a}
\end{equation}

The generator of trajectories for the system of $N$ particles
considered here, entails the additional \textit{assumption} that
only binary collisions occur between particles, i.e.,  there are no
collisions involving three or more particles simultaneously at
contact. Then, for a system of $N$ particles,
\begin{equation}
L=\sum_{r=1}^{N}{\bm v}_{r}\cdot \frac{\partial}{\partial {\bm
q}_{r}}+\frac{1}{2} \sum_{r}^{N} \sum_{s\neq r}^{N} T\left(
r,s\right). \label{a.19}
\end{equation}

\subsection{Adjoint Dynamics}

Next, consider the generator $\overline{L}$ defined by the second
equality in Eq. (\ref{a.9.a}). It can be identified from
\begin{eqnarray}
\int d\Gamma\,  A(\Gamma )\overline{L}\rho \left[ \Gamma \right]
&\equiv &-\int
d\Gamma \left[ LA(\Gamma )\right] \rho (\Gamma )  \nonumber \\
&=&-\sum_{r=1}^{N}\int d\Gamma\, \rho \left( \Gamma \right) {\bm
v}_{r} \cdot \frac{\partial}{\partial {\bm q}_{r}} A\left( \Gamma
\right) -\frac{1}{2} \sum_{r}^{N} \sum_{s \neq r}^{N} \int
d\Gamma\, \rho \left( \Gamma \right) \delta \left(
q_{rs}-\sigma \right)  \nonumber \\
&&\times \Theta \left( -\widehat{\bm q}_{rs}\cdot {\bm
g}_{rs}\right) \left| \widehat{\bm q}_{rs}\cdot {\bm
g}_{rs}\right| \left( b_{rs}-1\right) A\left( \Gamma \right) .
\label{a.20}
\end{eqnarray}
Define the inverse, $b_{rs}^{-1}$, of $b_{rs}$ by
$b_{rs}^{-1}b_{rs}=b_{rs}b_{rs}^{-1}=1$. Equation (\ref{a.8}) gives
directly
\begin{equation}
b_{rs}^{-1}{\bm g}_{rs}\equiv {\bm g}_{rs}^{\prime \prime }={\bm
g}_{rs}-\frac{1+\alpha}{ \alpha}\, \left( \widehat{\bm q}_{rs}
\cdot {\bm g}_{rs}\right) \widehat{\bm q}_{rs}. \label{a.23}
\end{equation}
A useful identity is given by
\begin{equation}
\int d\Gamma\,  X \left( \Gamma \right) b_{rs}Y\left( \Gamma
\right) =\int d\left( b_{rs}^{-1}\Gamma \right) X\left(
b_{rs}^{-1}\Gamma \right) Y\left( \Gamma \right) = \alpha ^{-1}
\int d\Gamma\, X\left( b_{rs}^{-1}\Gamma \right) Y\left( \Gamma
\right) , \label{a.24}
\end{equation}
for arbitrary functions $X(\Gamma)$ and $Y (\Gamma)$. The first
equality is obtained by changing integration variables from $ \left(
{\bm v}_{r},{\bm v}_{s}\right) $ to $\left( b_{rs} {\bm
v}_{r},b_{rs}{\bm v}_{s}\right) $. The factor $\alpha ^{-1}$ is the
Jacobian for this change of variables. Also,
\begin{equation}
b_{rs}^{-1}\left( \widehat{\bm q}_{rs}\cdot {\bm g}_{rs}\right)
=-\alpha ^{-1} \widehat{\bm q}_{rs} \cdot {\bm g}_{rs}.
\label{a.24a}
\end{equation}
Equation (\ref{a.20}) therefore can be rewritten as
\begin{eqnarray}
\int d\Gamma\,  A(\Gamma )\overline{L}\rho \left( \Gamma \right)
&=&\sum_{r=1}^{N} \int d\Gamma\, A \left( \Gamma \right) {\bm
v}_{r}\cdot \frac{\partial}{\partial {\bm q}_{r}} \rho \left(
\Gamma \right) - \sum_{r=1}^{N} \int_{S} d{\bm S}_{r} \cdot \int d
{\bm v}_{r } \int d\Gamma^{(r)}  {\bm v}_{r}\rho \left( \Gamma
\right) A\left( \Gamma \right) \nonumber \\
&&-\frac{1}{2}\sum_{r=1}^{N} \sum_{s \neq r}^{N} \int d\Gamma\,
A\left( \Gamma \right) \delta \left( q_{rs}-\sigma \right) \left|
\widehat{\bm q}_{rs}\cdot {\bm g}_{rs}\right|  \nonumber  \\
&&\times \left[ \Theta \left( \widehat{\bm q}_{rs}\cdot {\bm
g}_{rs}\right) \alpha ^{-2}b_{rs}^{-1}-\Theta \left( -\widehat{\bm
q}_{rs}\cdot {\bm g}_{rs}\right) \right] \rho \left( \Gamma
\right) .  \label{a.24b}
\end{eqnarray}
An integration by parts has been performed in the first term,
leaving a surface integral for which boundary conditions must be
specified. Then $d {\bm S}_{r}$ is the surface element vector
associated to ${\bm q}_{r}$ and $d\Gamma^{(r)}$ is the element of
volume of the $2(N-1)d$ phase space obtained by eliminating ${\bm
q}_{r}$ and ${\bm v}_{r}$ in the original one. This surface term
identically vanishes when, for instance, periodic boundary
conditions are considered for the system and $A(\Gamma)$ is a
compact function of the positions. In the following it will be
always considered that this is the case. The adjoint Liouville
operator is formally identified from the other non-surface terms as
\begin{equation}
\overline{L}=\sum_{i=r}^{N} {\bm v}_{r}\cdot
\frac{\partial}{\partial {\bm q}_{r}}-\frac{1}{2}\sum_{r}^{N}
\sum_{s \neq r}^{N}\overline{T}(r,s), \label{A20}
\end{equation}
where the new binary collision operator is
\begin{equation}
\overline{T}(r,s)=\delta (q_{rs}-\sigma )|\widehat{\bm
q}_{rs}\cdot {\bm g }_{rs}|\left[ \Theta \left( \widehat{\bm
q}_{rs}\cdot {\bm g}_{rs}\right) \alpha ^{-2}b_{rs}^{-1}-\Theta
\left( -\widehat{\bm q}_{rs}\cdot {\bm g}_{rs}\right) \right] .
\label{A21}
\end{equation}
The integral representation of this operator, similar to Eq.\
(\ref{a.16.a}), is
\begin{equation}
\overline{T}(r,s)= \sigma^{d-1} \int d \widehat{\bm \sigma}\,
\Theta (\widehat{\bm \sigma} \cdot {\bm g}_{rs} ) |\widehat{\bm
\sigma} \cdot {\bm g}_{rs}| \left[ \alpha^{-2} \delta ( {\bm
q}_{rs}-{\bm\sigma} ) b_{rs}^{-1}- \delta ({\bm q}_{rs}+{\bm
\sigma}) \right]. \label{A21a}
\end{equation}

\subsection{Time Correlation Functions and Time Reversed Generator}

The response functions in statistical mechanics take the form of
time correlation functions over the macrostate under consideration,
and can be written in two equivalent ways, corresponding to each of
the two representations in Eq.\ (\ref{a.9.a}),
\begin{equation}
C_{AB}(t)=\int d\Gamma\, \left[ e^{tL}A(\Gamma )\right]  B(\Gamma
)\rho (\Gamma )=\int d\Gamma\  A(\Gamma )e^{-t\overline{L}}\left[
\rho (\Gamma )B(\Gamma )\right], \label{a.2.5}
\end{equation}
where $A(\Gamma)$ and $B(\Gamma)$ are two phase functions. A third
representation can be identified in the form
\begin{equation}
C_{AB}(t)=\int d\Gamma\,  A(\Gamma )\left[ e^{-tL_{-}}B\left( \Gamma
\right) \right] e^{- t \overline{L}} \rho \left( \Gamma \right) .
\label{a.4}
\end{equation}
Comparison of Eqs.\ (\ref{a.2.5}) and (\ref{a.4}) shows that the
operator $L_{-}$ must verify the relation
\begin{equation}
\overline{L} (\rho B)=(\overline{L} \rho ) B+ \rho L_{-} B .
\label{A26}
\end{equation}
It follows that
\begin{equation}
\rho L_{-}B= \rho \sum_{r=1}^{N} {\bm v}_{r} \cdot \frac{\partial
B}{\partial {\bm q}_{r}}  -\frac{1}{2} \sum_{r=1}^{N} \sum_{s \neq
r}^{N} \left[ \overline{T}(r,s)(\rho B)-B \overline{T}(r,s) \rho
\right]. \label{A26.a}
\end{equation}
Next, consider the action of $\overline{T}(r,s)$ on $\rho B$ in
detail using the definition (\ref{A21}),
\begin{eqnarray}
\overline{T}(r,s)\left( \rho B \right) &=&\delta (q_{rs}-\sigma
)|\widehat{\bm  q}_{rs}\cdot{\bm g}_{rs}|\left[ \Theta \left(
\widehat{\bm q}_{rs} \cdot {\bm g}_{rs}\right) \alpha ^{-2}\left(
b_{rs}^{-1}\rho\right) \left( b_{rs}^{-1} B \right) \right.
\nonumber \\
&&\left. -\Theta \left( -\widehat{\bm q}_{rs}\cdot {\bm
g}_{rs}\right) \rho B \right] \nonumber \\
&=& B \overline{T}(r,s)\rho   \nonumber \\
&&+\delta (q_{rs}-\sigma )|\widehat{\bm q}_{rs }\cdot {\bm
g}_{rs}|\Theta \left( \widehat{\bm q}_{rs}\cdot {\bm
g}_{rs}\right) \left( \alpha
^{-2}b_{rs}^{-1}\rho \right) \left( b_{rs}^{-1}-1\right) B  \nonumber \\
&=&B\overline{T}(r,s)\rho +\rho \delta (q_{rs}-\sigma
)|\widehat{\bm q}_{rs}\cdot {\bm g}_{rs}|\Theta \left(
\widehat{\bm q}_{rs}\cdot {\bm g}_{rs} \right) \left(
b_{rs}^{-1}-1\right) B.  \label{A28}
\end{eqnarray}
In the last equality, use has been made of the identity
\begin{equation}
\delta (q_{rs}-\sigma )|\widehat{\bm q}_{rs}\cdot {\bm
g}_{rs}|\Theta \left( \widehat{\bm q}_{rs}\cdot {\bm
g}_{rs}\right) \alpha ^{-2}b_{rs}^{-1}\rho=\delta (q_{rs}-\sigma
)|\widehat{\bm q}_{rs}\cdot {\bm g}_{rs}|\Theta \left(
\widehat{\bm q}_{rs}\cdot {\bm g}_{rs}\right) \rho, \label{A27}
\end{equation}
valid for hard sphere or disk distribution functions
\cite{lutsko2}. This identity follows from the requirement that
the flux for a pair of particles at contact with relative velocity
${\bm g}_{rs}$ on the pre-collision hemisphere, should be the same
as that for ${\bm g} _{rs}^{\prime \prime }$ on the post-collision
hemisphere. See ref. \cite{lutsko2} for further details and
applications.

Use of Eq.\ (\ref{A28}) into Eq.\ (\ref{A26.a}) gives
\begin{equation}
\rho L_{-}B = \rho \sum_{r} {\bm v}_{r} \cdot \frac{\partial
B}{\partial {\bm q}_{r}} - \frac{\rho}{2} \sum_{r=1}^{N} \sum_{s
\neq r}^{N} \delta (q_{rs}-\sigma) | \widehat{\bm q}_{rs} \cdot {\bm
g}_{rs}| \Theta ( \widehat{\bm q}_{rs} \cdot {\bm g}_{rs})
(b_{rs}^{-1}-1). \label{A30}
\end{equation}
This in turn gives the identification
\begin{equation}
L_{-}=\sum_{r=1}^{N}{\bm v}_{r}\cdot \frac{\partial}{\partial {\bm
q}_{r}}- \frac{1}{2}\sum_{r=1}^{N}\sum_{s\neq r}^{N}T_{-}(r,s),
\label{A31}
\end{equation}
with a third binary collision operator $T_{-}(r,s)$ defined as
\begin{eqnarray}
T_{-}(r,s)& = & \delta (q_{rs}-\sigma )\Theta ( \widehat{\bm
q}_{rs} \cdot {\bm  q}_{rs})| \widehat{\bm q}_{rs} \cdot {\bm
q}_{rs}) |\left( b_{rs}^{-1}-1\right) \nonumber \\
& = & \sigma^{d-1} \int d \widehat{\bm \sigma}\, \Theta (
\widehat{\bm \sigma} \cdot {\bm g}_{rs}) |\widehat{\bm q}_{rs}
\cdot {\bm  q}_{rs})| \delta ({\bm q}_{rs} -{\bm \sigma})
(b_{rs}^{-1}-1). \label{A32}
\end{eqnarray}

It is a simple task using standard techniques to verify that Eq.\
(\ref{A26}) implies
\begin{equation}
e^{- t \overline{L}} (\rho B) = \left( e^{-t \overline{L}} \rho
\right)  e^{- t L_{-}} B, \label{A32.a}
\end{equation}
and Eq.\ (\ref{a.4}) follows. For the particular case
$\rho=\rho_{h}$, the HCS ensemble, it is
\begin{equation}
e^{-t \overline{L}} \rho_{h} \left[ \Gamma; T(0) \right] =
\rho_{h} \left[ \Gamma;T(t) \right] \label{A32.b}
\end{equation}
and, therefore, the correlation functions defined in Eq.\
(\ref{a.4}) can be expressed in the form
\begin{equation}
C_{AB}(t)= \int d\Gamma\, A(\Gamma) \left[ e^{-t L_{-}} B(\Gamma)
\right] \rho_{h} \left[ \Gamma; T(t) \right]. \label{A32.c}
\end{equation}

It is seen that $L_{-}$ is the same as the trajectory generator
identified in Eq. (\ref{a.19}), but with the restituting collision
rule in place of the direct collision rule. Therefore, $L_{-}$ is
the generator of time reversed trajectories in phase space.

\section{Dimensionless Representation}
\label{ap2}

The time dependence of the reference HCS ensemble has the scaling
form indicated in Eq.\ (\ref {1.2}). This means that the time
dependence can be removed entirely by a change of variables,
leaving a universal dimensionless function of the scaled
velocities. It is useful to introduce these same variables more
generally for other states as well, to partially account for
collisional cooling. The advantages will be apparent in the final
results. The set of dimensionless variables are chosen to be
\begin{equation}
{\bm q}_{r}^{\ast }=\frac{{\bm q}_{r}}{l }, \quad {\bm v}^{\ast}_{r}
=\frac{{\bm v}_{r}}{v_{0}(t)}, \quad ds= \frac{v_{0}(t)}{l }dt,
\label{B.1}
\end{equation}
where $v_{0}(t) \equiv v_{0}[T_{h}(t)]$ is defined in terms of the
temperature of a HCS reference state having the same initial total
energy as the actual system under consideration. The Liouville
equation then becomes
\begin{equation}
\left( \frac{\partial}{\partial s}+\overline{\mathcal{L}}^{\ast }
\right) \rho ^{\ast }(\Gamma^{\ast},s)=0, \label{B.2}
\end{equation}
with the reduced distribution function $\rho ^{\ast
}(\Gamma^{\ast},s)$ defined by
\begin{equation}
\rho ^{\ast }\left( \Gamma^{\ast},s \right) =[\ell v_{0}\left(
t\right) ]^{Nd}\rho (\Gamma ,t), \label{B.2a}
\end{equation}
being $\Gamma^{\ast} \equiv \left\{ {\bm q}_{r}^{\ast}, {\bm
v}_{r}^{\ast}; r=1, \ldots , N \right\}$ and the operator
$\overline{\mathcal{L}}^{\ast }$ given by\begin{equation}
\overline{\mathcal{L}}^{\ast }\rho ^{\ast }=\overline{L}^{\ast }\rho
^{\ast }+\frac{\zeta _{0}^{\ast
}}{2}\sum_{r=1}^{N}\frac{\partial}{\partial {\bm v}_{r}^{\ast}}
\cdot \left ( {\bm v}_{r}^{\ast} \rho ^{\ast } \right). \label{B.3b}
\end{equation}
Here,
\begin{equation}
\overline{L}^{\ast} = \frac{l}{v_{0}(t)} \overline{L} = [ \,
\overline{L} \, ]_{\Gamma = \Gamma^{\ast}} \label{B.3b.b}
\end{equation}
and $\zeta _{0}^{\ast }$ is the dimensionless cooling rate
\begin{equation}
\zeta _{0}^{\ast }=\frac{l \zeta_{0}(T)}{v_{0}(T)}. \label{B.3c}
\end{equation}
The first term, $\overline{L}^{\ast}$, of the generator
$\overline{\mathcal{L}}^{\ast}$ for dynamics in this dimensionless
Liouville equation, generates the usual hard sphere trajectories in
scaled variables. The second term rescales the velocities along the
trajectories, to represent their change due to the average cooling
associated with the HCS. A first advantage of this representation is
the fact that the HCS ensemble is a stationary solution of the
Liouville equation,
\begin{equation}
\overline{\mathcal{L}}^{\ast }\rho _{h}^{\ast }=0.  \label{B.4}
\end{equation}

Consider a phase function $A(\Gamma)$ whose dimensions scale out
as a factor with the above change of variables, i.e.,
\begin{equation}
A\left( \Gamma \right) =c_{A}\left[ v_{0}\left( t\right) \right]
A^{\ast }\left( \Gamma ^{\ast }\right) ,
\end{equation}
where $c_{A}\left[ v_{0}(t)\right] $ contains the dimensions of
$A\left( \Gamma \right) $ so that $A^{\ast }\left( \Gamma ^{\ast
}\right) $ is dimensionless. The ensemble average of $A\left(
\Gamma \right) $ at time $t$ is
\begin{equation}
\left\langle A\left( t\right) \right\rangle \equiv \int d\Gamma\,
\rho \left( \Gamma ,t\right) A\left( \Gamma \right) =c_{A}\left[
v_{0}\left( t\right) \right] \left\langle A^{\ast }\left( s\right)
\right\rangle ^{\ast }.  \label{B.5a}
\end{equation}
The dimensionless average is then given by
\begin{equation}
\left\langle A^{\ast }\left( s\right) \right\rangle ^{\ast }\equiv
\frac{ \left\langle A\left( t\right) \right\rangle }{c_{A}\left[
v_{0}\left( t\right) \right] }=\int d\Gamma ^{\ast }\, \left[
e^{-s \overline{\mathcal{L}} ^{\ast }}\rho ^{\ast }\left( \Gamma
^{\ast },0\right) \right] A^{\ast }\left( \Gamma ^{\ast }\right) .
\label{B.5e}
\end{equation}
Adjoint operators can be introduced just as in Appendix \ref{ap1}
for equivalent representations. For example,
\begin{equation}
\left\langle A^{\ast }\left( s\right) \right\rangle ^{\ast }=\int
d\Gamma ^{\ast }\, \rho ^{\ast }\left( \Gamma ^{\ast },0\right)
e^{s \mathcal{L}^{\ast }}A^{\ast }\left( \Gamma ^{\ast }\right)
\label{B.6}
\end{equation}
with
\begin{equation}
\mathcal{L}^{\ast }=L^{\ast }+\frac{\zeta_{0}^{\ast}}{2}
\sum_{r=1}^{N} {\bm v}_{r}^{\ast }\cdot \frac{\partial}{\partial
{\bm v}_{r}}; \label{B.6a}
\end{equation}
\begin{equation}
L^{\ast }=\frac{l }{v_{0}\left( t\right) }L = \left[ L
\right]_{\Gamma = \Gamma^{\ast}}. \label{B.6a.a}
\end{equation}
To illustrate the utility of this formulation, consider an average
over the HCS. Then Eq.\ (\ref{B.5e}) gives $\left\langle A^{\ast
}\left( s\right) \right\rangle ^{\ast }=\left\langle A^{\ast
}\left( 0\right) \right\rangle ^{\ast }$, since $\rho _{h}^{\ast
}$ is stationary. Using Eq.\ (\ref{B.6}), the time derivative at
$s=0$ is seen to be given by
\begin{equation}
\int d\Gamma ^{\ast }\, \rho _{h}^{\ast }\left( \Gamma ^{\ast
}\right) \mathcal{L}^{\ast }A^{\ast }\left( \Gamma ^{\ast
}\right)=0 .  \label{B.6b}
\end{equation}
Suppose now that $A^{\ast }$ be an arbitrary differentiable
function of the scaled total momentum of the system ${\bm
P}^{\ast} = \sum_{r} {\bm v}^{\ast}_{r}$, i.e.
$A^{\ast}=A^{\ast}({\bm P}^{\ast})$. By momentum conservation,
$L^{\ast }A^{\ast }(\mathbf{P}^{\ast })=0$ and Eq.\ (\ref{B.6b})
becomes
\begin{equation}
\frac{\zeta _{0}^{\ast }}{2} \int d\Gamma ^{\ast }\, \rho _{0}^{\ast
}\left( \Gamma ^{\ast }\right) \sum_{r=1}^{N}{\bm v}_{r}^{\ast
}\cdot \frac{\partial}{\partial {\bm v}_{r}}A^{\ast }\left( {\bm
P}^{\ast }\right) = \frac{\zeta _{0}^{\ast }}{2}\int d\Gamma ^{\ast
}\, \rho _{0}^{\ast }\left( \Gamma ^{\ast }\right) {\bm P}^{\ast
}\cdot \frac{\partial}{\partial {\bm P}^{\ast}} A^{\ast }\left({\bm
P}^{\ast }\right)=0. \label{B.6c}
\end{equation}
Since this holds for any function $A^{\ast}({\bm P}^{\ast})$, it
follows that $\rho _{h}^{\ast }(\Gamma^{\ast})$ must be
proportional to a delta function in the total momentum,
\begin{equation}
\rho _{h}^{\ast }\left( \Gamma ^{\ast }\right) =\delta \left( \mathbf{P}%
^{\ast }\right) \overline{\rho }_{h}^{\ast }\left( \Gamma ^{\ast
}\right). \label{B.6d}
\end{equation}
This is the result quoted in Eq.\ (\ref{1.3.4}) in the text.

The dimensionless form for correlation functions defined in Eq.\,
(\ref{a.2.5}) are obtained by using the representation in Eq.\
(\ref{B.6}),
\begin{equation}
C_{AB}^{\ast }(s)=\frac{C_{AB}(t)}{c_{A}\left[ v_{0}\left(
t\right) \right] c_{B}\left[ v_{0}\left( 0\right) \right] }=\int
d\Gamma ^{\ast }\, \left[ e^{ s\mathcal{L}^{\ast }}A^{\ast }\left(
\Gamma ^{\ast }\right) \right] B^{\ast }(\Gamma^{\ast})\rho ^{\ast
}\left( \Gamma ^{\ast },0\right).  \label{B.9}
\end{equation}
The adjoint representation follows directly as well,
\begin{equation}
C_{AB}^{\ast }\left( s\right) =\int d\Gamma ^{\ast }\, A^{\ast
}\left( \Gamma ^{\ast }\right) e^{-s\overline{\mathcal{L}}^{\ast
}}\left[ \rho^{\ast} \left( \Gamma ^{\ast },0\right)  B^{\ast
}\left( \Gamma ^{\ast }\right) \right] .  \label{B25BB}
\end{equation}

Finally, the representation in Eq.\  (\ref{a.4}) gives the third
equivalent form:
\begin{equation}
C_{AB}^{\ast }\left( s\right) =\int d\Gamma ^{\ast }\,
A^{\ast}(\Gamma^{\ast}) \left[ e^{-s
\overline{\mathcal{L}^{\ast}}} \rho^{\ast}(\Gamma^{\ast},0)
\right] e^{-s \mathcal{L}^{\ast}_{-}} B^{\ast}(\Gamma^{\ast}),
\label{B30A}
\end{equation}
with
\begin{equation}
\mathcal{L}_{-}^{\ast }=L_{-}^{\ast
}+\frac{\zeta_{0}^{\ast}}{2}\sum_{r=1}^{N}{\bm v}_{r}^{\ast }\cdot
\frac{\partial}{\partial {\bm v}_{r}^{\ast}} \label{B31}
\end{equation}
and
\begin{equation}
L_{-}^{\ast}= \frac{l}{v_{0}(t)} L_{-} = \left[ L_{-}
\right]_{\Gamma = \Gamma^{\ast}}. \label{B31.a}
\end{equation}
For the particular case of the HCS, in Eqs.\,
(\ref{B.9})-(\ref{B30A}) it is
\begin{equation}
e^{-s \overline{\mathcal{L}}^{\ast}} \rho_{h}^{\ast} \left(
\Gamma^{\ast},0 \right)= \rho_{h}^{\ast}(\Gamma^{\ast},0)
=\rho_{h}^{\ast}(\Gamma^{\ast}),
\end{equation}
due to the stationarity of the distribution of the HCS in the
reduced scales.

The results of this Appendix, starting from hard sphere dynamics,
agree with those obtained in ref. \cite{DBB06}, starting from more
general interaction potentials and taking the scaling limit (see
Sec. 8 in ref. \cite{DBB06}).

\section{Hydrodynamic Modes}
\label{ap3} In this Appendix, the eigenvalues of the hydrodynamic
transport matrix $ \mathcal{K}^{\ast hyd }\left( {\bm k}^{\ast
}\right) $, whose expression is given by  Eqs. (\ref{1.4.1a})-
(\ref{1.4.3}), are presented and some comments made on their
implication for transport in a granular fluid. Similar results,
but restricted to the low density limit have been reported in
\cite{ByD05}. The eigenvalues of the matrix $\mathcal{K} ^{\ast
hyd }$ are the solutions $\gamma_{\alpha}$ to the cubic equation
\begin{eqnarray}
\gamma ^{3}& - & \left[  \frac{2(d-1)\eta ^{\ast }}{d}+\kappa
^{\ast }+ \frac{2\lambda ^{\ast }}{d}-\zeta^{ \ast T }\right]
k^{\ast 2} \gamma^{2}  \nonumber \\
& - & \left[ \frac{\zeta _{0}^{\ast 2}}{4} - \left\{ \frac{\zeta
_{0}^{\ast }}{2} \left[ \frac{2(d-1)\eta ^{\ast
}}{d}+\kappa^{\ast} -\frac{2 \lambda^{\ast}}{d}+\zeta^{ \ast T}
\right] -\frac{p^{\ast}_{h}}{2}\left( \frac{2p^{\ast }_{h}}{d}
+\zeta^{ \ast U } \right) -\frac {p^{\ast }_{h}}{2} \frac{\partial
\ln p_{h}}{\partial \ln n_{h}}\right\} k^{\ast 2} \right] \gamma
\nonumber \\
& + & \frac{\zeta _{0}^{\ast } p_{h}^{\ast}}{4} \left(
\frac{\partial}{\partial \ln n_{h}}\, \ln
\frac{\zeta_{0}^{2}}{p_{h}} \right)k^{\ast 2}=0, \label{c.3.2}
\end{eqnarray}
plus the $(d-1)$-fold degenerated shear modes
\begin{equation}
\gamma_{\perp} =- \frac{\zeta _{0}^{\ast }}{2}+\eta ^{\ast
}k^{\ast 2}. \label{c.3.3}
\end{equation}
If the limit $\alpha \rightarrow 1$ is taken for this equation,
$\zeta _{0}^{\ast }\rightarrow 0$ and the solutions to order
$k^{\ast 2}$ give the familiar hydrodynamic modes associated with
normal fluids: the two propagating sound modes, the heat mode and
the $d-1$ transverse shear modes \cite{McLbook}. But, when the
solution to the above equation is considered at finite $\alpha $,
the modes to order $k^{\ast 2}$ are quite different,
\begin{equation}
\gamma_{1}(k^{\ast})=\frac{p^{\ast }_{h}}{\zeta _{0}^{\ast
}}\left( \frac{\partial }{\partial \ln n_{h}}\ln \frac{\zeta
_{0}^{2}}{p_{h}}\right) k^{\ast 2}, \label{c.3.4}
\end{equation}
\begin{equation}
\gamma _{2}(k^{\ast})=\frac{\zeta _{0}^{\ast }}{2}-\left[
\frac{p^{\ast }_{h}}{2\zeta ^{\ast }_{0}}\left( \frac{ 2p^{\ast
}_{h}}{d}+2\frac{\partial \ln \zeta _{0}}{\partial \ln
n_{h}}+\zeta ^{ \ast U }\right) -\frac{ 2 \lambda^{\ast}}{d} +
\zeta ^{ \ast T} \right] k^{\ast 2}, \label{c.3.5}
\end{equation}
\begin{eqnarray}
\gamma_{3} (k^{\ast}) \equiv \gamma_{\parallel}(k^{\ast})
&=&-\frac{\zeta _{0}^{\ast }}{2}+ \left[ \frac{ p^{\ast
}_{h}}{2\zeta _{0}^{\ast }} \left( \frac{2p^{\ast }_{h}}{d}-
2\frac{\partial \ln \zeta _{0}}{
\partial \ln n_{h}}+\zeta ^{ \ast U} + 2\frac{\partial \ln p_{h}}{\partial \ln
n_{h}}\right)  \right. \nonumber \\
&&+ \left. \frac{2(d-1)\eta ^{\ast }}{d}+\kappa ^{\ast
}\right]k^{\ast 2}.
\end{eqnarray}
All eigenvalues are real and hence there are no propagating modes.
Also, the limit $\alpha \rightarrow 1$ is singular so these modes
in this limit do not represent the familiar hydrodynamic
excitations of a normal fluid. The drastic difference in the
nature of the hydrodynamic modes obtained as the elastic limit of
the above eigenvalues, is due to the non analyticity of the
eigenvalues and eigenvectors about the point $\alpha =1$ and
$k=0$. Close to the elastic limit, both $\zeta _{0}^{\ast }
\propto \left( 1-\alpha ^{2}\right) $ and $ k$ are small
parameters, and the type of modes obtained depends on how these
parameters approach zero \cite{ByD05}. This is an indication of
the fact that the inelasticity, even when small, gives rise to
drastically different transport in the fluid. For the purposes at
hand, attention is restricted here to the $\alpha \neq 1$ forms of
these modes.

There exists a critical wavelength $k_{\perp}^{\ast c}$ defined by
\begin{equation}
k^{\ast c}_{\perp}=\frac{\zeta _{0}^{\ast }}{2 \eta ^{\ast }},
\label{c.3.8}
\end{equation}
such that for $k^{\ast} < k_{\perp}^{\ast c}$ the shear modes become
unstable. Similarly, there exists a threshold wavelength for the
$\gamma_{\parallel}$ mode such that it becomes unstable as well.
This means that the homogeneous state characterized by these
hydrodynamic equations, is unstable to sufficiently long wavelength
perturbations that excite these modes. This instability of the HCS
has been well established in the literature \cite{Mcnamara,
IGZanetti}. This implies physically that the response due to the
unstable modes grows until such time as the linear theory breaks
down, and further analysis of the dynamics has to be carried out
using the full non-linear hydrodynamic equations. It is still an
open question as to the nature of the final state \cite{MJJavIS}.

Further insight into the nature of the hydrodynamic response of this fluid
can be obtained by looking at the corresponding eigenvectors. To lowest
order in $k$ these are found to be
\begin{equation*}
\varphi_{1}({\bm k}^{\ast})=\left(
\begin{array}{c}
1 \\
-2 \frac{\partial \ln \zeta_{0}}{\partial \ln n_{h}} \\
0 \\
{\bm 0}
\end{array}
\right) ,\quad {\bm \varphi}_{2}({\bm k}^{\ast})=\left(
\begin{array}{c}
0 \\
1 \\
0 \\
{\bm 0}
\end{array}
\right)
\end{equation*}
\begin{equation}
 \varphi_{\parallel}({\bm k}^{\ast})=\left(
\begin{array}{c}
0 \\
0 \\
1 \\
{\bm 0}
\end{array}
\right) , \quad { \varphi}_{\perp,i}({\bm k}^{\ast})=\left(
\begin{array}{c}
0 \\
0 \\
0 \\
{\widehat{\bm i}}
\end{array}
\right).   \label{c.3.10}
\end{equation}
Here, ${\bm 0}=0$ and $\widehat{\bm i}=\widehat{\bm 1}=1$, for
$d=2$, while ${\bm 0}= \left( \begin{array}{c} 0 \\ 0 \end{array}
\right)$ and $i=1,2$, with $\widehat{\bm 1}= \left( \begin{array}{c} 1 \\
0\end{array} \right)$  and $\widehat{\bm 2}= \left(
\begin{array}{c} 0 \\ 1 \end{array} \right)$, for $d=3$.

The first of these modes is excited by the condition
\begin{equation}
\delta T^{\ast }=-2\frac{\partial \ln \zeta _{0}}{\partial \ln
n_{h}}\, \delta n^{\ast }  \label{c.3.11}
\end{equation}
and zero flow velocity. This can be interpreted as follows. The
cooling rate has the form $\zeta_{0}(n_{h},T_{h}) = T_{h}^{1/2}
\overline{\zeta }_{0}(n_{h})$. It then follows that this condition
for exciting the first mode corresponds to variations in the
temperature and density that leave the cooling rate constant. The
second mode in Eq.(\ref {c.3.10}) is due to a temperature
perturbation at constant density and also zero velocity, while the
third one is due to a longitudinal velocity perturbation at
constant temperature and density. The last $d-1$ modes are the
response to a transverse velocity perturbation, again at constant
temperature and density.

\section{Microscopic Conservation Laws}
\label{ap4}

The aim here is to derive  the conservation laws (more precisely,
balance equations) for the microscopic observables of interest,
namely, the number density $\mathcal{N}(\Gamma;{\bm r})$, energy
density $\mathcal{E} (\Gamma; {\bm r})$, and momentum density
$\bm{\mathcal{G}}(\Gamma;{\bm r})$. Their ensemble averages give
the hydrodynamic fields. One set of balance equations is
associated with the dynamics for $t>0$, whose generator is the $L$
operator given by Eq.\ (\ref{a.19}). Another set of balance
equations describes the dynamics for $t<0$, as generated by
$L_{-}$ defined in Eq.\ (\ref{A31}). Both dynamics lead to forms
of fluxes that appear in the Green-Kubo expressions for both
elastic and inelastic hard sphere transport coefficients (see
appendix \ref{ap8}).

Let  $ \left\{ \mathcal{A}_{\alpha} ( \Gamma; {\bm r},t) \right\} $
denote the set of phase functions $\left\{ \mathcal{N}, \mathcal{E},
\bm{\mathcal G} \right\}$. Consider first $t>0$. The  phase
functions obey the equation
\begin{equation}
\frac{\partial {\mathcal A}_{\alpha} \left( \Gamma; {\bm
r},t\right) }{\partial t}-L(\Gamma) {\mathcal A}_{\alpha}
(\Gamma;{\bm r},t)=0,  \label{D.1}
\end{equation}
where $L(\Gamma)$ is given in Eq.\ (\ref{a.19}). The aim here is
to evaluate the action of the $L$ operator on the density
${\mathcal A}_{\alpha}$ to identify a balance equation of the form
\begin{equation}
\frac{\partial {\mathcal A} \left( \Gamma; {\bm r},t\right)
}{\partial t}+\bm{\nabla} \cdot {\bm j}_{\alpha} \left(\Gamma;
{\bm r},t\right) =- w_{\alpha} \left( \Gamma; {\bm r},t\right) .
\label{D.1a}
\end{equation}
Here, ${\bm j}_{\alpha}$ is the flux associated with the density
${\mathcal A}_{\alpha}$, and $w_{\alpha}$ is a source that
signifies a local loss contribution in this density that cannot be
expressed as a divergence. It is non zero only in the equation
for the energy density.

The number density is defined as
\begin{equation}
{\mathcal N}\left( \Gamma;{\bm r}\right)  \equiv \sum_{r=1}^{N}
\delta \left({\bm r}- {\bm q}_{r}\right) ,  \label{D.1b}
\end{equation}
so
\begin{eqnarray}
L {\mathcal N} (\Gamma;{\bm r}) &=&\sum_{r=1}^{N}{\bm v}_{r}\cdot
\frac{\partial}{\partial {\bm q}_{r}} \delta \left( {\bm r}-{\bm
q}_{r}\right) = -\bm{\nabla} \cdot \sum_{r=1}^{N} {\bm
v}_{r}\delta \left( {\bm r}-{\bm q}_{r}\right) .
\end{eqnarray}
This gives the microscopic continuity equation,
\begin{equation}
\frac{\partial {\mathcal N}\left(\Gamma; {\bm r},t\right)
}{\partial t}+ \bm{\nabla} \cdot \frac{ \bm{\mathcal
G}\left(\Gamma; {\bm r},t\right) }{m}=0. \label{D.2b}
\end{equation}
where the number flux density is seen to be proportional to the
momentum density
\begin{equation}
\bm{\mathcal G} \left( \Gamma; {\bm r}\right) \equiv
\sum_{r=1}^{N} m {\bm v}_{r}\delta \left( {\bm r}-{\bm
q}_{r}\right) . \label{D.2a}
\end{equation}

Consider now the equation for this density. It is
\begin{eqnarray}
L \bm{\mathcal G} \left(\Gamma;{\bm r}\right) &=&\sum_{r=1}^{N} m
{\bm v}_{r } {\bm v}_{r} \cdot \frac{\partial}{\partial {\bm q}_{r}}
\delta \left({\bm r}- {\bm q}_{r}\right) +\frac{m}{2}\sum_{r=1}^{N}
\sum_{s \neq r}^{N}T \left( r,s\right) \left[ {\bm v}_{r} \delta
\left( {\bm r}-{\bm q}_{r}\right) + {\bm
v}_{s} \delta ({\bm r}-{\bm q}_{s}) \right] \nonumber \\
& = & - \bm{\nabla} \cdot \left[ \sum_{r=1}^{N} m {\bm v}_{r} {\bm
v}_{r}\delta \left( {\bm r}-{\bm q}_{r}\right) +\frac{(1+\alpha)m
\sigma}{4} \sum_{r=1}^{N} \sum_{s \neq r}^{N} \delta \left(
q_{rs}-\sigma \right) \Theta \left( -
\widehat{\bm q}_{rs}\cdot {\bm g}_{rs}\right) \right. \nonumber \\
&& \left. \times \left| \widehat{\bm q}_{rs}\cdot {\bm
g}_{rs}\right| ^{2} \widehat{\bm q}_{rs} \widehat{\bm q}_{rs}
\int_{0}^{1}d\gamma\,
 \delta \left( {\bm r}-{\bm q}_{r}+\gamma{\bm q}_{rs}\right) \right].  \label{D1}
\end{eqnarray}
In the second equality, use has been made of the relation
\begin{equation}
(b_{rs}-1){\bm v}_{r}=-(b_{rs}-1){\bm v}_{s}=- \frac{1+\alpha}{2}
\left( \widehat{\bm q}_{rs} \cdot {\bm g}_{rs}\right) \widehat{
\bm q}_{rs}  \label{D7}
\end{equation}
and the identity
\begin{equation}
\delta ({\bm r}-{\bm q}_{r})-\delta ({\bm r}-{\bm q} _{s})=-
\int_{0}^{1}d\gamma\,  \frac{\partial }{\partial \gamma }\delta
\left( {\bm r}-{\bm q}_{r}+\gamma {\bm q}_{rs}\right) =- {\bm
q}_{rs} \cdot \bm{\nabla} \int_{0}^{1}d\gamma\,  \delta \left(
{\bm r}-{\bm q}_{r}+\gamma {\bm q}_{rs}\right) . \label{D9}
\end{equation}
Equation (\ref{D1}) gives the conservation law for the momentum,
\begin{equation}
\frac{\partial}{\partial t} \bm{\mathcal G}\left( \Gamma; {\bm
r},t\right)+ \bm{\nabla} \cdot {\sf h} \left( \Gamma;{\bm r}, t
\right) =0,  \label{D13}
\end{equation}
with the tensor momentum flux density ${\sf h}$ identified as
\begin{eqnarray}
{\sf h} \left(\Gamma; {\bm r}\right) &=&\sum_{r=1}^{N} m {\bm v}_{r}
{\bm v}_{r }\delta ({\bm r}-{\bm q}_{r})+\frac{m( 1+\alpha)
\sigma}{4}  \sum_{r=1}^{N} \sum_{s\neq r}^{N} \delta \left(
q_{rs}-\sigma \right)
\notag \\
&&\times \Theta \left( - \widehat{\bm q}_{rs}\cdot {\bm
g}_{rs}\right) \left( \widehat{\bm q}_{rs}\cdot {\bm g}_{rs}\right)
^{2}\widehat{\bm q}_{rs} \widehat{\bm q}_{rs} \int_{0}^{1}d\gamma\,
\delta \left({\bm r}-{\bm q}_{r}+\gamma {\bm q}_{rs}\right) .
\label{D14}
\end{eqnarray}
The first term on the right hand side is the called kinetic part
of the momentum flux, while the second term is the collisional
transfer part.

Finally, the energy density is given by
\begin{equation}
\mathcal{E}\left(\Gamma; {\bm r}\right) =\sum_{r=1}^{N}
\frac{mv_{r}^{2}}{2} \delta ({\bm r}-{\bm q}_{r}). \label{D15}
\end{equation}
Proceeding in a similar way as for the momentum density,
\begin{eqnarray}
L {\mathcal E} \left( \Gamma; {\bm r}\right) &=&-\bm{\nabla} \cdot
\sum_{r=1}^{N} \frac{ mv_{r}^{2}}{2} {\bm v}_{r} \delta ({\bm
r}-{\bm q}_{r})+\frac{m}{4} \sum_{r=1}^{N} \sum_{s \neq r}^{N}
T\left(r,s \right) \left[ v_{r}^{2}\delta \left( {\bm r}- {\bm
q}_{r}\right)
+v_{s}^{2}\delta \left( {\bm r}-{\bm q}_{s}\right) \right]  \nonumber \\
&=&- \bm{\nabla} \cdot \left[ \sum_{r=1}^{N} \frac{ mv_{r}^{2}}{2}
{\bm v}_{r} \delta ({\bm r}-{\bm q}_{r}) + \frac{m(1+\alpha
)\sigma}{4} \sum_{r=1}^{N} \sum_{s \neq r}^{N} \delta
(q_{rs}-\sigma )\Theta \left( -\widehat{\bm q}_{rs} \cdot
{\bm g}_{rs}\right) \right.  \notag \\
&&\left. \times (\widehat{\bm q}_{rs} \cdot {\bm g}_{rs})^{2}
\widehat{\bm q}_{rs} \cdot {\bm G}_{rs} \widehat{\bm q}
_{rs}\int_{0}^{1}d\gamma\,  \delta \left( {\bm r}-{\bm q}
_{r}+\gamma {\bm q}_{rs}\right) \right]  \nonumber \\
&&-\frac{m(1-\alpha ^{2})}{8} \sum_{r=1}^{N} \sum_{s \neq
r}^{N}\delta (q_{rs}-\sigma )\Theta \left( - \widehat{\bm
q}_{rs}\cdot {\bm g}_{rs}\right) |\widehat{\bm q}_{rs}\cdot {\bm
g}_{rs}|^{3}\delta \left( {\bm r}-{\bm q}_{r}\right). \label{D16}
\end{eqnarray}
This consists of two parts, one which can be expressed as a
gradient and a second one that is inherently local, characterizing
the loss in energy due to the inelastic character of collisions.
The energy balance equation becomes
\begin{equation}
\frac{\partial}{\partial t} {\mathcal E} (\Gamma;{\bm r},t) +
\bm{\nabla} \cdot {\bm s}\left(\Gamma;  {\bm r},t\right) =-
w\left(\Gamma;  {\bm r} ,t\right) ,  \label{D20}
\end{equation}
with the energy flux density
\begin{eqnarray}
{\bm s}\left(\Gamma; {\bm r}\right) &=&\sum_{r=1}^{N}
\frac{mv_{r}^{2}}{2} {\bm v} _{r} \delta ({\bm r}-{\bm q}_{r})+
\frac{m(1+\alpha )\sigma}{4}
\sum_{r=1}^{N} \sum_{s \neq r}^{N} \delta (q_{rs}-\sigma )  \nonumber \\
&&\times \Theta \left( - \widehat{\bm q}_{rs}\cdot
\mathbf{g}_{ij}\right) (\widehat{\bm q}_{rs} \cdot {\bm
g}_{rs})^{2}\left( \widehat{\bm q}_{rs} \cdot {\bm G}_{rs} \right)
\widehat{\bm q}_{rs}\int_{0}^{1}d \gamma\, \delta \left( {\bm
r}-{\bm q}_{r}+\gamma {\bm q}_{rs}\right) , \label{D21}
\end{eqnarray}
and the energy source
\begin{equation}
w\left( \Gamma; {\bm r}\right) = \frac{m(1-\alpha^{2})}{8}
\sum_{r=1}^{N} \sum_{s \neq r}^{N} \delta (q_{rs}-\sigma )\Theta
\left( -\widehat{\bm q}_{rs}\cdot {\bm g} _{rs}\right)
|\widehat{\bm q}_{rs} \cdot {\bm g}_{rs}|^{3}\delta \left( {\bm
r}-{\bm q}_{r}\right).  \label{D22}
\end{equation}
This completes the identification of the flux and source terms in
the microscopic balance equations for the forward dynamics.

A similar analysis can be done for the balance equations
associated with the backward dynamics of the phase functions,
described by the equations
\begin{equation}
\frac{\partial \mathcal{A}_{\alpha}\left(\Gamma; {\bm r},-t\right)
}{\partial \left( -t\right) } -L_{-} \mathcal{A}_{\alpha} \left(
\Gamma; {\bm r},-t\right) =0,  \label{D.B2}
\end{equation}
where $t>0$. The form of the conservation laws now is
\begin{equation}
\frac{\partial \mathcal{A}_{\alpha}\left(\Gamma; {\bm r},-t\right)
}{\partial \left( -t\right) } +\bm{\nabla} \cdot {\bm
j}_{\alpha}^{-}\left(\Gamma; {\bm r},-t \right)
=w_{\alpha}^{-}\left(\Gamma; {\bm r},-t \right).  \label{D17}
\end{equation}
If the generators $L$ and $L_{-}$ were the same, then this would
just be the time reversal of Eq.\ (\ref{D.1a}). However, the
contributions for positive times from the binary collision operators
$T\left( i,j\right) $ are replaced  by those from $ T_{-}\left(
i,j\right)$ when considering negative times. Repeating the above
analysis for this case, gives the time reversed fluxes as
\begin{eqnarray}
{\sf h}^{-}(\Gamma;{\bm r}) &=&m \sum_{r=1}^{N} {\bm v}_{r} {\bm
v}_{r} \delta ({\bm r}-{\bm q}_{r})+\frac{m(1+\alpha )\sigma}{4
\alpha } \sum_{r=1}^{N} \sum_{s \neq r}^{N} \delta (q_{rs}-\sigma)
\nonumber \\
& & \times \Theta (\widehat{\bm q}_{rs} \cdot {\bm g}_{rs}) (
\widehat{\bm q}_{rs} \cdot {\bm g}_{rs})^{2} \widehat{\bm q}_{rs}
\widehat{\bm q}_{rs}  \int_{0}^{1} d\gamma\, \delta ( {\bm r}-{\bm
q}_{r}+\gamma {\bm q}_{rs} ), \label{D.B.3}
\end{eqnarray}
\begin{eqnarray}
{\bm s}^{-}( \Gamma; {\bm r}) &=&\frac{m}{2}\sum_{r=1}^{N} v_{r}^{2}
{\bm v}_{r}\delta ({\bm r}-{\bm q}_{r})+ \frac{m (1+\alpha
)\sigma}{4 \alpha } \sum_{r=1}^{N} \sum_{s \neq r}^{N} \delta (q_{rs}-\sigma )  \nonumber \\
&&\times \Theta ( \widehat{\bm q}_{rs} \cdot {\bm g}_{rs})
(\widehat{\bm q}_{rs} \cdot {\bm g}_{rs} )^{2}    ( \widehat{\bm
q}_{rs} \cdot {\bm G}_{rs}) \widehat{\bm
q}_{rs}\int_{0}^{1}d\gamma\, \delta \left( {\bm r}-{\bm
q}_{r}+\gamma {\bm q}_{rs}\right) , \label{D.B.4}
\end{eqnarray}
while the expression for the reverse energy source is
\begin{equation}
w^{-}(\Gamma; {\bm r})=\frac{m(1-\alpha ^{2})}{8 \alpha^{2} }
\sum_{r=1}^{N} \sum_{s \neq r}^{N} \delta (q_{rs} - \sigma )
\Theta ( \widehat{\bm q}_{rs} \cdot {\bm g}_{rs} ) (\widehat{\bm
q}_{rs} \cdot {\bm g}_{rs})^{3} \delta \left( {\bm r}-{\bm
q}_{r}\right) . \label{D.B.5}
\end{equation}
The superscript $-$ is used here to indicate that the quantities
are associated with the time reversed equations. The source term
$w^{-}$ in the balance equation for the energy is now positive,
accounting for collisional increase in energy on moving backwards
along a trajectory originally generated forward in time. In the
elastic limit, this latter effect vanishes. However, the fluxes
still differ from those for $t>0$, with the collisional transfer
contributions being defined on different pre-collision
hemispheres.

\subsection{Dimensionless Balance Equations}

The purpose of this subsection is two fold. First, the
dimensionless forms of the various fluxes that determine the
hydrodynamic parameters are identified. Second, a special property
of the source term in the equation for the energy noted in ref.
\cite {DBB06}, namely that it is orthogonal to the set of
conjugate densities $\left\{ \widetilde{\psi }_{\alpha }^{\ast
}\right\} $ in the homogeneous limit, is shortly reviewed here.

Dimensionless forms for the microscopic densities $\mathcal{N}$,
$\bm{\mathcal G}$, and $\mathcal{E}$ are defined by (recall the
choice $l =n_{h}^{-1/d}$ made for the length scale)
\begin{equation}
\mathcal{N}^{\ast}(\Gamma^{\ast};{\bm r}^\ast) \equiv
\frac{{\mathcal N}(\Gamma;{\bm r})}{n_{h}}= \sum_{r=1}^{N} \delta
({\bm r}^{\ast}-{\bm q}^{\ast}_{r}), \label{E3.2}
\end{equation}
\begin{equation}
\mathcal{E}^{\ast}(\Gamma^{\ast};{\bm r}^\ast) \equiv
\frac{{\mathcal E}(\Gamma;{\bm r})}{n_{h} T_{h}}= \sum_{r=1}^{N}
v_{r}^{\ast 2} \delta ({\bm r}^{\ast}-{\bm q}^{\ast}_{r}),
\label{E3.1}
\end{equation}
\begin{equation}
\bm{\mathcal{G}}^{\ast}(\Gamma^{\ast};{\bm r}^\ast) \equiv
\frac{\bm{\mathcal G}(\Gamma;{\bm r})}{m n_{h} v_{0}(T_{h})}=
\sum_{r=1}^{N} {\bm v}^{\ast}_{r} \delta ({\bm r}^{\ast}-{\bm
q}^{\ast}_{r}). \label{E3}
\end{equation}
With this choice, the balance equations in the dimensionless form
become
\begin{equation}
\frac{\partial {\mathcal N}^{\ast }}{\partial
s}+\frac{\partial}{\partial {\bm r}^{\ast}} \cdot \bm{\mathcal
G}^{\ast }=0,  \label{E6}
\end{equation}
\begin{equation}
\left( \frac{\partial }{\partial s} -\frac{\zeta_{0}^{\ast}}{2}
\right) \bm{\mathcal G}^{\ast} +\frac{\partial}{\partial {\bm
q}_{r}^{\ast}} \cdot {\sf h}^{\ast}=0, \label{E7}
\end{equation}
\begin{equation}
\left( \frac{\partial}{\partial s}-\zeta _{0}^{\ast } \right)
{\mathcal E}^{\ast} +\frac{\partial}{\partial {\bm r}^{\ast}}
\cdot {\bm s}^{\ast}= -w^{\ast}. \label{E9}
\end{equation}
The functional forms of the fluxes and source terms are related to
those given above by
\begin{equation}
{\sf h}^{\ast}(\Gamma^{\ast};{\bm r}^{\ast})= \frac{{\sf
h}(\Gamma;{\bm r})}{m n_{h} v_{0}^{2}(T_{h})}= \left[ {\sf h} (
\Gamma;{\bm r}) \right]_{\Gamma = \Gamma^{\ast}, {\bm r}={\bm
r}^{\ast}}, \label{E9.0}
\end{equation}
\begin{equation}
{\bm s}^{\ast}( \Gamma^{\ast};{\bm r}^{\ast})= \frac{{\bm
s}(\Gamma;{\bm r})}{n_{h}T_{h} v_{0}(T_{h})} = 2 \left[ {\bm s}
(\Gamma;{\bm r}) \right]_{\Gamma = \Gamma^{\ast}, {\bm r}={\bm
r}^{\ast}}, \label{E9.1}
\end{equation}
\begin{equation}
w^{\ast}(\Gamma^{\ast}; {\bm r}^{\ast})= \frac{l w(\Gamma;{\bm
r})}{n_{h} T_{h} v_{0}(T_{h})} = 2 \left[ w(\Gamma ;{\bm r} )
\right]_{\Gamma = \Gamma^{\ast}, {\bm r}={\bm r}^{\ast}}.
\label{E9.2}
\end{equation}
It is understood that, in addition to the indicated substitutions,
the changes $\sigma \rightarrow \sigma^{\ast}=\sigma /l$ and $ m
\rightarrow 1$, are made. The direct densities functions
$a^{\ast}_{\alpha}(\Gamma^{\ast};{\bm r}^{\ast})$ appearing in the
response functions are defined by Eq.\ (\ref{1.5.2}). In the case
of $a_{2}^{\ast }$, the balance equations give
\begin{equation}
\frac{\partial a_{2}^{\ast }}{\partial s}-\frac{2  \zeta
_{0}^{\ast }}{d} {\mathcal E}^{\ast } + \frac{\partial}{\partial
{\bm r}^{\ast}} \cdot  \left( \frac{2{\bm s}^{\ast}}{d}
-\bm{\mathcal G}^{\ast }\right) =- \frac{2 w^{\ast }}{d}.
\end{equation}
For reasons that become apparent below, add a term $\sum_{\alpha}
\mathcal{K}_{2\alpha }^{\ast hyd }\left( {\bm 0}\right) a_{\alpha
}^{\ast }$ on both sides of the above equation to write
\begin{equation}
\frac{\partial a_{2}^{\ast }}{\partial s}+ \sum_{\alpha}
\mathcal{K}_{2\alpha }^{\ast hyd }\left( {\bm 0}\right) a_{\alpha
}^{\ast }+ \frac{\partial}{\partial {\bm r}^{\ast}} \cdot \left(
\frac{2 {\bm s}^{\ast }}{d}-\bm{\mathcal G}^{\ast }\right) =-
\ell^{\ast }, \label{E10Z}
\end{equation}
where the new source $\ell^{\ast }$ is
\begin{equation}
\ell^{\ast }(\Gamma^{\ast};{\bm r}^{\ast},s) \equiv \frac{2
w^{\ast }}{d}-\zeta _{0}^{\ast }\left[ \frac{3}{2} a_{2}^{\ast
}+\left( \frac{\partial \ln \zeta _{0}}{\partial \ln n_{h}}
+1\right) a_{1}^{\ast }\right] . \label{E10a}
\end{equation}
Evidently, the Fourier representation of Eq.\ (\ref{E10Z}) is
\begin{equation}
\frac{\partial \widetilde{a}_{2}^{\ast }}{\partial s}+
\sum_{\alpha} \mathcal{K}_{2\alpha }^{\ast hyd }\left( {\bm
0}\right) \widetilde{a}_{\alpha }-i{\bm k}^{\ast} \cdot \left(
\frac{2 \widetilde{\bm s}^{\ast }}{d}- \widetilde{\bm {\mathcal
G}}^{\ast} \right) =\widetilde{\ell}^{\ast }\left( {\bm
k}^{\ast},s\right) . \label{E11}
\end{equation}

The homogeneous limit of the source term, $\widetilde{l}^{\ast
}\left( {\bm 0},s\right) $, has an important orthogonality
property, which is the reason for adding the contribution
$\sum_{\alpha} \mathcal{K}_{2\alpha }^{hyd\ast }\left( {\bm
0}\right) a_{\alpha }^{\ast }$ above. To see this, the Fourier
transform for ${\bm k}^{\ast}={\bm 0}$ of the second term on the
right hand side of Eq.\ (\ref{E10a}) is written in the equivalent
form
\begin{eqnarray}
\zeta _{0}^{\ast }\left[ \frac{3}{2}\widetilde{a}_{2}^{\ast
}\left({\bm 0,}s\right) +\left( \frac{\partial \ln \zeta
_{0}}{\partial \ln n_{h}} +1\right) \widetilde{a}_{1}^{\ast }\left(
{\bm 0}, s\right) \right] &=&e^{s \mathcal{L}^{\ast }}\left[
\frac{3}{2}\zeta _{0}^{\ast }\widetilde{a} _{2}^{\ast }\left( {\bm
0}\right) +\zeta _{0}^{\ast }\left( \frac{
\partial \ln \zeta _{0}}{\partial \ln n_{h}}+1\right) \widetilde{a}
_{1}^{\ast }\left( {\bm 0}\right) \right]  \nonumber \\
&=&e^{s \mathcal{L}^{\ast }}\left[ \sum_{\gamma}
\widetilde{a}_{\gamma }^{\ast }\left( {\bm 0}\right)
\frac{2}{V^{\ast} d}\int d\Gamma ^{\ast }\ \widetilde{w}^{\ast
}\left( {\bm 0}\right) \widetilde{\psi }_{\gamma }^{\ast
}\left({\bm 0}\right) \right]. \nonumber \\ \label{E12a}
\end{eqnarray}
The last equality is obtained as follows. From the definition of
$\widetilde{\psi }_{\alpha}^{\ast}$ in Eq.\ (\ref{1.5.4}) it is
\begin{eqnarray}
\frac{2}{V^{\ast }d}\int d\Gamma ^{\ast }\,  \widetilde{w}^{\ast
}\left( {\bm 0}\right) \widetilde{\psi }_{\alpha }^{\ast }\left(
{\bm 0 }\right) &=&\frac{2}{d} \left[ l v_{0}(T_{h})
\right]^{-Nd}\frac{N_{\alpha} l}{ n_{h}T_{h} v_{0}(T_{h})V} \int
d\Gamma\, \widetilde{w}(\Gamma;{\bm 0}) \nonumber \\
& & \times \left[ \frac{\partial
\rho_{h}(\Gamma;n_{h},T_{h})}{\partial n_{h}} \delta_{\alpha
1}+\frac{\partial \rho (\Gamma;n_{h},T_{h})}{\partial T_{h}}
\delta_{ \alpha 2} \right] \nonumber \\
&=& \frac{ 2 l}{ n_{h} T_{h} v_{0}(T_{h}) V d}\left(
\delta_{\alpha 1} n_{h} \frac{\partial}{\partial n_{h}} -
\delta_{\alpha 2} T_{h}
\frac{\partial}{\partial T_{h}} \right) \nonumber \\
& & \times \int d \Gamma\, \widetilde{w} (\Gamma;{\bm 0})
\rho_{h}(\Gamma; n_{h},T_{h}). \label{E13a}
\end{eqnarray}
Moreover, from Eq.\ (\ref{D20}),
\begin{equation}
\int d \Gamma\, \widetilde{w}(\Gamma;{\bm 0})\rho_{h}
(\Gamma;n_{h},T_{h})=\frac{d}{2} n_{h}V \zeta_{0}(n_{h},T_{h})
T_{h},
\end{equation}
so that Eq.\ (\ref{E13a}) can be rewritten as
\begin{eqnarray}
\frac{2}{V^{\ast }d}\int d\Gamma ^{\ast }\,  \widetilde{w}^{\ast
}\left( {\bm 0}\right) \widetilde{\psi }_{\alpha }^{\ast }\left(
{\bm 0 }\right) =  \zeta_{0}^{\ast} \left\{ \delta_{\alpha 1}
\left[ 1+ \frac{\partial \ln \zeta_{0}(n_{h},T_{h})}{\partial \ln
n_{h}} \right] +\delta_{\alpha 2} \frac{3}{2} \right\}.
\end{eqnarray}
This proves Eq.\ (\ref{E12a}). Use of it into the Fourier transform
of Eq.\ (\ref{E10a}) particularized for ${\bm k}^{\ast}={\bm 0}$
yields
\begin{equation}
\widetilde{\ell}^{\ast }\left(\Gamma^{\ast}; {\bm 0},s\right) =e^{s
\mathcal{L}^{\ast }}\left( 1-P^{\dagger }\right)
\frac{2}{d}\widetilde{w}^{\ast }\left( {\bm 0}\right) \label{E14}
\end{equation}
with the operator ${P}^{\dagger }$ given by
\begin{equation}
P^{\dagger }X\left( \Gamma ^{\ast }\right) = V^{\ast  -1}
\sum_{\alpha}\widetilde{a}_{\alpha }^{\ast }\left( \Gamma^{\ast};
{\bm 0}\right) \int d\Gamma ^{\ast}\, \widetilde{\psi }_{\alpha
}^{\ast }\left( \Gamma^{\ast};{\bm 0}\right) X\left( \Gamma ^{\ast
}\right) . \label{E15}
\end{equation}
This is a projection operator since the quantities $\left\{
\widetilde{a}_{\alpha }^{\ast}(\Gamma^{\ast };{\bm 0}) \right\} $
and $\left\{ \widetilde{\psi }_{\beta }^{\ast
}\left(\Gamma^{\ast}; {\bm 0}\right) \right\}$ form a biorthogonal
set, in the sense that
\begin{equation}
V^{\ast -1} \int d\Gamma ^{\ast }\, \widetilde{a}_{\alpha }^{\ast
}\left(\Gamma^{\ast}; {\bm 0}\right) \widetilde{\psi }_{\beta
}^{\ast }\left( \Gamma^{\ast}; {\bm 0} \right) =\delta _{\alpha
\beta },
\end{equation}
as shown in ref. \cite{DBB06}. The utility of this result is that
$ \widetilde{\ell}^{\ast }\left( {\bm 0},s\right) $ is the source
phase function for the transport coefficient $\zeta^{U}$, as shown
in Eq.\ ({\ref{F2}). Then,  the presence of this orthogonal
projection is essential for the existence of the large $s$ limit
in the Green-Kubo representation for $\zeta^{U}$, as discussed in
reference \cite{DBB06}. A similar orthogonal projection occurs for
all of the direct fluxes $F^{S,f}$ occurring in the representation
(\ref{5.4}) for transport coefficients.

In summary, the conservation laws associated with the chosen phase
functions $\left\{ a_{\alpha }\right\} $ in dimensionless variables
take the form
\begin{equation}
\frac{\partial \widetilde{a}_{\alpha }^{\ast }\left(\Gamma^{\ast};
{\bm k}^{\ast},s\right) }{
\partial s}+ \sum_{\beta} \mathcal{K}_{\alpha \beta }^{\ast hyd }
\left( \Gamma^{\ast}; {\bm 0}\right) \widetilde{a}_{\beta}^{\ast
}\left( \Gamma^{\ast}; {\bm k}^{\ast},s\right) -i {\bm k}^{\ast}
\cdot \widetilde{\bm f}_{\alpha}^{\ast }\left( \Gamma^{\ast}; {\bm
k}^{\ast},s\right) =\delta _{\alpha 2} \widetilde{\ell}^{\ast
}\left( \Gamma^{\ast}; {\bm k}^{\ast} ,s\right), \label{E17}
\end{equation}
where $\mathcal{K}_{\alpha \beta}^{ \ast hyd}(\Gamma^{\ast};{\bm
0})$ is the ${\bm k}^{\ast}={\bm 0}$ limit of the matrix given in
Eqs. (\ref{1.4.1a})-(\ref{1.4.3}) and
\begin{equation}
{\bm f}_{1}^{\ast}(\Gamma^{\ast};{\bm r}^{\ast})= \bm{\mathcal
G}^{\ast}( \Gamma^{\ast};{\bm r}^{\ast}), \label{E18aa}
\end{equation}
\begin{equation}
{\bm f}_{2}^{\ast}(\Gamma^{\ast};{\bm r}^{\ast})= \frac{2}{d} {\bm
s}^{\ast}(\Gamma^{\ast};{\bm r}^{\ast})- \bm{\mathcal G}^{\ast}(
\Gamma^{\ast};{\bm r}^{\ast}), \label{E18a}
\end{equation}
\begin{equation}
{\sf f}_{3,ij}^{\ast}(\Gamma^{\ast};{\bm r}^{\ast})= {\sf
h}_{ij}^{\ast}( \Gamma^{\ast};{\bm r}^{\ast}). \label{E18}
\end{equation}
The last equation above gives the tensor flux appearing in the
equation for the dimensionless momentum $\bm{\mathcal
G}^{\ast}(\Gamma^{\ast};{\bm r}^{\ast}) \equiv {\bm a}_{3}^{\ast}
(\Gamma^{\ast};{\bm r}^{\ast})$.

\section{Conjugate Functions and Conservation Laws}
\label{ap5} Consider the conjugate functions $\widetilde{\psi
}_{\alpha }^{\ast }\left( {\bm k}^{\ast} \right) $ that occur in
the response function (\ref{1.5.1}), obtained from the linear
response analysis, and are defined by Eq.\ (\ref{1.5.4}). The
dimensionless dynamical equation obeyed by these functions is
\begin{equation}
\left( \partial _{s}+\overline{\mathcal{L}}^{\ast }\right) \widetilde{\psi }%
_{\alpha }^{\ast }\left( {\bm k}^{\ast},s\right) =0.  \label{E21}
\end{equation}
In the homogeneous limit ${\bm k}^{\ast}={\bm 0}$, these functions
verify \cite {DBB06}
\begin{equation}
\overline{\mathcal{L}}^{\ast }\widetilde{\psi }_{\alpha }^{\ast
}\left( {\bm 0}\right) = \sum_{\beta}  \mathcal{K}^{\ast hyd
}_{\beta \alpha } ({\bm 0}) \widetilde{\psi }_{\beta }^{\ast
}\left({\bm 0}\right), \label{E20}
\end{equation}
where $ \mathcal{K}^{\ast hyd}\left({\bm 0}\right)$ is the
hydrodynamic transport matrix given by Eqs.\  (\ref{1.4.1a}) and
(\ref{1.4.3}) in the main text. This property allows the
construction of a new set of ``conservation laws'':
\begin{equation}
\frac{\partial}{\partial s} \widetilde{\psi }_{\alpha }^{\ast
}\left( {\bm k}^{\ast},s\right) + \sum_{\beta} \mathcal{K}_{\beta
\alpha }^{\ast hyd}\left({\bm 0}\right) \widetilde{\psi } _{\beta
}^{\ast }\left({\bm k}^{\ast},s\right) -i {\bm k}^{\ast}\cdot
\widetilde{\bm  \gamma }_{\alpha}^{\ast} \left( {\bm
k}^{\ast},s\right) =0, \label{E22}
\end{equation}
with the ``conjugate fluxes'' $\widetilde{\bm \gamma }_{\alpha
}^{\ast} \left( {\bm k} ,s\right) $ defined by
\begin{equation}
i {\bm k}^{\ast} \cdot \widetilde{\bm \gamma }_{\alpha }^{\ast}
\left({\bm k}^{\ast} ,s\right) \equiv - \left(
\overline{\mathcal{L}}^{\ast }-\sum_{\beta} \mathcal{K}_{\beta
\alpha }^{\ast hyd}\left( {\bm 0}\right) \right) \widetilde{\psi
}_{\beta }^{\ast }\left( {\bm k}^{\ast},s\right) . \label{E23}
\end{equation}
The property (\ref{E20}) assures that $i {\bm k}^{\ast} \cdot
\widetilde{\bm \gamma }_{\alpha }^{\ast} $ vanishes at ${\bm
k}^{\ast}=0,$ justifying the introduction of $\widetilde{\bm
\gamma } _{\alpha }^{\ast}$ as a flux. It is the conjugate
quantity occurring in the Green-Kubo expressions of the transport
coefficients, Eq.\ (\ref{5.6}). More precisely, $\Upsilon^{\ast}$
in that expression is defined by
\begin{equation}
\Upsilon_{\alpha}^{\ast}  \equiv \widehat{\bm k}^{\ast} \cdot
\widetilde {\bm \gamma} _{\alpha }^{\ast} \left( {\bm 0} \right) .
\label{E24}
\end{equation}

\section{Transport Coefficients: Some details}
\label{ap6}

The linear response analysis of ref. \cite{DBB06}, identified the
transport coefficients in the intermediate Helfand representation
by Eqs. (118)-(120). The transformation of those results to the
dimensionless, hard sphere form is outlined in Sec.\  VIII of that
reference. Some further details for the expressions of the
dimensionless transport coefficients are described briefly here.

The dimensionless response functions have a dynamics generated by
$\overline{ \mathcal{L}}^{\ast }$, as in Eq.\ (\ref{B25BB}).
However, the transport coefficients are given in terms of
correlation functions whose dynamics is generated by
$\overline{\mathcal{L}}^{\ast }-\left[ \mathcal{K}^{\ast hyd
}\left({\bm 0}\right) \right] ^{T}$, where the superscript $T$
indicates transposed. As discussed in ref. \cite{DBB06} and also in
the main text here, this generator compensates for both the cooling
of the reference state (through the change of $\overline{L}^{\ast }$
by  $\overline{\mathcal{L}}^{\ast }$) and the dynamics of
homogeneous perturbations of that state (through the subtraction of
$ \mathcal{K} ^{\ast hyd }\left( {\bm 0}\right)  $).

Consider first the Euler transport coefficient $\zeta ^{U}$ given
in the intermediate Helfand representation by Eq.\ (118) of ref.\
\cite{DBB06}, whose dimensionless form is
\begin{equation}
\zeta ^{\ast U }=-\lim \widehat{\bm k}^{\ast} \cdot \overline{\bm
S}_{23}^{\ast (1) }(s),  \label{F1}
\end{equation}
where $\overline{\bm S}_{23}^{\ast (1)}(s)$ is obtained from
\begin{eqnarray}
\overline{\bm S}_{23}^{\ast }({\bm k}^{\ast}, s) &=& \frac{
v_{0}\left[T_{h}(t) \right]}{v_{0} \left[T_{h}(0) \right]} V^{\ast
-1} \int d \Gamma^{\ast}\, \widetilde{\ell}^{\ast}
(\Gamma^{\ast},{\bm k}^{\ast} ) e^{-s \overline{\mathcal
L}^{\ast}} \widetilde{\psi}_{\parallel}^{*} (\Gamma^{\ast};-{\bm k}^{\ast}) \nonumber \\
&=& V^{\ast -1} \int d\Gamma ^{\ast }\, \widetilde{\ell}^{\ast
}\left( \Gamma^{\ast}; {\bm k}^{\ast} \right) e^{- s \left(
\overline{\mathcal{L}}^{\ast }+\frac{\zeta _{0}^{\ast }}{2}
\right) }\widetilde{\psi }_{\parallel}^{\ast }\left(
\Gamma^{\ast}; -{\bm k}^{\ast} \right), \label{F2}
\end{eqnarray}
as explained below. In this expression, $\widetilde{\ell}^{\ast
}\left(\Gamma^{\ast}; {\bm k}^{\ast} \right) $ is the source term
in the balance equation for the energy given in Eq. (\ref{E10a})
and $\ \widetilde{\psi }_{\parallel}^{\ast }(\Gamma^{\ast};{\bm
k}^{\ast})$ is the conjugate density associated with the
longitudinal component of the flow velocity. Equation (\ref{F2})
is easily derived by using Eq.\ (168) in ref. \cite{DBB06}. It can
also be expressed as
\begin{equation}
\overline{\bm S}_{23}^{\ast }({\bm k}^{\ast}, s)=V^{\ast -1} \int
d \Gamma^{\ast}\, \widetilde{\ell}^{\ast} (\Gamma^{\ast},{\bm
k}^{\ast} ) \widetilde{\psi}_{\parallel}^{\ast} (\Gamma^{\ast};-
{\bm k}^{\ast},s), \label{F2.1}
\end{equation}
with
\begin{equation}
\widetilde{\psi}_{\parallel}^{\ast} (\Gamma^{\ast};{\bm k}^{\ast},s)
\equiv  \sum_{\alpha} \mathcal{U}^{\ast}_{\parallel \alpha}(s)
\widetilde{\psi}_{\alpha}^{\ast} (\Gamma^{\ast};{\bm k}^{\ast}),
\label{F2.2}
\end{equation}
the matrix evolution operator $\mathcal{U}^{\ast}(s)$ being given
by
\begin{equation}
\mathcal{U}^{\ast}_{\alpha \beta}(s) \equiv \widetilde{C}^{\ast
hyd -1}_{\beta \alpha} e^{-s \overline{\mathcal{L}}^{\ast}} =
\left[ e^{s \mathcal{K}^{\ast hyd}({\bm 0})} \right]_{\beta
\alpha} e^{-s \overline{\mathcal{L}^{\ast}}}. \label{F2.3}
\end{equation}
Here, $\mathcal{K}^{\ast hyd} ({\bm k}^{\ast})$ is as always the
matrix given in Eqs.\ (\ref{1.4.1a})-(\ref{1.4.3}). It will be
seen in the following that, consistently  with the description
given at the beginning of this Section, going to the dimensionless
expressions of the transport coefficients implies replacing the
matrix evolution $\mathcal{U}(t,T)$ used in ref. \cite{DBB06} by
$\mathcal{U}^{\ast}(s)$ defined above.

The quantity $\overline{\bm S}^{\ast (1)}_{23}(s)$ in Eq.\
(\ref{F1}) is generated from $\overline{S}^{\ast}_{23}({\bm
k}^{\ast},s)$ by means of the Taylor series expansion around ${\bm
k}^{\ast}={\bm 0}$, that for any function $X({\bm k}^{\ast})$ is
defined as
\begin{equation}
X({\bm k}^{\ast})=X({\bm 0})+i{\bm k}^{\ast} \cdot {\bm X}^{(1)}-
{\bm k}^{\ast} {\bm k}^{\ast} : {\sf X}^{(2)} + \ldots. \label{F4}
\end{equation}
Then, it is
\begin{equation}
{\bm S}_{23}^{\ast (1)}(s)  =  - V^{\ast-1} \int d \Gamma^{\ast}\,
\widetilde{\ell}^{\ast} (\Gamma^{\ast};{\bm 0})  e^{-s \left(
\overline{\mathcal{L}}^{\ast }+\frac{\zeta _{0}^{\ast }}{2} \right)
} \widetilde{\bm \psi}^{\ast (1)}(\Gamma^{\ast}), \label{F4.1}
\end{equation}
with
\begin{equation}
\widetilde{\bm \psi}^{\ast (1)}(\Gamma^{\ast})= - \sum_{r=1}^{N}
{\bm q}^{\ast}_{r} \frac{\partial}{\partial v^{\ast}_{s,
\parallel}} \rho^{\ast}_{h}(\Gamma^{\ast}). \label{F4.2}
\end{equation}
Upon writing Eq.\ (\ref{F4.1}), a contribution involving
$\widetilde{\bm \ell}^{\ast (1)} (\Gamma^{\ast})$ has been omitted
since it vanishes due to the symmetry of
$\widetilde{\psi}_{\parallel}^{\ast}(\Gamma^{\ast};{\bm 0})$. Taking
now into account Eq.\ (\ref{E14}) and the definitions given in
(\ref{E9.2}) and (\ref{5.1AAB}), it is easily verified that
\begin{equation}
\widetilde{\ell}^{\ast}(\Gamma^{\ast};{\bm 0})=(1-P^{\dagger})
W^{\ast}(\Gamma^{\ast}). \label{F4.3}
\end{equation}
Use of this into Eq.\ (\ref{F4.1}) and later substitution of the
result into Eq.\ (\ref{F1}) leads to Eqs.\
(\ref{5.3.A})-(\ref{5.3.5a}), after using the symmetry of the
integrand.

The Navier-Stokes transport coefficients at order $k^{2}$
associated with the heat and momentum flux, are given by Eq. (119)
of ref. \cite{DBB06} and, more explicitly, they are identified in
Appendix F of that reference. In dimensionless form, they are the
elements of the matrix
\begin{equation}
\Lambda^{\ast }=\lim \widehat{\bm k}^{\ast} \widehat{\bm k}^{\ast}
: \left[ \overline{\sf D}^{\ast (1) }(s)-\overline{\bm D}^{\ast
}({\bm 0},0)\overline{\bm C}^{(1)\ast }(s)\right] , \label{F5}
\end{equation}
where the  correlation functions $\overline{C}^{\ast }({\bm
k}^{\ast}, s) $ and $\overline{\bm D}^{\ast }({\bm k}^{\ast}, s)$
are defined by
\begin{equation}
\overline{C}_{\alpha \beta }^{\ast }({\bm k}^{\ast},s)= V^{\ast -1}
\int d\Gamma ^{\ast }\, \widetilde{a}_{\alpha }^{\ast }\left(
\Gamma^{\ast}; {\bm k}^{\ast} \right) \widetilde{\psi }^{\ast
}_{\beta} \left( \Gamma^{\ast}; -{\bm k}^{\ast},s \right).
\label{F6}
\end{equation}
\begin{equation}
\overline{\bm D}_{\alpha \beta }^{\ast }({\bm k}^{\ast}, s)=
V^{\ast -1} \int d\Gamma ^{\ast }\, \widetilde{\bm f}_{\alpha
}^{\ast }\left(\Gamma^{\ast}; {\bm k}^{\ast} \right)
\widetilde{\psi }^{\ast }_{\beta} \left( \Gamma^{\ast}; -{\bm
k}^{\ast},s \right), \label{F7}
\end{equation}
with the direct densities $\widetilde{a}_{\alpha }^{\ast }$ given in
Eq.\ (\ref {1.5.2}) and the associated fluxes $\widetilde{\bm
f}_{\alpha }^{\ast } $ given in Eqs.\ (\ref{E18aa})-(\ref{E18}).

The explicit form of the first order in $k^{\ast}$ expanded
correlation matrix $\overline{\bm C}^{ \ast \left( 1\right)}
\left( s \right) $ above is
\begin{eqnarray}
\overline{\bm C}_{\alpha \beta}^{\ast \left( 1\right)
}\left(s\right) &=&V^{\ast -1 }\int d\Gamma ^{\ast }\,
\widetilde{\bm a}_{\alpha}^{\ast (1)} (\Gamma^{\ast})
\widetilde{\psi }_{\beta}^{\ast } \left( \Gamma^{\ast}; {\bm
0},s \right) \nonumber \\
&&-V^{\ast -1}\int d\Gamma ^{\ast }\, \widetilde{a}_{\alpha}^{\ast
}\left(
\Gamma^{\ast}; {\bm 0}\right) \widetilde{\bm \psi}_{\beta}^{\ast (1)} (\Gamma^{\ast};s)  \nonumber \\
&=&-V^{\ast -1}\int d\Gamma ^{\ast }\,
\widetilde{a}_{\alpha}^{\ast }\left( \Gamma^{\ast}; {\bm 0}\right)
\widetilde{\bm \psi }^{\ast (1) }_{\beta} \left( \Gamma^{\ast};s
\right). \label{F8}
\end{eqnarray}
The contribution involving $\widetilde{\bm a}_{\alpha }^{\ast \left(
1\right) } (\Gamma^{\ast})$ is proportional to the center of mass
coordinates and vanishes from symmetry. A similar analysis of ${\sf
D}^{\ast \left( 1\right)}\left(s\right) $ gives
\begin{equation}
\overline{\sf D}_{\alpha \eta }^{\ast \left( 1\right)
}\left(s\right) =- V^{\ast -1}\int d\Gamma ^{\ast }\,
\widetilde{\bm f}_{\alpha }^{\ast }\left( \Gamma^{\ast}; {\bm
0}\right) \widetilde{\bm \psi }^{\ast (1) }_{\beta} \left(
\Gamma^{\ast};s \right). \label{F9}
\end{equation}
The two terms on the right side of Eq. (\ref{F5}) now combine and
the transport matrix $\Lambda ^{\ast }$ takes the simple form
\begin{equation}
\Lambda _{\alpha \beta }^{\ast }= - \lim \widehat{\bm k}^{\ast}
\widehat{\bm k}^{\ast} : V^{\ast -1} \int d\Gamma ^{\ast }\,
\widetilde{\bm f}^{\ast}(\Gamma^{\ast};{\bm 0}) (1-P) \widetilde{\bm
\psi }^{\ast (1) }_{\beta} \left( \Gamma^{\ast};s \right),
\label{F10a}
\end{equation}
where $P$ is the projection operator
\begin{equation}
PX(\Gamma^{\ast}) \equiv  \sum_{\alpha}
\psi^{\ast}_{\alpha}(\Gamma^{\ast};{\bm 0}) V^{\ast -1} \int
d\Gamma^{\ast} \widetilde{a}_{\alpha}^{\ast} (\Gamma^{\ast};0)
X(\Gamma^{\ast }). \label{F10aa}
\end{equation}
Since the operator $1-P$ is orthogonal to the invariants of
$\mathcal{U}^{\ast}(s)$, as shown in ref. \cite{DBB06}, Eq.\
(\ref{F10a}) is equivalent to
\begin{equation}
\Lambda^{\ast}_{\alpha \beta} = - \lim  V^{\ast -1} \int d
\Gamma^{\ast} F_{\alpha}^{\ast f} (\Gamma^{\ast})  \widehat {\bm
k}^{\ast} \cdot \widetilde{\bm \psi}_{\beta}^{\ast
(1)}(\Gamma^{\ast};s), \label{F.10}
\end{equation}
where $F_{\alpha}^{\ast f}$ is the projected flux
\begin{equation}
F_{\alpha }^{\ast f}(\Gamma^{\ast})=\left( 1-P^{\dagger }\right)
\widehat{\bm k}^{\ast} \cdot \widetilde{\bm f}_{\alpha }^{\ast
}\left(\Gamma^{\ast}; {\bm 0}\right) . \label{F11}
\end{equation}
The projection operator $P^{\dagger }$ is the adjoint of $P$, and
it is the same as that given by Eq.\ (\ref{E15}).

\subsection{Shear Viscosity}

The dimensionless shear viscosity is determined from
$\Lambda_{44}^{\ast}$ through
\begin{eqnarray}
\eta ^{\ast } &=& \Lambda _{44}^{\ast }= - \lim V^{\ast -1}
\widehat{\bm k}^{\ast} \widehat{\bm k}^{\ast} :  \int d\Gamma
^{\ast }\, F_{4}^{\ast f} (\Gamma^{\ast}) \left[
\mathcal{U}^{\ast}(s) \widetilde{\bm \psi}^{\ast (1)}
(\Gamma^{\ast}) \right]_{4}
\nonumber \\
&=&- \lim  V^{\ast -1} \widehat{\bm k} \widehat{\bm k} : \int
d\Gamma ^{\ast }\, \widetilde{\bm f}_{4}^{\ast }\left(
\Gamma^{\ast};{\bm 0}) \right)  e^{- s\left(
\overline{\mathcal{L}}^{\ast }+\frac{\zeta _{0}^{\ast }}{2}\right)
} \widetilde{\bm \psi }_{4}^{\ast \left( 1\right) }\left(
\Gamma^{\ast} \right). \label{F12}
\end{eqnarray}
The projection operator in the definition of the projected fluxes,
Eq.\ (\ref{F11}), does not contribute in this case due to the
symmetry of the integrand. The conjugate moment, $\widetilde{\bm
\psi }_{4}^{\ast \left( 1\right) }\left( \Gamma^{\ast} \right)$,
is easily obtained  from the definition in Eq.\ (\ref{F4}) and
Eq.\ (\ref{1.6.1}),
\begin{equation}
\widehat{\bm k}^{\ast} \cdot \widetilde{\bm \psi }_{4}^{\ast
\left( 1\right) }\left( \Gamma^{\ast} \right) =-\sum_{r=1}^{N}
\left( \widehat{\bm k}^{\ast} \cdot {\bm q}_{r}^{\ast }\right)
\frac{\partial}{\partial v^{\ast}_{r,\perp 1}} \rho
_{h}^{\ast}\left( \Gamma ^{\ast }\right),  \label{F13}
\end{equation}
where $v^{\ast}_{r,\perp 1}$ is the first of the transversal
components of ${\bm v}_{r}$.  Moreover, using Eq.\ (\ref{E18}),
\begin{equation}
\widehat{\bm k} \cdot \widetilde{\bm
f}^{\ast}_{4}(\Gamma^{\ast};{\bm 0})= {\sf H}^{\ast}_{\parallel
\perp 1} (\Gamma^{\ast}), \label{F13a}
\end {equation}
so Eq.\ (\ref{F12}) becomes
\begin{equation}
\eta^{\ast}= -\lim V^{\ast -1} \int d\Gamma^{\ast}\, {\sf
H}_{xy}^{\ast}(\Gamma^{\ast}) e^{-s \left( \mathcal{L}^{\ast}
+\frac{\zeta_{0}^{\ast}}{2} \right)} \mathcal{M}_{\eta}^{\ast}
(\Gamma^{\ast}), \label{F13b}
\end{equation}
with
\begin{equation}
\mathcal{M}_{\eta}^{\ast} (\Gamma^{\ast})=-\sum_{r=1}^{N}
q^{\ast}_{r,x} \frac{\partial}{\partial v^{\ast}_{r,y}}
\rho_{h}^{\ast}(\Gamma^{\ast}). \label{F13c}
\end{equation}
The $x$ and $y$ components in the above expression are due to the
arbitrary choice of the $x$ axis along $\widehat{\bm k}$ and the
$y$ axis along the first tranverse direction.

An alternative expression for the shear viscosity can be derived
as follows. Define
\begin{equation}
\mathcal{M}^{\ast}_{\eta,ij}(\Gamma^{\ast})= -\frac{1}{2}
\sum_{r=1}^{N} \left( q^{\ast}_{r,i} \frac{\partial}{\partial
v^{\ast}_{r,j}}+ q^{\ast}_{r,j} \frac{\partial}{\partial
v^{\ast}_{r,i}} -\frac{2}{d} \delta_{ij} {\bm q}^{\ast}_{r} \cdot
\frac{\partial}{\partial {\bm v}^{\ast}_{r}} \right) \rho^{\ast}
(\Gamma^{\ast}), \label{F13d}
\end{equation}
where $i$ and $j$ are spatial coordinates. The symmetry of Eq.\
(\ref{F13b}) implies that it is equivalent to
\begin{equation}
\eta^{\ast} =-\frac{1}{d^{2}+d+2} \lim V^{\ast -1} \sum_{i=1}^{d}
\sum_{j=1}^{d} \int d \Gamma^{\ast} {\sf H}_{ij}^{\ast}
(\Gamma^{\ast}) e^{-s \left( \overline{\mathcal
L}^{\ast}+\frac{\zeta_{0}^{\ast}}{2} \right)} \mathcal{M}_{\eta ,
ij}^{\ast} (\Gamma ^{\ast} ). \label{F13e}
\end{equation}
This is the expression presented in Eqs.\ (\ref{5.10}) and
(\ref{5.11}) in the main text.

\subsection{Bulk Viscosity}

The bulk viscosity $\kappa^{\ast}$ occurs in the $33$ matrix element
of the hydrodynamic transport matrix, namely it is
\begin{equation}
\kappa^{\ast}+\frac{2(d-1) \eta^{\ast}}{d} = \Lambda_{33}^{\ast
}=-\lim V^{\ast -1}\int d\Gamma ^{\ast }\, F_{3}^{\ast f}e^{-s
\left( \overline{\mathcal{L}}^{\ast }+\frac{ \zeta _{0}^{\ast
}}{2}\right)} \mathcal{M}_{3}^{\ast}(\Gamma^{\ast}) . \label{F14}
\end{equation}
The projected flux is found to be
\begin{equation}
F_{3}^{f}(\Gamma^{\ast}) =\widehat{\bm k}^{\ast} \cdot
\widetilde{\bm f}_{3}^{\ast }\left( \Gamma^{\ast};{\bm 0}\right)
-\frac{p^{\ast }_{h}}{2}\frac{\partial \ln p_{h}}{\partial \ln
n_{h}}\, \widetilde{a}_{1}\left( \Gamma^{\ast}; {\bm 0}\right)
-\frac{p^{\ast }_{h}}{2}\,  \widetilde{a}_{2}\left(\Gamma^{\ast};
{\bm 0}\right) \label{F15}
\end{equation}
and the conjugate moment is
\begin{equation}
\mathcal{M}_{3}^{\ast} (\Gamma^{\ast})= \widehat{\bm k}^{\ast} \cdot
{\bm \psi}_{3}^{\ast (1)} (\Gamma^{\ast}) =-\sum_{r=1}^{N} \left(
\widehat{\bm k }^{\ast} \cdot{\bm q}_{r}^{\ast }\right) \widehat{\bm
k}^{\ast} \cdot  \frac{\partial}{\partial {\bm v}^{\ast}_{r}} \rho
_{h}\left( \Gamma ^{\ast }\right). \label{F16}
\end{equation}
Taking into account that $\widehat{\bm k}^{\ast} \cdot
\widetilde{\bm f}_{3}^{\ast }\left( \Gamma^{\ast};{\bm 0}\right)
={\sf H}_{33}^{\ast} (\Gamma^{\ast})$ and the symmetry of the
tensor ${\sf H}^{\ast}$, the expressions used in Sec. \ref{s3},
Eqs.\ (\ref{5.20}) and (\ref{5.21}), follow directly.

\subsection{Thermal conductivity}

The dimensionless thermal conductivity $\lambda^{\ast}$ is given by
\begin{eqnarray}
\lambda ^{\ast } &=& \frac{d}{2}\Lambda _{22}^{\ast }= -
\frac{d}{2} \lim V^{\ast -1}\int d\Gamma ^{\ast }\, F_{2}^{\ast f}
\left[ \mathcal{U}(s) \widehat{\bm k}^{\ast} \cdot \widetilde{\bm
\psi}^{\ast (1)} (\Gamma^{\ast}) \right]_{2}
\nonumber \\
&=& - \frac{d}{2}\lim V^{\ast -1}\int d\Gamma ^{\ast }\,
F_{2}^{\ast f}e^{- s \left( \overline{\mathcal{L}}^{\ast
}-\frac{\zeta _{0}^{\ast }}{2}\right) } \widehat{\bm k} \cdot
\widetilde{\bm \psi }_{2}^{\ast \left( 1\right) }\left(
\Gamma^{\ast} \right) . \label{F17}
\end{eqnarray}
The projected flux is
\begin{eqnarray}
F_{2}^{\ast f} & = & \widehat{\bm k}^{\ast}  \cdot \left[
\widetilde{{\bm f}}_{2}^{\ast
}\left( \Gamma^{\ast}; {\bm 0}\right) -\frac{2 p^{\ast }_{h}}{d}\  {\bm P}^{\ast} \right] \nonumber \\
& = & \frac{2}{d} \widehat{\bm k}\cdot \left[ \widetilde{\bm
s}^{\ast }\left(\Gamma^{\ast}; {\bm 0}\right) -\left( p^{\ast
}_{h}+\frac{d}{2}\right) {\bm P}^{\ast} \right] , \label{F18}
\end{eqnarray}
with
\begin{equation}
{\bm P}^{\ast}= \sum_{r=1}^{N} {\bm v}^{\ast}_{r},
\end{equation}
and the conjugate moment is
\begin{equation}
\widehat{\bm k}^{\ast} \cdot \widetilde{\bm \psi}_{2}^{\ast \left(
1\right) }\left( \Gamma^{\ast} \right) =-\frac{1}{2}\sum_{r=1}^{N}
\left( \widehat{\bm k}^{\ast} \cdot {\bm q}_{r}^{\ast }\right)
\frac{\partial}{\partial {\bm v}^{\ast}_{r}} \cdot \left[ {\bm
v}^{\ast}_{r} \rho_{h}^{\ast} (\Gamma^{\ast}) \right]. \label{F19}
\end{equation}
The above expressions are equivalent to Eqs.\ (\ref{5.26}) and
(\ref{5.27}).

\subsection{The $\protect\mu $ Coefficient}

The expression of the coefficient $\mu^{\ast} $  from linear
response reads
\begin{eqnarray}
\mu ^{\ast } &=& \frac{d}{2} \Lambda _{21}^{\ast } \nonumber \\
&=& - \frac{d}{2} \lim V^{\ast -1} \int d\Gamma ^{\ast }\,
F_{2}^{\ast f} \left[ \mathcal{U}^{\ast}(s) \widehat{\bm k}^{\ast}
\cdot \widetilde{\bm \psi}^{\ast (1)} (\Gamma^{\ast})
\right]_{1} \nonumber \\
& = & - \frac{d}{2} \lim V^{\ast -1} \int d \Gamma^{\ast}\,
F_{2}^{\ast f} e^{- s \mathcal{L}^{\ast}} \left[ \widehat{\bm
k}^{\ast} \cdot \widetilde{\bm \psi}_{1}^{\ast (1)}
(\Gamma^{\ast}) -2 \frac{\partial \ln \zeta_{0}}{\partial \ln
n_{h}} \widehat{\bm k}^{\ast} \cdot \widetilde{\bm \psi}_{2}^{\ast
(1)} (\Gamma^{\ast}) \right] \nonumber \\
&& - d \frac{\partial \ln \zeta_{0}}{\partial \ln n_{h}} \lim
V^{\ast -1} \int d\Gamma^{\ast} F_{2}^{\ast f} e^{-s \left(
\widetilde{\mathcal{L}}^{\ast} -\frac{\zeta_{0}^{\ast}}{2}
\right)} \widehat{\bm k}^{\ast} \cdot \widetilde{\bm \psi}^{\ast
(1)}_{2} (\Gamma^{\ast}). \label{F20}
\end{eqnarray}
Since the projected flux is the same as in Eq.\ (\ref{F17}), the
second term on the right hand side of Eq.\ (\ref{F20}) is
proportional to $\lambda ^{\ast }$, and the expression can be
written
\begin{eqnarray}
\overline{\mu }^{\ast } & \equiv & \mu ^{\ast }-2 \frac{\partial
\ln \zeta _{0}}{\partial \ln n_{h}}\lambda ^{\ast } \nonumber \\
& = & -\frac{d}{2} \lim V^{\ast -1} \int d\Gamma ^{\ast }\,
F_{2}^{\ast f} e^{-s \overline{ \mathcal{L}}^{\ast }}\widehat{\bm
k}^{\ast} \cdot \left[ \widetilde{\bm \psi }_{1}^{\ast \left(
1\right) }\left( \Gamma^{\ast} \right) -2 \frac{\partial \ln
\zeta_{0}}{\partial \ln n_{h}} \widehat{\bm k}^{\ast} \cdot
\widetilde{\bm \psi}_{2}^{\ast (1)} (\Gamma^{\ast}) \right].
\label{F21}
\end{eqnarray}
Therefore, the involved conjugate moments are
\begin{equation}
\widehat{\bm k}^{\ast} \cdot \widetilde{\bm \psi }_{1}^{\ast
\left( 1\right) }\left( \Gamma^{\ast} \right) =\left[ l
v_{0}\left( T \right) \right]^{Nd} n \int d{\bm r}\, \widehat{\bm
k}^{\ast} \cdot{\bm  r}^{\ast}  \left[  \frac{\delta \rho _{lh}
(\Gamma|\left\{ y_{\alpha} \right\})}{\delta n ( {\bm r})}
\right]_{\left\{y_{\alpha} \right\}= \left\{ n,T,{\bm 0}
\right\}},\label{F22}
\end{equation}
\begin{equation}
\widehat{\bm k}^{\ast} \cdot \widetilde{\bm \psi }_{2}^{\ast
\left( 1\right) }\left( \Gamma^{\ast} \right) =\left[ l
v_{0}\left( T \right) \right]^{Nd} T \int d{\bm r}\, \widehat{\bm
k}^{\ast} \cdot{\bm  r}^{\ast}  \left[  \frac{\delta \rho _{lh}
(\Gamma|\left\{ y_{\alpha} \right\})}{\delta T ( {\bm r})}
\right]_{\left\{y_{\alpha} \right\}= \left\{ n,T,{\bm 0}
\right\}}.\label{F22a}
\end{equation}
Substitution of Eqs.\ (\ref{F22}) and (\ref{F22a}) into Eq.\
(\ref{F21}), and use of the symmetry of the integrand leads to
Eqs.\ (\ref{5.31})-(\ref{5.34}).

\section{Elastic Hard Spheres}
\label{ap7}

The aim of this Appendix is to give the Green-Kubo expressions for
the transport coefficients in the case of elastic hard spheres or
disks. The linear response method used in reference \cite{DBB06},
was specifically applied to initial perturbations of a local HCS.
The results apply as well in the elastic case for perturbations of a
local equilibrium distribution, with only  the replacement of the
local HCS ensemble by a local equilibrium  Gibbs ensemble. Here, for
simplicity and to compare with standard results in the literature, a
local grand canonical ensemble will be used. Strictly speaking, this
is not the $\alpha \rightarrow 1$ limit of the results derived in
this paper for a granular fluid, since the HCS ensemble in the
elastic limit goes over to a microcanonical ensemble with constant
total energy and momentum, i.e., the so-called MD ensemble. The
results obtained by using different ensembles are known to be
equivalent up to fluctuations that do not contribute in the
thermodynamic limit, but the ensemble used affects the forms of the
projected fluxes appearing in the Green-Kubo expressions.

The form of the representations for all transport coefficients is,
of course, the same as in the inelastic case, but with some
simplifications. The most obvious ones are the absence of cooling,
implying that $ \overline{\mathcal{L}}^{\ast }$ reduces to $
\overline{L}^{\ast }$, and the absence of any dynamics for
homogeneous perturbations, that reflects itself in that $
\mathcal{K}^{\ast hyd }\left( {\bm 0}\right) = 0 $. Of course, both
simplifications are closely related. Moreover, the dimensionless
time $s$ is now simply proportional to $t$. The direct fluxes $
F^{\ast S}$ in Eq.\ (\ref{5.6}) vanish since all the the source
terms are zero and, consequently, also all the transport
coefficients related to the cooling rate. On the other hand, the
direct fluxes $ F^{\ast f}$ are unchanged. The generic transport
coefficient in Eq.\ (\ref{6.1.1}) then becomes
\begin{equation}
\omega _{el}=\lim V^{\ast -1}\int d\Gamma ^{\ast }\ F^{f}\left(
\Gamma ^{\ast }\right) e^{-s \overline{L}^{\ast }}\mathcal{M}^{\ast}
\left( \Gamma ^{\ast }\right) .  \label{H1}
\end{equation}
It remains only to determine the moments $\mathcal{M}^{\ast} \left(
\Gamma ^{\ast }\right) $, and afterwards the Green-Kubo
representation can be constructed directly as done in the main text
for the general inelastic case.

The distribution of the local grand canonical ensemble is
\cite{McLbook},
\begin{equation}
\rho _{lGC}\left( \Gamma \right) = \Xi (\Gamma) \exp \left\{
-Q_{L}+\int d{\bm r}\left[ \varrho \left( {\bm r}\right)
\mathcal{N}\left( \Gamma; {\bm r}\right) -\frac{ \mathcal{E}(
\Gamma;{\bm r})}{T({\bm r})} +{\bm Y} \left( {\bm r}\right) \cdot
\bm{\mathcal{G}} \left(\Gamma; {\bm r}\right) \right] \right\},
\label{H1d}
\end{equation}
where the  parameters of the ensemble are defined as
\begin{equation}
\varrho({\bm r}) \equiv  \frac{1}{T({\bm r})} \left[\varsigma ({\bm
r})- \frac{ mU^{2}({\bm r})}{2 T({\bm r})} \right],
\end{equation}
\begin{equation}
{\bm Y} ({\bm r}) \equiv  \frac{n({\bm r}) m {\bm U}({\bm
r})}{T({\bm r})},\label{H1b}
\end{equation}
$\varsigma $ being the chemical potential, ${\bm U}$ the flow
velocity, $T$ the temperature in units where the Boltzmann constant
is unity, and $Q_{L}$ the normalization constant. Moreover, $ \Xi
(\Gamma)$  is the overlap function that is zero for any
configuration of overlapping pairs and unity otherwise.

Consider first the conjugate fluxes defined in Eq.\ (\ref{1.5.4})
with the substitution of $\rho_{lh}$ by $\rho_{lGC}$, i. e.,
\begin{equation}
\widetilde{\psi }_{\alpha }^{\ast }\left( \Gamma^{\ast}; {\bm
k}^{\ast} \right) = M_{\alpha } \int d {\bm r}\, e^{i{\bm k}\cdot
{\bm r}} \left[ \frac{\delta \rho _{lGC}\left[ \Gamma |\left\{
y_{\beta } \right\} \right]}{\delta y_{\alpha }\left( {\bm r}
\right) }\right]_{\{ y_{\beta}\}=\left\{ n,T,{\bm 0}\right\}}.
\label{H2}
\end{equation}
The functional differentiations can be performed explicitly in this
case, with the results
\begin{equation}
\widetilde{\psi }_{1}^{\ast }\left( \Gamma^{\ast}; {\bm k}^{\ast
}\right) =\left[ 1+n_{h} \widetilde{h}_{e}\left({\bm k}\right)
\right] ^{-1} \widetilde{\mathcal{N}}^{\ast} \left( \Gamma^{\ast};
{\bm k}^{\ast }\right) \rho _{GC}^{\ast } (\Gamma^{\ast}),
\label{H3}
\end{equation}
\begin{equation}
\widetilde{\psi }_{2}^{\ast }\left( \Gamma^{\ast};  {\bm k}^{\ast
}\right) =\left[ \widetilde{\mathcal{E}}^{\ast}(\Gamma^{\ast};{\bm
k}^{\ast})- \frac{d}{2}
\widetilde{\mathcal{N}}^{\ast}(\Gamma^{\ast};{\bm k}^{\ast}) \right]
\rho_{GC}^{\ast}(\Gamma^{\ast}), \label{H4}
\end{equation}
\begin{equation}
\widetilde{\bm{\psi }}_{3}^{\ast }\left(\Gamma^{\ast};{\bm
k}^{\ast}\right) = 2
\widehat{\bm{\mathcal{G}}}^{\ast}(\Gamma^{\ast};{\bm k}^{\ast}) \rho
_{GC}^{\ast }(\Gamma^{\ast}). \label{H5}
\end{equation}
Here $\widetilde{h}_{e}\left( {\bm k}\right) $ is the Fourier
transform of $ g_{e}(r)-1$, where $g_{e}(r)$ is the equilibrium
radial distribution function. The notation
$\widetilde{\bm{\psi}}^{\ast}_{3} \equiv \left\{
\widetilde{\psi}_{\parallel}^{\ast},
\widetilde{\bm{\psi}}^{\ast}_{\perp} \right\}$ has been used. It is
seen that the $\widetilde{\psi }_{\alpha }\left(\Gamma^{\ast}; {\bm
k}\right)$'s are linear combinations of the direct fluxes $
\widetilde{a }_{\alpha }\left( \Gamma^{\ast};{\bm k}\right) $
multiplied by the grand canonical ensemble dimensionless
distribution $\rho_{GC}^{\ast}(\Gamma^{\ast})$.

The moments $\mathcal{M}^{\ast}\left( \Gamma ^{\ast }\right) $
appearing in the Helfand expressions, are the coefficients of $i
k^{\ast}$ in the expansions of the $ \widetilde{\psi }_{\alpha
}\left( {\bm k}^{\ast}\right)$'s, as indicated in Eq. (\ref{5.4a}),
\begin{equation}
\mathcal{M}^{\ast} \left( \Gamma ^{\ast }\right) = \widehat{\bm
k}^{\ast} \cdot \widetilde{\bm{\psi }}^{\ast \left( 1\right)}
(\Gamma^{\ast}).
\end{equation}
In detail, they are found to be
\begin{equation}
\widehat{\bm k}^{\ast} \cdot \widetilde{\bm{\psi} }_{1}^{\ast \left(
1\right)} (\Gamma^{\ast})=\left[ 1+n_{h} \widetilde{h}_{e}\left(
{\bm 0}\right) \right] ^{-1}\sum_{r=1}^{N} \widehat{\bm k}^{\ast}
\cdot {\bm q}_{r}^{\ast }\rho _{GC}^{\ast } (\Gamma^{\ast}),
\label{H6}
\end{equation}
\begin{equation}
\widehat{\bm k}^{\ast} \cdot \widetilde{\bm{\psi}}_{2}^{\ast \left(
1\right) } (\Gamma^{\ast})=\sum_{r=1}^{N} \widehat{\bm k}^{\ast}
\cdot {\bm q}_{r}^{\ast }\left( v_{r}^{\ast 2}-\frac{d }{2}\right)
\rho _{GC}^{\ast }(\Gamma^{\ast}),  \label{H7}
\end{equation}
\begin{equation}
\widehat{\bm k}^{\ast} \cdot \widetilde{\bm{\psi}}_{\parallel}^{\ast
\left( 1\right)} (\Gamma^{\ast})= 2 \sum_{r=1}^{N} \left(
\widehat{\bm k}^{\ast} \cdot {\bm q}_{r}^{\ast }\right) \left(
\widehat{\bm k}^{\ast} \cdot {\bm v}_{r}^{\ast }\right) \rho
_{GC}^{\ast }(\Gamma^{\ast}), \label{H8}
\end{equation}
\begin{equation}
\widehat{\bm k}^{\ast} \cdot \widetilde{\bm{\psi}}_{\perp i}^{\ast
\left( 1\right)}(\Gamma^{\ast})= 2 \sum_{r=1}^{N} \left(
\widehat{\bm k}^{\ast} \cdot {\bm q}_{r}^{\ast }\right)
v^{\ast}_{r,\perp i} \rho _{GC}^{\ast }(\Gamma^{\ast}). \label{H9}
\end{equation}
Once these moments have been identified, the various transport
coefficients for a system of elastic hard spheres or disks can be
directly written.

\subsection{Shear and Bulk Viscosities}

The Helfand forms in Sec.\ \ref{s3} for the shear and bulk
viscosities of the granular model, in the elastic limit become
\begin{equation}
\eta ^{\ast }=\lim \Omega _{H}^{\eta }\left( s\right) ,\quad
\kappa^{\ast }=\lim \Omega _{H}^{\kappa }\left( s\right),
\label{H10}
\end{equation}
respectively, where $\Omega _{H}^{\eta }\left( s\right) $ and
$\Omega _{H}^{\kappa }\left( s\right) $ are now equilibrium time
correlation functions,
\begin{equation}
\Omega _{H}^{\eta }\left( s\right) = - \frac{V^{\ast -1}}{d^{2}+d-2}
\sum_{i=1}^{d} \sum_{j=1}^{d} \int d\Gamma ^{\ast }\ {\sf
H}_{ij}^{\ast }(\Gamma^{\ast} ) e^{-s \overline{L}^{\ast }}\left(
M_{\eta ,ij}^{\ast }\rho _{GC}^{\ast }\right), \label{H11}
\end{equation}
\begin{equation}
\Omega _{H}^{\kappa }\left( s\right) = -2 (V^{\ast} d^{2} )^{-1}
\int d\Gamma ^{\ast }\, \text{tr } {\sf H}^{\ast f} e^{-s
\overline{L}^{\ast }}\left( M_{\kappa }^{\ast }\rho _{GC}^{\ast
}\right),   \label{H12}
\end{equation}
with ${\sf H}_{ij}^{\ast }$ given by Eq.\ (\ref{5.2.B}) and ${\sf
H}_{ij}^{\ast }$  by Eq.\ (\ref{5.5.DF}). Moreover,
\begin{equation}
M_{\eta ,ij}^{\ast }=\sum_{r=1}^{N} \left( q_{r,i}^{\ast
}v_{r,j}^{\ast }+q_{r,j}^{\ast }v_{r,i}^{\ast }-\frac{2}{d}\delta
_{ij}{\bm q}_{r}^{\ast }\cdot {\bm v}_{r}^{\ast }\right),
\end{equation}
\begin{equation}
M_{\kappa }^{\ast }=\sum_{r=1}^{N} {\bm q}_{r}^{\ast }\cdot {\bm
v}_{r}^{\ast }.  \label{H13}
\end{equation}
These are the usual well-known results for systems of elastic hard
spheres or disks, in dimensionless form. The corresponding
Green-Kubo expressions are
\begin{equation}
\eta^{\ast }=\Omega _{0}^{\eta }+\lim \int_{0}^{s}ds^{\prime }\Omega
_{G}^{\eta }\left( s^{\prime }\right) ,
\end{equation}
\begin{equation}
\kappa ^{\ast }=\Omega _{0}^{\kappa }+\lim \int_{0}^{s}ds^{\prime
}\Omega _{G}^{\kappa }\left( s^{\prime }\right).   \label{H.14}
\end{equation}
The instantaneous parts, $\Omega_{0}^{\eta}=\Omega_{H}^{\eta}(0)$
and $\Omega_{0}^{\kappa}=\Omega_{H}^{\kappa}(0)$, now can be
evaluated exactly with the result
\begin{equation}
\Omega _{0}^{\kappa }=\frac{d+2}{d}\, \Omega _{0}^{\eta
}=\frac{\sqrt{2} \pi^{(d-1)/2}\sigma^{\ast d+1}}{\Gamma(d/2) d^{2}}
g_{e}\left( \sigma \right).  \label{H15}
\end{equation}
The calculation of the conjugate fluxes in the expressions of
$\Omega _{G}^{\eta }\left( s\right) $ and $\Omega _{G}^{\kappa
}\left( s\right) $ requires some care. The calculation of
$\Upsilon_{\eta,ij}^{\ast}$ has been discussed in detail in Sec.
\ref{s3}, and the result is given by Eq.\ (\ref{5.16}). The
subsequent expression for the elastic shear viscosity is presented
in Eq.\ (\ref{5.17}). In a completely similar way, it is obtained
that the integrand in the Green-Kubo expression for the bulk
viscosity is
\begin{equation}
\Omega _{G}^{\kappa }\left( s\right) = (V^{\ast}d^{2})^{-1} \int
d\Gamma ^{\ast }\ \text{tr } {\sf H}^{\ast f} (\Gamma^{\ast})
e^{-s \overline{L}^{\ast }}\left[ \text{tr } {\sf H}^{\ast -}
(\Gamma^{\ast} ) \rho _{GC}^{\ast } (\Gamma^{\ast} ) \right] .
\label{H20}
\end{equation}

\subsection{Thermal Conductivity}

The expression of the thermal conductivity in the Helhand
representation reads
\begin{equation}
\lambda^{\ast} = \lim \Omega _{H}^{\lambda }\left( s\right) ,
\label{H21}
\end{equation}
where for a system of elastic hard spheres or disks,
$\Omega_{H}^{\lambda}$ is the time equilibrium correlation
function
\begin{equation}
\Omega _{H}^{\lambda }\left( s\right) = - (V^{\ast}d)^{-1} \int
d\Gamma ^{\ast }\, {\bm S}^{\ast f} \cdot e^{-s \overline{L}^{\ast
}}\left( {\bm M}^{\ast}_{\lambda } \rho _{GC}^{\ast} \right).
\label{H22}
\end{equation}
The moment ${\bm M}^{\ast}_{\lambda }$ is identified from Eq.\
(\ref{H7}) as
\begin{equation}
{\bm M}^{\ast}_{\lambda }=\sum_{r=1}^{N} {\bm q}_{r}^{\ast }\left(
v_{r}^{\ast 2}-\frac{d}{2}\right). \label{H23}
\end{equation}
The corresponding Green-Kubo representation is
\begin{equation}
\lambda ^{\ast }=\Omega _{0}^{\lambda }+\lim
\int_{0}^{s}ds^{\prime }\, \Omega _{G}^{\lambda }\left( s^{\prime
}\right) .  \label{H24}
\end{equation}
Again, the instantaneous contribution $\Omega_{0}^{\lambda} =
\Omega_{H}^{\lambda}(0)$ can be calculated exactly,
\begin{equation*}
\Omega _{0}^{\lambda }= \frac{\pi^{(d-1)/2} \sigma^{\ast
d+1}}{\sqrt{2} \Gamma (d/2) d}\, g_{e}(\sigma).
\end{equation*}
The analysis of the Green-Kubo flux $\bm{\Upsilon}_{\lambda }$ is
similar to that for the viscosities, leading to the result
\begin{equation}
\Omega _{G}^{\lambda }\left( s\right) = (V^{\ast}d)^{-1}  \int
d\Gamma^{\ast }\, {\bm S}^{\ast f}\cdot e^{-s \overline{L}^{\ast
}}\left( {\bm S} ^{\ast -}\rho _{GC}^{\ast }\right) ,  \label{H25}
\end{equation}
where ${\bm S}^{\ast -}(\Gamma^{\ast})$ is the volume integrated
energy flux for the time reversed microscopic laws identified in
Appendix \ref{ap4} as
\begin{eqnarray}
{\bm S}^{\ast -} (\Gamma^{\ast}) & = & \sum_{r=1}^{N} v^{\ast
2}_{r} {\bm v}^{\ast}_{r} \nonumber \\
& & + \sigma^{\ast} \sum_{r=1}^{N} \sum_{s \neq r}^{N} \delta (
q^{\ast}_{rs} -\sigma^{\ast} ) \Theta \left( \widehat{\bm
q}_{rs}^{\ast} \cdot {\bm g}_{rs}^{\ast} \right) \left( \widehat{\bm
q}_{rs}^{\ast} \cdot {\bm g}_{rs}^{\ast} \right)^{2} (\widehat{\bm
q}_{rs}^{\ast} \cdot {\bm G}^{\ast}_{rs})   \widehat{\bm
q}^{\ast}_{rs}.
\end{eqnarray}

\subsection{$\protect\bm{\mu} $ Coefficient}

It has been mentioned several time that the transport coefficient
$\mu$ vanishes in the elastic case, and it is instructive to see
this at the formally exact level of linear response theory. The
elastic limit of the Helfand form for $\mu^{\ast}$ in Eq.\
(\ref{F20}) is
\begin{equation}
\mu ^{\ast }=-  \frac{d}{2} \lim V^{\ast -1} \int d\Gamma ^{\ast
}\, {\bm S}^{\ast f} \cdot e^{-s \overline{L}^{\ast }}\left( {\bm
M}_{\mu }^{\ast }\rho _{GC}^{\ast }\right),  \label{H26}
\end{equation}
where
\begin{equation}
{\bm M}_{\mu }^{\ast } \equiv \widetilde{\bm{\psi}} ^{\ast
(1)}_{1} = \left[ 1+n_{h}\widetilde{h}\left({\bm 0} \right)
\right]^{-1} \sum_{r=1}^{N}  {\bm q}_{r}^{\ast }. \label{H27}
\end{equation}
This is proportional to the center of mass position of the system.
Its time evolution is, therefore, proportional to the total momentum
${\bm P}^{\ast}$. But, by definition, ${\bm S}^{\ast f}$ is
projected orthogonal to $ {\bm P}^{\ast}$. As, moreover, it is
easily seen that the average on the right hand side of Eq.\
(\ref{H26}) vanishes for $s=0$,  it is concluded that it vanishes
for all times.

\section{Stationary Representation for MD simulations}
\label{ap8} The transformation to dimensionless velocities defined
in the text, is based on scaling relative to $v_{0}(t)\equiv \left[
2 T_{h}(t)/m \right]^{1/2}$. This was done so as to be able to pose
theoretical questions of interest in an elegant self consistent
form. However, this is inconvenient in practice since the cooling
rate is given implicitly in terms of the stationary HCS. Instead,
the same analysis can be performed by scaling with a known function
$\xi (t)$ instead of $v_{0}(t)$ \cite{Diff1}, to get a Liouville
equation in the form
\begin{equation}
\frac{\partial}{\partial t} \rho^{**} (\Gamma^{**},t) -
\frac{\partial \ln \xi (t)}{\partial t} \sum_{r=1}^{N}
\frac{\partial}{\partial {\bm v}^{**}} \cdot \left[ {\bm v}^{**}
\rho^{**}(\Gamma^{**},t) \right]+\frac{\xi(t)}{l}
\overline{L}^{**}(\Gamma^{**}) \rho^{**}(\Gamma^{**},t)=0,
\label{B2.1}
\end{equation}
where  $\Gamma^{**} \equiv \left\{ {\bm q}^{**},{\bm
v}_{r}^{**}\right\} \equiv \left\{ {\bm q}_{r}/l; {\bm v}_{r}/\xi(t)
\right\}$ . Here, as earlier, $l $ is a constant characteristic
length in the system,
\begin{equation}
\rho^{**}(\Gamma^{**},t) \equiv \left[ \ell \chi (t)\right] ^{Nd
}\rho \left( \Gamma ,t\right),  \label{B2.2}
\end{equation}
and
\begin{equation}
\overline{L}^{**} (\Gamma^{**})\equiv  \frac{l}{\xi(t)}
\overline{L}(\Gamma)= \left[ \overline L (\Gamma) \right]_{\Gamma=
\Gamma^{**}}. \label{B2.2.2}
\end{equation}
Next define a new time variable $\tau$ by
\begin{equation}
d \tau =\frac{\xi (t)dt}{l}  \label{B2.3}
\end{equation}
and choose $\xi (t)$ to make the coefficients of the Liouville
equation (\ref{B2.1}) independent of $\tau$, namely verifying
\begin{equation}
\frac{d \xi^{-1}(t)}{d t} = \frac{\xi_{0}}{2 l}
\end{equation}
where $\xi_{0}$ is an arbitrary dimensionless constant that can be
picked for convenience. Then, the choice $\xi(t)=2l/ \xi_{0} t $ is
made and Eq.\ (\ref{B2.3}) becomes
\begin{equation}
d \tau = \frac{2 dt}{\xi_{0}t}. \label{B3}
\end{equation}
The scaled Liouville equation (\ref{B2.1}) takes the form
\begin{equation}
\frac{\partial}{\partial \tau} {\rho}^{**}+\frac{\xi_{0}}{2}
\sum_{r=1}^{N} \frac{\partial}{\partial {\bm v}^{**}_{r}} \cdot (
{\bm v}_{r}^{**} \rho^{**})+\overline{L}^{**}\rho^{**}=0. \label{B4}
\end{equation}
This is formally the same as Eq.\ (\ref{B.2}), except that here the
cooling rate has been replaced by the arbitrary constant $\xi_{0}$.

There is a stationary solution $\rho^{**}_{st}$ to Eq.\ (\ref{B4})
determined  by
\begin{equation}
\overline{\mathcal{L}}^{**} (\Gamma^{**}; \xi_{0}) \rho_{st}^{**}
(\Gamma^{**}; \xi_{0})=0,
\end{equation}
\begin{equation}
\overline{\mathcal{L}}^{**}\left( \Gamma^{**}; \xi_{0} \right)
=\frac{\xi_{0}}{2} \sum_{r=1}^{N} \frac{\partial}{\partial {\bm
v}^{*}_{r}}\cdot {\bm v}^{*}_{r} +\overline{L}^{**}. \label{B5}
\end{equation}
Clearly $\rho^{**}(\Gamma^{**};\xi_{0}) $ is the same function as
$\rho _{h}^{\ast }(\Gamma^{\ast})$ with only the unknown value
$\zeta _{0}^{\ast }$ replaced by the constant $\xi_{0}$ .
Interestingly, it is possible to determine $\zeta _{0}^{\ast }$ from
the chosen value of $\xi_{0}$ and the measured value of the
temperature $T_{st}^{**}$  of the steady state, defined by
\begin{equation}
T_{st}^{**}= \frac{1}{d} \int d \Gamma^{**}\,  v_{1}^{**2}
\rho_{st}^{**} (\Gamma^{**}; \xi_{0}). \label{B5.a}
\end{equation}
The three mentioned quantities are related by
\begin{equation}
\zeta _{0}^{\ast }=\frac{\xi_{0}}{\sqrt{2\tilde{T}_{st}^{**}}}.
\label{B6}
\end{equation}
This relationship may be derived as follows. Define for homogeneous
systems in general the temperature
\begin{equation}
T^{**}(\tau) = \frac{1}{d} \int d \Gamma^{**}\,  v_{1}^{**2}
\rho^{**} (\Gamma^{**},\tau; \xi_{0})= \frac{T(t)}{m \chi^{2}(t)}
\label{B5.aa}
\end{equation}
i.e., the temperature $T^{**}$ is the actual temperature but
expressed in the arbitrary scaling variables. Now consider the
dynamical equation associated with this scaled temperature, that
follows directly from Eq.\ (\ref{B5.a}),
\begin{equation}
\left( \frac{\partial }{\partial \tau}-\xi_{0}\right) T^{**}(\tau)
=-\frac{l \zeta_{0}(t)T^{**}(\tau)}{\xi (t)}. \label{B7}
\end{equation}
Using now the scaling property  of the distribution function of the
HCS  and, therefore, of $\rho^{**} (\Gamma^{**}; \xi_{0})$ this
becomes
\begin{equation}
\left( \frac{\partial }{\partial \tau}-\xi_{0} \right) T^{**} (
\tau)=-\sqrt{2}\zeta _{0}^{\ast } T^{** 3/2}(\tau) \label{B8}
\end{equation}
whose solution is
\begin{equation}
T^{**}(\tau) =\frac{\xi_{0}^{2}}{2\zeta _{0}^{\ast 2}}\left[
1+\left( \frac{\xi_{0}}{\zeta _{0}^{\ast }\sqrt{2 T^{**}(0)}-1}
\right) e^{- \tau \xi_{0}/2} \right]^{-2} ,  \label{B9}
\end{equation}
which in the long time limit goes to Eq.\ (\ref{B6}) above.
Therefore, in practice, one imagines measuring $T^{**}_{st}$ rather
than solving for $\zeta _{0}^{\ast }$ self-consistently in the HCS
state. Also, the different generators defined earlier and the
stationary representation of two-time correlation functions over the
HCS ensemble can be translated into this language of arbitrary
scaling \cite{Diff1,BRyM04}.


\begin{thebibliography}{99}
\bibitem{chemE} Hydrodynamics derived from hard sphere models have found
widespread use in multi-phase computational fluid dynamics methods to
describe numerous industrial processes involving solid particulates in gas
fluidized systems.
\end{thebibliography}
\end{document}